\documentclass[pdflatex,sn-mathphys-num]{sn-jnl}%

\usepackage{graphicx}%
\usepackage{eso-pic}
\usepackage{multirow}%
\usepackage{amsmath,amssymb,amsfonts}%
\usepackage{amsthm}%
\usepackage{array}
\usepackage{mathrsfs}%
\usepackage[title]{appendix}%
\usepackage{xcolor}%
\usepackage{textcomp}%
\usepackage{manyfoot}%
\usepackage[most]{tcolorbox}
\usepackage{booktabs}%
\usepackage{algorithm}%
\usepackage{algorithmicx}%
\usepackage{algpseudocode}%
\usepackage{listings}%
\usepackage{lmodern}
\usepackage{courier}
\usepackage{makecell}
\newtcolorbox{promptbox}[1][]{
  colbacktitle=black!60,
  coltitle=white,
  colback=gray!5,
  fontupper=\footnotesize\small\ttfamily,
  boxsep=5pt,
  left=5pt,
  right=5pt,
  top=5pt,
  bottom=5pt,
  boxrule=1pt,
  title={#1},
  width=0.95\textwidth, %
  enhanced,
  breakable,            %
}

\usepackage{xcolor}
\usepackage{listings}
\colorlet{punct}{red!60!black}
\definecolor{background}{HTML}{EEEEEE}
\definecolor{delim}{RGB}{20,105,176}
\colorlet{numb}{magenta!60!black}
\lstdefinelanguage{json}{
   basicstyle=\normalfont\ttfamily,
   numbers=left,
   numberstyle=\scriptsize,
   stepnumber=1,
   numbersep=8pt,
   showstringspaces=false,
   breaklines=true,
   frame=lines,
   backgroundcolor=\color{background},
   literate=
    {:}{{{\color{punct}{:}}}}{1}
     {,}{{{\color{punct}{,}}}}{1}
     {\{}{{{\color{delim}{\{}}}}{1}
     {\}}{{{\color{delim}{\}}}}}{1}
     {[}{{{\color{delim}{[}}}}{1}
     {]}{{{\color{delim}{]}}}}{1},
}

\usepackage{xcolor}
\usepackage{listings}
\usepackage{tcolorbox}
\tcbuselibrary{skins, breakable, listings} %

\colorlet{punct}{red!60!black}
\definecolor{delim}{RGB}{20,105,176}
\colorlet{numb}{magenta!60!black}

\lstdefinelanguage{json}{
   literate=
    {:}{{{\color{punct}{:}}}}{1}
     {,}{{{\color{punct}{,}}}}{1}
     {\{}{{{\color{delim}{\{}}}}{1}
     {\}}{{{\color{delim}{\}}}}}{1}
     {[}{{{\color{delim}{[}}}}{1}
     {]}{{{\color{delim}{]}}}}{1},
}

\newtcblisting{jsonbox}[1][]{
  colbacktitle=black!60,    %
  coltitle=white,           %
  colback=gray!5,           %
  boxsep=5pt,
  left=18pt,                %
  right=5pt,
  top=5pt,
  bottom=5pt,
  boxrule=1pt,              %
  title={#1},               %
  width=0.95\textwidth,     %
  enhanced,
  breakable,                %
  listing only,             %
  listing options={         %
    language=json,
    basicstyle=\footnotesize\ttfamily, %
    numbers=left,
    numberstyle=\footnotesize\color{gray},
    stepnumber=1,
    numbersep=8pt,
    showstringspaces=false,
    breaklines=true,
    lineskip=0pt          %
  }
}

\theoremstyle{thmstyleone}%

\theoremstyle{thmstyletwo}%

\theoremstyle{thmstylethree}%

\begin{document}
\AddToShipoutPictureFG*{%
  \AtPageUpperLeft{%
    \put(\LenToUnit{\dimexpr1in+\oddsidemargin\relax},-\LenToUnit{32mm}){%
      \makebox(0,0)[lb]{\includegraphics[trim={76.8bp 68.4bp 73.44bp 73.2bp},clip,width=45mm]{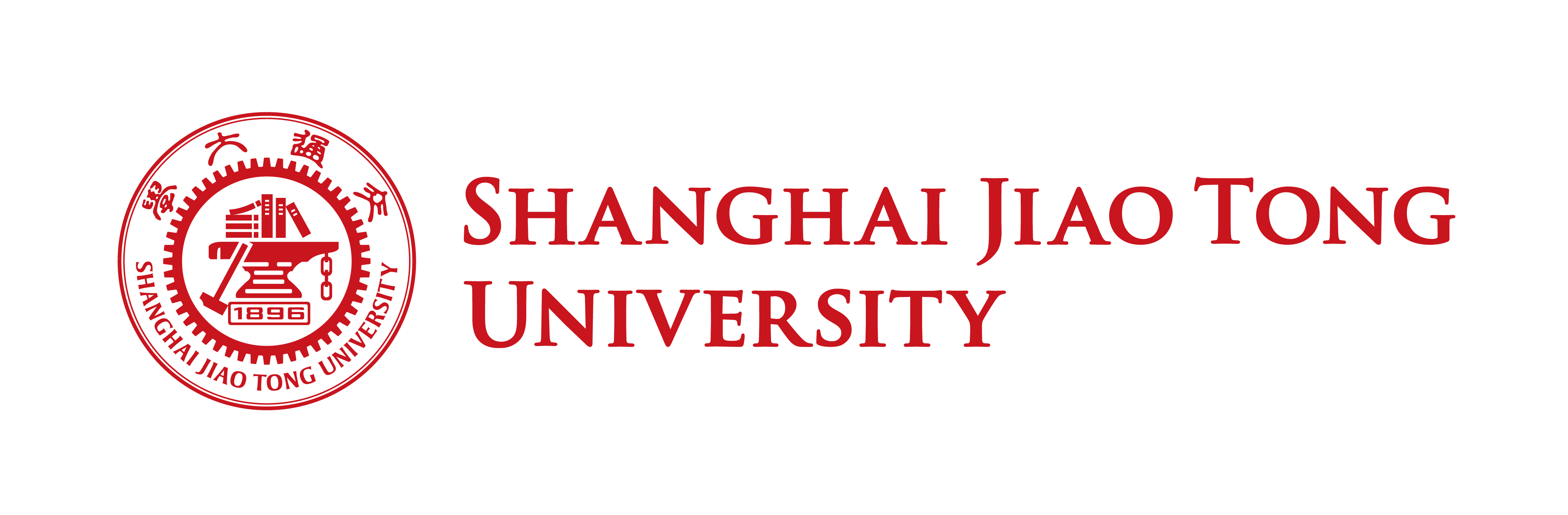}}%
    }%
    \put(\LenToUnit{\dimexpr1in+\oddsidemargin+\textwidth\relax},-\LenToUnit{30mm}){%
      \makebox(0,0)[rb]{\includegraphics[width=35mm]{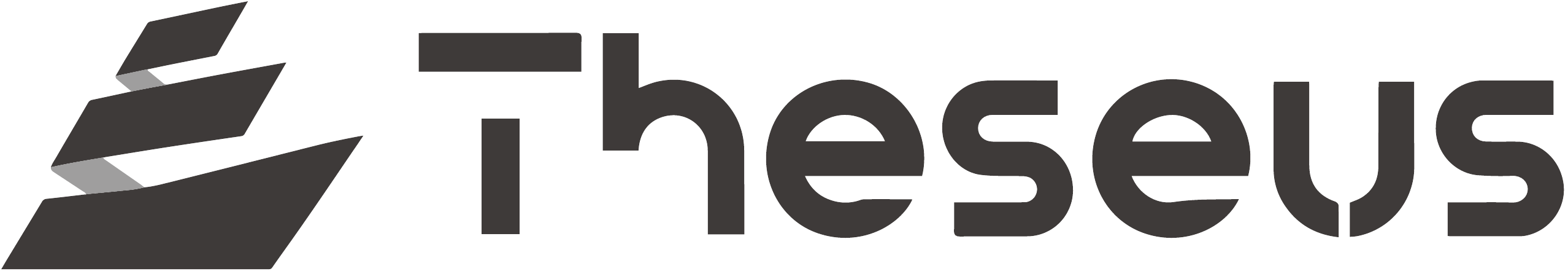}}%
    }%
  }%
}

\title[Article Title]{Generative reasoning emulating expert curation moves protein functional annotation beyond pattern matching at scale}

\author[1,2]{\fnm{Yumou} \sur{Liu}}
\equalcont{These authors contributed equally to this work.}

\author[3]{\fnm{Lingdong} \sur{Shen}}
\equalcont{These authors contributed equally to this work.}

\author[2]{\fnm{Yujun} \sur{Wu}}
\equalcont{These authors contributed equally to this work.}

\author[2]{\fnm{Tongyue} \sur{Xu}}
\equalcont{These authors contributed equally to this work.}

\author[2]{\fnm{Jinhang} \sur{Xu}}
\author[5]{\fnm{Chenkai} \sur{Pan}}
\author[3]{\fnm{Qi} \sur{Liu}}
\author[2]{\fnm{Jiabao} \sur{Pan}}
\author[6]{\fnm{Yijie} \sur{Zhang}}
\author[4]{\fnm{Jingxuan} \sur{Wei}}
\author[5]{\fnm{Siyuan} \sur{Li}}
\author[5]{\fnm{Jintao} \sur{Chen}}

\author*[1]{\fnm{Xuanhe} \sur{Zhou}}\email{zhouxuanhe@sjtu.edu.cn}
\author*[2]{\fnm{Conghui} \sur{He}}\email{heconghui@pjlab.org.cn}
\author*[3]{\fnm{Jinren} \sur{Ni}}\email{nijinren@iee.pku.edu.cn}
\author*[1,2]{\fnm{Cheng} \sur{Tan}}\email{chengtan9907@gmail.com}

\affil[1]{Shanghai Jiao Tong University} 
\affil[2]{Shanghai Artificial Intelligence Laboratory} 
\affil[3]{Peking University} 
\affil[4]{University of Chinese Academy of Sciences}
\affil[5]{Zhejiang University}
\affil[6]{McGill University}

\abstract{
The exponential growth of genomic sequencing and structural prediction has unveiled a vast protein universe, yet functional annotation remains a critical bottleneck. High-fidelity characterization currently relies on expert manual curation to synthesize heterogeneous computational evidence and resolve biological conflicts. Although there are some automated pattern-matching tools, they produce fragmented labels lacking mechanistic context, leaving millions of proteins as uncharacterized functional dark matter. 
Here we ask whether the integrative reasoning performed by expert database curators—weighing heterogeneous evidence, reconciling contradictions, and synthesizing mechanistic explanations—can be emulated computationally at scale. 
We introduce SeqStudio, a generative system that orchestrates sequence homology, domain architecture, three-dimensional structural similarity, and membrane topology within a large language model reasoning framework. Benchmarked against the expert-curated Swiss-Prot database, SeqStudio achieves over 91\% annotation consistency; performance is maintained on a temporal benchmark of 1,057 Swiss-Prot entries first released in 2025. 
Applied to over 20 million proteins from UniProtKB—the majority with minimal or absent functional characterization—SeqStudio elevates median annotation completeness from 1–3 to 5–6 functional fields per protein, transforming fragmented computational signals into coherent mechanistic explanations. Crucially, by leveraging its confidence calibration for an autonomous quality audit, SeqStudio reduces expert annotation costs by 99.6\% and achieves a greater than 2,300-fold speedup compared to traditional manual curation.
The approach further reveals its capacity to weigh cross-modal evidence by functional relevance, reconstruct multi-domain functional logic for complex architectures, and detect remote homologs invisible to sequence-based search. 
These results establish that evidence-grounded generative reasoning can extend expert-level protein annotation to scales inaccessible to manual curation, opening a path toward functional characterization of the sequenced but uncharacterized proteome.
}


\maketitle

\section{Introduction}
\label{sec1}

The exponential growth of genomic sequencing has unveiled a vast protein universe, yet our capacity to decipher its functional landscape remains a critical bottleneck. While deep learning has effectively resolved the protein structure prediction problem to near-experimental accuracy~\cite{jumper2021highly,baek2021accurate,jang2024accurate}, the functional annotations for the vast majority remain fragmentary or absent~\cite{unsal2022learning,durairaj2023uncovering}. Crucially, this disparity is no longer primarily a problem of insufficient data or inadequate computational tools. Sequence homology search, domain databases, structural comparisons, and rule-based systems now provide multiple, overlapping lines of molecular evidence for a large and growing fraction of sequences in UniProtKB/TrEMBL~\cite{boutet2007uniprotkb}. Rather, the fundamental bottleneck lies in cognitive integration: current automated methods detect isolated signals effectively but lack the capacity to reconcile heterogeneous, sometimes contradictory inputs into coherent mechanistic interpretations~\cite{jeffery2023current,bonetta2020machine}. Bridging this gap has historically required the integrative reasoning of expert manual curation—a process that produces the trusted, high-fidelity annotations of Swiss-Prot~\cite{uniprot2025uniprot} but cannot scale to the expanding sequence space. Consequently, over 200 million proteins in TrEMBL remain without coherent functional characterization~\cite{varadi2022alphafold,escudeiro2022functional}, limiting their utility in hypothesis-driven biomedical research and rational protein engineering.

For decades, automated annotation has been dominated by pattern-matching paradigms. Foundational tools such as BLAST~\cite{blast} and HMMER~\cite{finn2011hmmer} retrieve evolutionary homologs through sequence similarity, while InterProScan~\cite{jones2014interproscan} maps conserved domain signatures and functional motifs. Rule-based systems, including UniRule~\cite{unirule} and ARBA~\cite{saidi2017rule}, then propagate annotations downstream through rigid, expert-defined logic gates~\cite{kilinc2023improved,bonetta2020machine}. These methods are effective when query sequences share high identity with experimentally characterized references, but their outputs become increasingly fragmented and unreliable as similarity decreases~\cite{durairaj2023uncovering}. A typical output is a disjointed list of domain identifiers and ontology terms—individual signals that are correct in isolation but lack the mechanistic context of how these parts interact within a functioning molecule~\cite{jeffery2023current,bonetta2020machine}. This limitation becomes acute at the boundaries of existing knowledge. When sequence identity falls into the evolutionary twilight zone, or when sequence homology implies one protein family while structural comparison supports another, rule-based systems have no mechanism to adjudicate. Consequently, they default to vague labels or abstain entirely. Resolving these ambiguities currently requires a form of judgment that no existing pipeline provides computationally. 

In parallel, supervised deep learning methods have moved beyond explicit pattern matching by learning to predict Gene Ontology terms directly from protein sequences or structures~\cite{kulmanov2020deepgoplus,gligorijevic2021structure,you2019netgo}. More recently, protein language models pre-trained on large sequence databases~\cite{elnaggar2021prottrans,rives2021biological,lin2023evolutionary} have yielded learned representations that further improve GO term prediction on established benchmarks. These approaches have measurably advanced the state of the art in community benchmarks such as CAFA~\cite{zhou2019cafa,jiang2016expanded}, yet their integration into large-scale database curation pipelines such as UniProt's has remained limited~\cite{uniprot2025uniprot}. A central reason is that their output, ranked lists of ontology labels, addresses the classification dimension of annotation but does not provide the mechanistic narratives, evidence reconciliation, or explanatory context that characterize expert curation and that are essential for downstream applications such as hypothesis generation and protein engineering. Taken together, these developments suggest that the persistent bottleneck in functional annotation lies not in the accuracy of individual label predictions. Rather, it lies in the integrative reasoning needed to synthesize disparate signals into coherent functional narratives—\textbf{the very capability that distinguishes expert curation from automated classification}.

In practice, this integrative reasoning is performed by a small community of expert database curators. When annotating a protein for a reference database such as UniProtKB/Swiss-Prot~\cite{uniprot2025uniprot}, a curator does not rely on any single tool. Instead, they orchestrate multiple lines of evidence—sequence homology, domain architecture, structural fold, membrane topology, and genomic context—and then weigh the relative reliability of each, reconcile contradictions, and synthesize a coherent mechanistic explanation of how the protein operates within its biological setting. This expert reasoning process is what elevates Swiss-Prot from a collection of computational hits to a trusted, manually reviewed knowledge base. However, expert curation is inherently limited by human throughput: Swiss-Prot currently contains approximately 570,000 reviewed entries, whereas TrEMBL holds over 200 million unreviewed sequences~\cite{uniprot2025uniprot}. This gap of roughly three orders of magnitude quantifies the central tension in functional genomics: \textbf{the reasoning that produces reliable annotation cannot scale, and the automation that scales cannot reason}.

The rapid maturation of large language models (LLMs)~\cite{achiam2023gpt,singh2025openaigpt5card,yang2025qwen3,liu2024deepseek} offers a potential path toward resolving this tension. Trained on large-scale text corpora that encompass substantial scientific literature, LLMs encode broad biochemical knowledge and exhibit capabilities that parallel key aspects of the curator's workflow: weighing heterogeneous inputs through extended-context reasoning, reconciling contradictions via multi-step inference, and synthesizing explanations in structured natural language~\cite{brown2020language,wei2022chain,lewis2020retrieval}. However, applying generic LLMs directly to protein annotation risks hallucination, particularly when the model operates without explicit physical or evolutionary constraints~\cite{kaminski2023plm,ji2023survey,huo2024multi}. One established strategy to mitigate this risk is to ground language models in the outputs of domain-specific computational tools~\cite{boiko2023autonomous,m2024augmenting}, an approach that has shown promise in several areas of the biological sciences~\cite{huang2025biomni,gao2025democratizing,ma2026toolsgenie}. However, for protein functional annotation, no systematic assessment has established whether tool-grounded language models can match the consistency of expert curation rather than merely produce plausible-sounding outputs. Without such evidence, it remains unclear whether these approaches offer genuine integrative reasoning or merely fluent reformulation of retrieved signals. The question we address, therefore, is not whether language models should replace established bioinformatics tools, but \textbf{whether an LLM, when grounded in the outputs of these tools, can emulate the integrative reasoning of expert curators at an unprecedented scale}.

\begin{figure}[!ht]
  \centering
  \includegraphics[width=\linewidth]{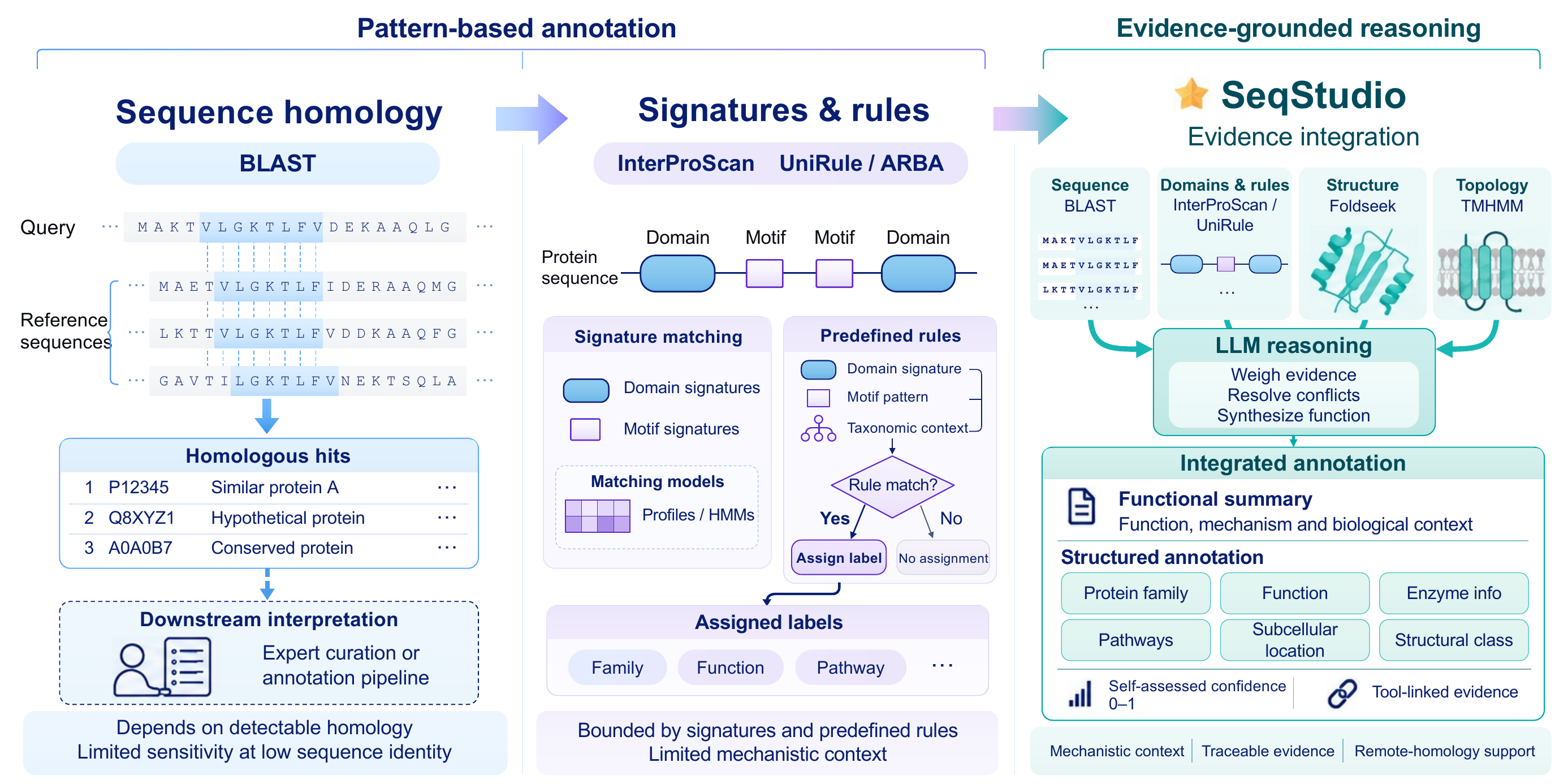}
  \caption{\textbf{From pattern-based annotation to evidence-grounded reasoning.} Left, BLAST retrieves homologous sequences for downstream interpretation by expert curators or annotation pipelines. Middle, InterProScan identifies domain and motif signatures, while UniRule/ARBA apply predefined rules to assign labels when their conditions are met. Right, SeqStudio integrates sequence homology, domains and rules, structural similarity, and membrane topology to generate a functional summary and six-field structured annotation with self-assessed confidence and tool-linked evidence. The schematic contrasts the roles of retrieval, rule-based assignment, and evidence synthesis; it does not imply that downstream curation is absent from conventional workflows.}
  \label{fig:teaser}
\end{figure}

Here we report an empirical investigation of this question. We introduce SeqStudio, a generative annotation system designed to operationalize the expert curation workflow described above. SeqStudio orchestrates the same classes of evidence that curators rely on—sequence homology (BLAST~\cite{blast}), domain architecture and rule-based assignments (InterProScan~\cite{jones2014interproscan} and UniRule~\cite{unirule}), three-dimensional structural similarity (Foldseek~\cite{van2024fast}), and membrane topology (TMHMM~\cite{hallgren2022deeptmhmm})—and integrates their outputs within an LLM-driven reasoning framework. Specifically, we ask whether this emulation can approach the consistency of expert curation, whether it maintains performance on newly released Swiss-Prot entries, and what functional patterns emerge when the system is applied at a scale no manual effort could reach.

Compared with the Swiss-Prot database curated by experts, SeqStudio achieves a consistency of annotation of over 91\% in functional dimensions, and ablation analysis further reveals that each class of molecular evidence makes distinct and complementary contributions to different aspects of functional prediction. Critically, this performance is maintained on a temporal benchmark of 1,057 Swiss-Prot entries first released in 2025, supporting its applicability to newly released entries without establishing their absence from model training corpora. Applied to millions of minimally annotated TrEMBL sequences, SeqStudio elevates the median completeness of annotation from 1–3 to 5–6 functional fields per protein, transforming fragmentary computational signals into coherent, multi-dimensional functional profiles. Beyond aggregate enrichment, the system demonstrates its capacity to weigh and synthesize evidence across analytical modalities, reconstruct multi-domain functional logic for complex architectures, and detect remote homologs that elude sequence-based search. Furthermore, we demonstrate that this reasoning-driven framework provides a highly cost-efficient solution for scaling annotation. By establishing an autonomous human-in-the-loop triage system based on predictive confidence, SeqStudio accelerates the annotation workflow by over 2,300-fold and reduces expert curation costs by 99.6\% without sacrificing fidelity. Taken together, these results provide empirical evidence that the integrative reasoning historically confined to human experts can be computationally emulated and extended at scale.

\section{Results}\label{sec2}

\subsection{Emulated reasoning achieves curator-level consistency on the Swiss-Prot gold standard}
\label{sec:swissprot}

\subsubsection{Systematic evaluation across taxonomic groups and functions}
\label{subsec:systematic_evalution}

To assess whether emulated reasoning can achieve curator-level annotation quality, we systematically annotated the entire Swiss-Prot database~\cite{uniprot2025uniprot} and compared the outputs against its expert-curated records. To prevent trivial self-retrieval, each query sequence was excluded from its own search results during all evidence-gathering steps. The evaluation examines annotation completeness across taxonomic groups, confidence behavior across functional categories, predictive consistency at both the taxonomic and functional-dimension level, the relationship between predicted confidence and empirical accuracy, and semantic alignment between generated and expert narratives.

\begin{figure}[!htbp]
  \centering
  \vspace{-4mm}
  \includegraphics[width=\linewidth]{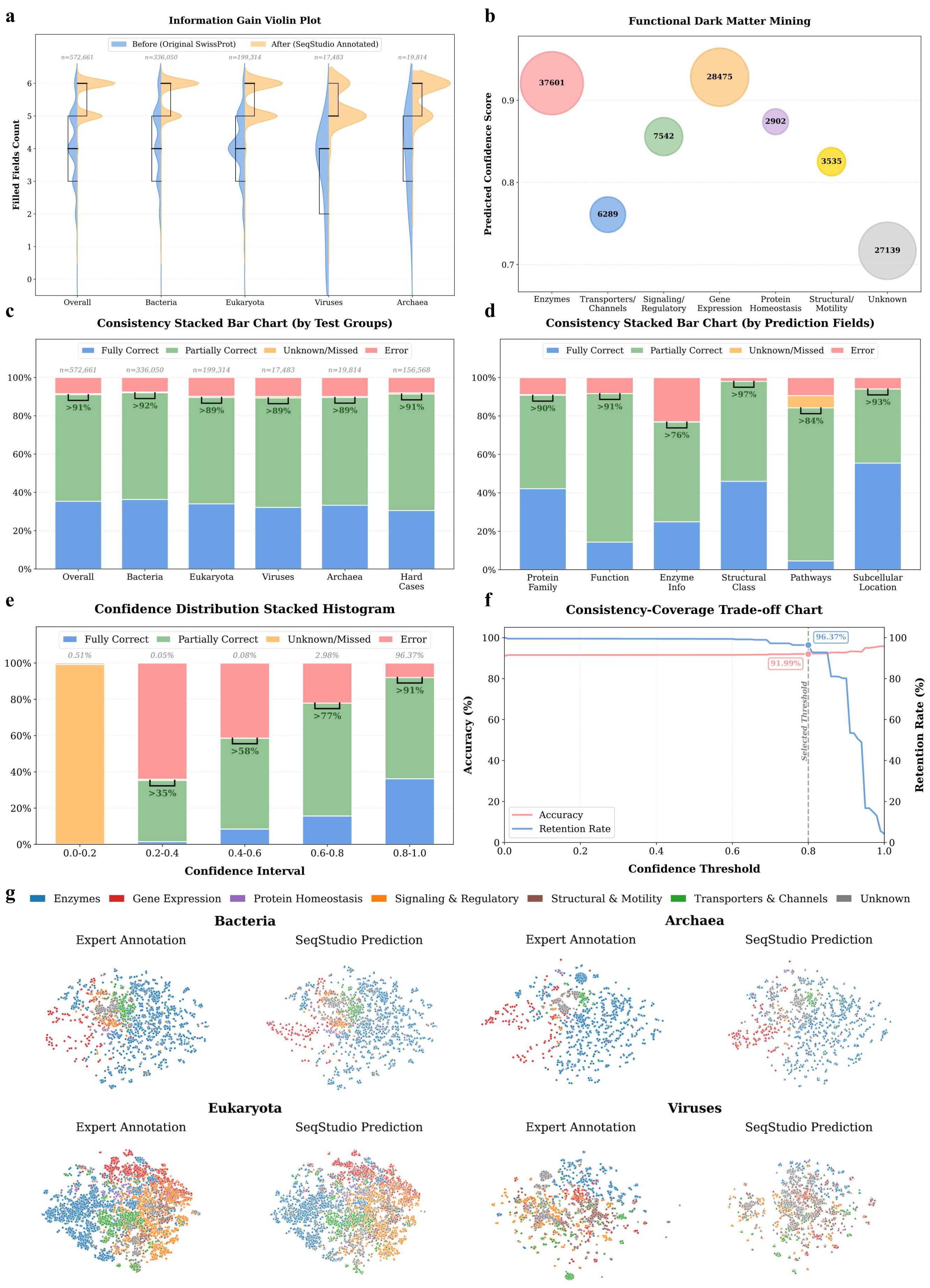}
  \vspace{-4mm}
  \caption{\textbf{SeqStudio performance evaluation on Swiss-Prot.}
    \textbf{a}, Distribution of filled functional fields per protein before (Swiss-Prot records) and after (SeqStudio annotation), stratified by taxonomic group.
    \textbf{b}, Functional category distribution of newly annotated proteins (bubble size: count; vertical axis: average confidence).
    \textbf{c-e}, Consistency by test groups, prediction fields, and confidence intervals.
    \textbf{f}, Consistency-coverage trade-off versus confidence threshold.
    \textbf{g}, t-SNE visualization comparing expert and SeqStudio annotations.
  }
  \label{fig:swissprot_bench}
\end{figure}

\textbf{Annotation completeness increases consistently across all taxonomic groups.} We quantified completeness as the number of filled functional fields per protein across six dimensions: Protein Family, Function, Enzyme Info, Structural Class, Pathways, and Subcellular Location. Baseline Swiss-Prot entries predominantly cluster between 3 and 4 filled fields (Figure~\ref{fig:swissprot_bench}a), with Viruses and Archaea exhibiting the most severe gaps—median counts of approximately 3, reflecting the limited experimental characterization available for these domains. Following SeqStudio annotation, all taxonomic groups exhibit a distinct upward shift in information density, with median field counts increasing by 1–2 units and consistently stabilizing above 5 out of 6 possible fields. The effect is most pronounced for taxonomic groups with the sparsest baseline records, suggesting that the system derives functional insights from primary computational evidence rather than merely propagating existing database metadata.

\textbf{Functional role assignments exhibit category-dependent confidence that separates reliable predictions from genuine ambiguity.} Across functional categories (Figure~\ref{fig:swissprot_bench}b), 37,601 enzymes and 28,475 gene expression regulators receive the highest average confidence $>0.92$, consistent with the rich homology and domain signatures available for these well-characterized classes. Membrane transporters (6,289) show moderate confidence ($\sim$0.76), likely reflecting the inherent complexity of multi-pass transmembrane topology. Notably, proteins assigned to the 'Unknown' category cluster at distinctly lower confidence ($\sim$0.71) indicating that the system withholds high-confidence labels when evidence is insufficient rather than forcing a classification. This separation suggests that predicted confidence can serve as a practical triage criterion, flagging genuinely ambiguous cases for expert review.

\textbf{Consistency exceeds 91\% across taxonomic groups and remains robust for sparsely annotated proteins.} We define consistency as the fraction of predicted fields rated as either fully correct or partially correct when compared against the corresponding Swiss-Prot record (see Appendix \ref{app:swissprot_setup}). Overall consistency with expert curation exceeds 91\% and remains above 89\% across Bacteria, Eukaryota, Viruses, and Archaea (Figure~\ref{fig:swissprot_bench}c). A particularly informative test comes from ``hard cases''—the 130,598 proteins whose ground-truth records contain three or fewer filled fields—where consistency still exceeds 91\%. Because these entries offer minimal surface-level metadata to echo, this result provides evidence that annotation quality derives from evidence synthesis rather than replication of existing database content.

\textbf{Performance varies systematically across functional dimensions, mirroring the known difficulty gradient in manual curation.} Structural Class and Subcellular Location achieve the highest consistency (Figure~\ref{fig:swissprot_bench}d), likely because structural similarity search (Foldseek) and topology prediction (TMHMM) provide strong, direct evidence for these fields. Enzyme Info presents greater difficulty ($>76\%$), consistent with the challenge of inferring catalytic mechanism from sequence and structure alone. This dimension-specific pattern mirrors the known difficulty gradient in manual curation, where enzymatic activity often requires experimental confirmation. The confidence distribution further supports the reliability of these results (Figure~\ref{fig:swissprot_bench}e): 96.37\% of predictions fall in the highest confidence interval (0.8–1.0), where empirical consistency remains above 91\%. Low-confidence predictions concentrate among 'Unknown', reinforcing the pattern observed in Figure~\ref{fig:swissprot_bench}b that the system reserves low confidence for genuinely uncertain cases rather than distributing errors uniformly.

\textbf{Predicted confidence correlates monotonically with empirical consistency, enabling principled quality filtering.} Even without confidence filtering, baseline consistency exceeds 90\% (Figure~\ref{fig:swissprot_bench}f), and consistency increases monotonically with the confidence threshold. Critically, the retention curve remains nearly flat from 0 to 0.8, then drops steeply—indicating that the vast majority of predictions cluster at high confidence, with relatively few occupying the intermediate range. This distribution shape means that moderate filtering eliminates a small number of uncertain predictions while preserving the bulk of the annotation volume. At the 0.8 cutoff, 96.37\% of annotations are preserved at 91.99\% consistency. We therefore adopt the 0.8 threshold for all subsequent large-scale analyses, as it represents the elbow point beyond which further filtering yields diminishing consistency gains at substantial coverage cost. Over half of all predictions (approximately 55\%) exceed 0.9 confidence, placing them well within the high-reliability regime.

\textbf{Generated descriptions cluster with expert annotations in semantic embedding space.} To assess whether the generated descriptions capture the same functional distinctions as expert annotations—beyond surface-level keyword overlap—we projected both into a shared semantic embedding space using Qwen3-Embedding~\cite{qwen3embedding} and visualized with t-SNE~\cite{van2008visualizing} (Figure~\ref{fig:swissprot_bench}g). Generated summaries (triangles) and Swiss-Prot records (circles) consistently co-cluster by functional category across all taxonomic groups. This co-localization indicates that the generated text encodes similar functional semantics to expert narratives, rather than producing superficially plausible but semantically divergent descriptions. 

\subsubsection{Performance on Swiss-Prot entries first released in 2025}
\label{sec:ablation2}

A central concern for any LLM-based system evaluated on curated databases is that high performance may reflect memorization of training data rather than genuine reasoning. The backbone LLM may have encountered Swiss-Prot records or related biological knowledge during training, so accuracy on established entries cannot entirely distinguish reasoning from recall. To assess performance on newly released records, we constructed a temporal benchmark comprising 1,057 Swiss-Prot entries first released in 2025. Entry creation dates do not establish whether the underlying sequences or related annotations appeared in other sources before model training; this benchmark therefore does not by itself rule out memorization.

\begin{figure}[h]
  \centering
  \vspace{-2mm}
  \includegraphics[width=\linewidth]{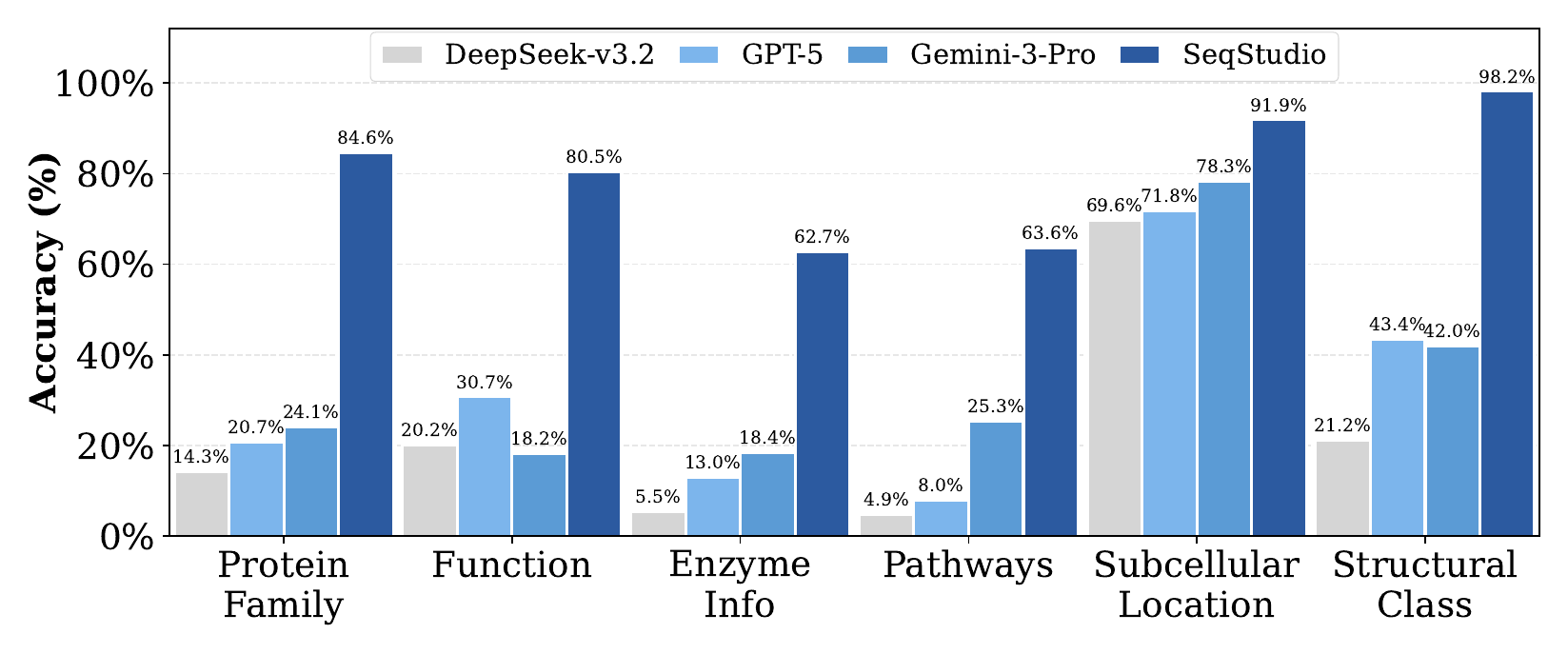}
  \vspace{-4mm}
  \caption{\textbf{Evaluation on Swiss-Prot entries first released in 2025.} Accuracy of SeqStudio and three frontier foundation models across six functional dimensions on 1,057 Swiss-Prot entries first released in 2025. SeqStudio's per-dimension accuracy on this 2025 subset closely tracks its full-database performance (Section~\ref{subsec:systematic_evalution}), while LLMs operating without tool augmentation remain substantially below curator-level consistency.}
  \label{fig:ablation2}
\end{figure}

Across all six functional dimensions, SeqStudio's consistency on the 2025 subset closely tracks its performance on the full Swiss-Prot benchmark: Structural Class reaches 98.2\% (full database: $\sim$97\%), Protein Family 84.6\% (full database: $\sim$90\%), and Subcellular Location 91.9\% (full database: $\sim$93\%). These results show that the observed performance extends to this set of newly released Swiss-Prot entries, although they do not isolate the contribution of prior training-data exposure. To further contextualize these findings, we evaluated three frontier LLMs including GPT-5, Gemini-3-Pro, and DeepSeek-v3.2, providing only the raw amino acid sequence as input. Without access to evidence from external bioinformatics tools, even the strongest model GPT-5 achieved only 24.1\% on Protein Family and 18.4\% on Enzyme Info (Figure~\ref{fig:ablation2}). These results show that sequence-only LLM prediction under the tested settings falls well below SeqStudio on the same entries. This comparison supports the utility of tool-derived evidence but does not establish the absence of training-data exposure.

\subsection{Anatomy of the reasoning process}

\subsubsection{Each evidence modality makes distinct contributions to different functional dimensions}

\begin{figure}[!htbp]
  \centering
  \includegraphics[width=0.94\linewidth]{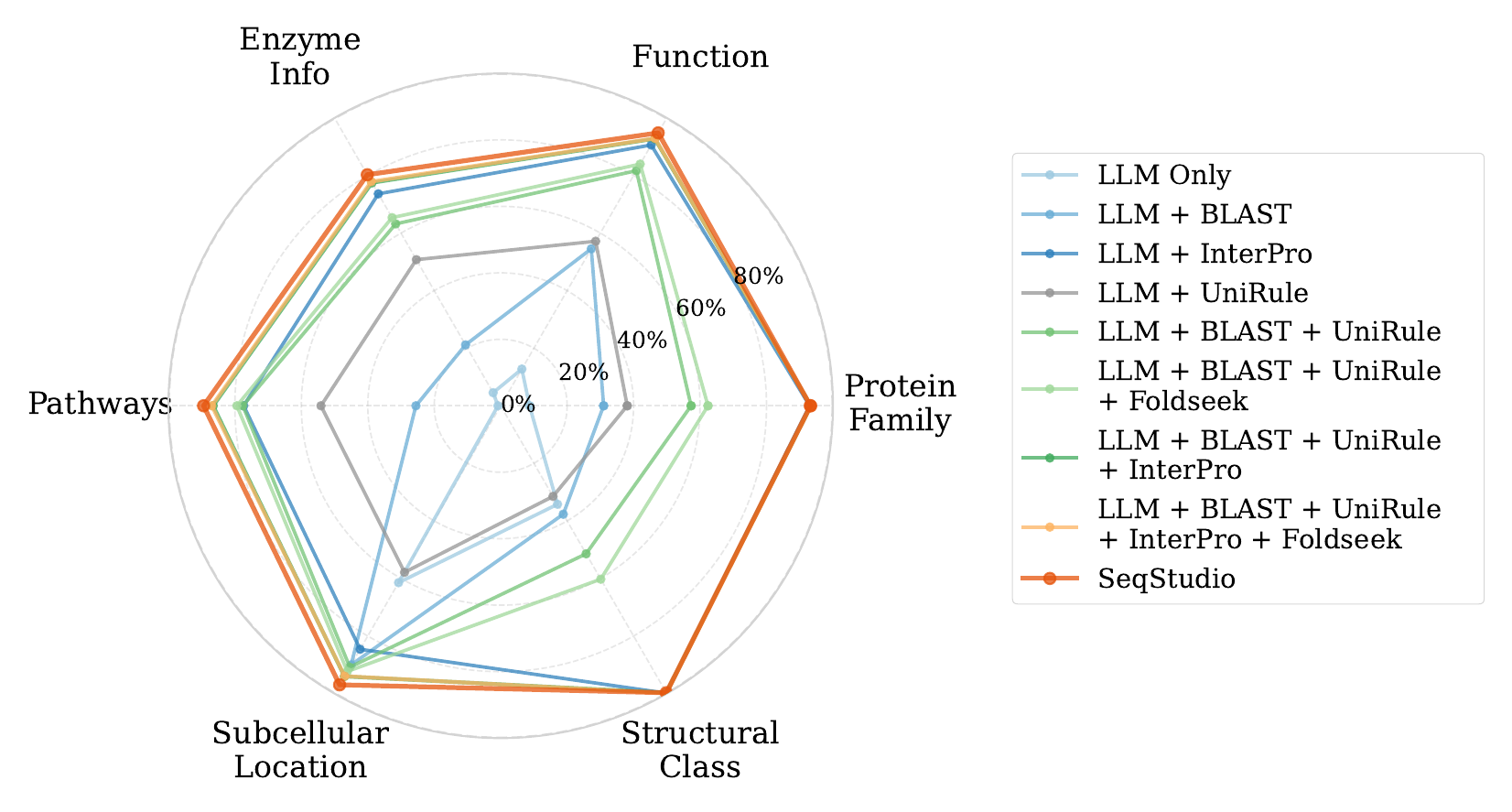}
  \caption{\textbf{Ablation analysis reveals asymmetric contributions of evidence modalities across functional dimensions.} Radar chart showing annotation consistency as bioinformatics modules are progressively added to a naive LLM baseline. InterProScan alone accounts for the majority of performance on Protein Family and Structural Class, while Enzyme Info and Pathways require the combination of multiple tools. Structure-based tools (Foldseek, TMHMM) provide targeted improvements on Subcellular Location and Structural Class.}
  \label{fig:ablation1}
\end{figure}

To determine whether curator-level consistency arises from a single dominant evidence source or requires the integration of multiple modalities, we performed a systematic ablation analysis. Starting from a naive LLM baseline exemplified by DeepSeek-V3.2 that receives only the protein sequence, we progressively incorporated bioinformatics evidence sources and measured consistency across functional dimensions (Figure~\ref{fig:ablation1}). All configurations described below include the LLM as the reasoning backbone; \textit{LLM + BLAST only}, for instance, denotes the LLM operating with BLAST outputs as its sole external evidence.

\textbf{Parametric knowledge alone is near-absent for mechanistic annotation.} The most striking initial finding is the severity of the LLM-only baseline's failure on mechanistic dimensions: 4.54\% on Enzyme Info and 0.82\% on Pathways, indicating that parametric knowledge contains almost no actionable information about catalytic mechanisms or metabolic context. The sole partial exception is Subcellular Location (61.45\%), suggesting that localization-related associations are comparatively well-represented in biomedical training corpora, likely because subcellular compartment names co-occur frequently with protein names in the literature. These results provide direct quantitative evidence that LLM-based annotation without external tool grounding is fundamentally insufficient for protein functional characterization.

\textbf{InterProScan accounts for the majority of performance on identity- and architecture-defined dimensions.} Domain and motif signatures from InterProScan dominate both Protein Family and Structural Class. The near-saturation effect has a clear biological rationale: conserved domain signatures are definitional for protein family membership and structural architecture. A protein is classified as a kinase because it contains a protein kinase domain. Consequently, additional evidence sources like BLAST homology provide less than 1 percentage point of marginal improvement on these two dimensions once InterProScan is included. This finding carries a practical implication: for large-scale family classification and structural typing, domain annotation alone approaches the informational ceiling.

\textbf{Mechanistic dimensions such as Enzyme Info and Pathways require progressive accumulation of multiple evidence streams. } In sharp contrast to the single-source dominance observed above, no individual tool achieves more than 77\% consistency on Enzyme Info or 78\% on Pathways. Performance on these dimensions increases gradually with each added evidence layer: from LLM-only (4.54\% / 0.82\%) through BLAST+UniRule (63.19\% / 77.60\%) to the full configuration (80.27\% / 89.41\%). This gradual accumulation pattern reflects the biological reality that enzymatic mechanism and pathway membership cannot be inferred from any single molecular feature. Catalytic activity depends on active-site geometry captured partially by domain signatures, evolutionary conservation of catalytic residues captured by homology search, and metabolic context captured by rule-based systems. None of them is sufficient alone. The observation that Enzyme Info remains the lowest-performing dimension even in the full configuration (80.27\%) mirrors the known difficulty gradient in expert curation, where enzymatic activity assignment frequently requires experimental confirmation beyond computational inference.

\textbf{Evidence sources interact non-additively, with both synergistic and saturating regimes.} The ablation data reveal that the contributions of individual tools are not simply additive. On Pathways, BLAST alone achieves 25.51\% and UniRule alone 54.09\%, yet their combination reaches 77.60\%. A synergistic interaction wherein homology-derived pathway candidates and rule-based classification logic mutually constrain each other. Conversely, on Protein Family, InterProScan alone (92.98\%) already saturates the performance space; adding it to a BLAST+UniRule base (57.30\%) yields 93.37\%, with virtually no residual contribution from the preceding tools. These contrasting interaction patterns—synergistic for integrative dimensions and saturating for definitional dimensions—suggest that the value of multi-source integration is concentrated precisely where no single evidence type is definitionally sufficient.

\textbf{Structure-based and topology tools provide targeted, dimension-specific contributions.} Foldseek and TMHMM, while contributing modestly to aggregate performance, provide non-redundant information on specific dimensions. Foldseek improves Structural Class from 51.43\% to 60.24\% in the BLAST+UniRule context, consistent with the unique capacity of three-dimensional fold comparison to resolve structural classification when sequence-based evidence is ambiguous. TMHMM, incorporated in the final step (SeqStudio Full vs. the preceding four-tool configuration), contributes a 3.0\% improvement on Subcellular Location (93.94\% to 96.94\%), reflecting the relationship between transmembrane helix topology and protein localization. 

\subsubsection{Context-dependent weighting and synthesis of cross-modal evidence}

\begin{figure}[htbp]
    \centering
    \includegraphics[width=\linewidth]{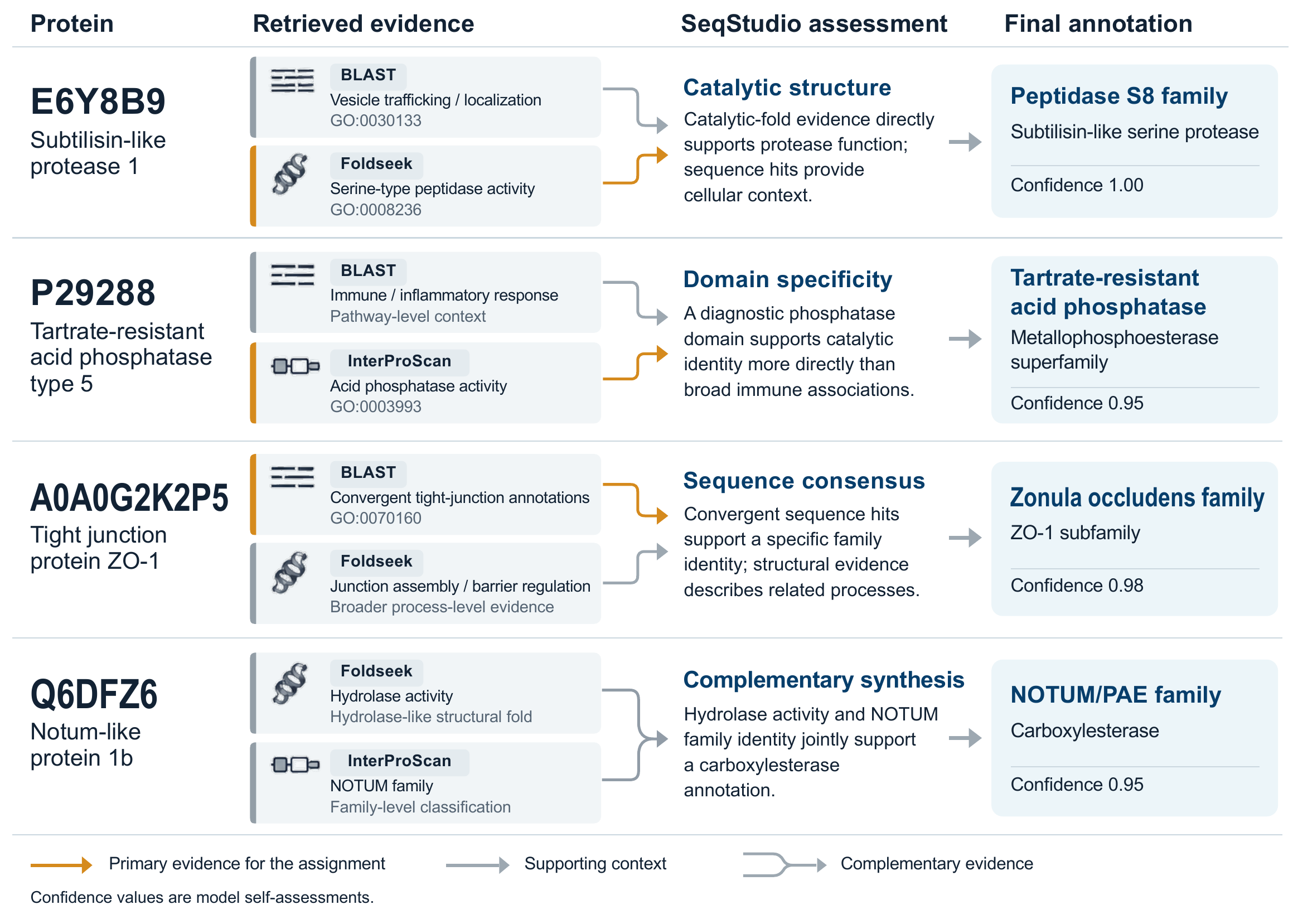}
    \caption{\textbf{Context-dependent evidence weighting in four protein annotations.} Each row links the protein, retrieved evidence, SeqStudio assessment, and final annotation. E6Y8B9 emphasizes a catalytic structural match; P29288 emphasizes a diagnostic phosphatase domain; A0A0G2K2P5 emphasizes convergent sequence-based family assignments; and Q6DFZ6 combines structural activity evidence with a domain-based family classification. Orange arrows indicate the primary evidence for the displayed assignment, gray arrows indicate supporting context, and merging gray arrows indicate complementary evidence. Confidence values are model self-assessments, not independently measured probabilities of correctness.}
    \label{fig:case}
\end{figure}

Sequence homology, structural similarity, and domain signatures describe different aspects of a protein, so their outputs can differ in specificity without being mutually contradictory. We examined four representative cases to understand how SeqStudio weighs evidence for the annotation being assigned (Figure~\ref{fig:case}). The cases illustrate catalytic structural evidence, domain specificity, sequence consensus, and complementary synthesis. They provide qualitative examples of evidence use rather than a controlled test of a universal ranking among tools.

\textbf{Catalytic structural evidence supports the molecular-function assignment.} For subtilisin-like protease 1 (E6Y8B9), BLAST retrieves annotations associated with vesicle trafficking and localization (GO:0030133), whereas Foldseek retrieves serine-type peptidase activity annotations (GO:0008236) from structural matches. SeqStudio uses the catalytic-fold evidence to support a Peptidase S8 family assignment and a subtilisin-like serine protease annotation (self-assessed confidence: 1.00). The sequence-derived annotations provide cellular context and need not conflict with protease activity. Structural similarity supports this functional inference but does not by itself establish catalytic activity experimentally.

\textbf{Domain specificity supports catalytic identity.} For tartrate-resistant acid phosphatase type 5 (P29288), BLAST hits provide immune and inflammatory response context, while InterProScan supplies acid phosphatase activity evidence (GO:0003993) and metallophosphoesterase superfamily membership. SeqStudio prioritizes the diagnostic domain for the acid phosphatase assignment (self-assessed confidence: 0.95), while retaining the broader biological context. The distinction is between the specificity of the evidence for catalytic identity and its relevance to a pathway, rather than a contradiction between the two descriptions.

\textbf{Sequence consensus supports a specific family assignment.} For tight junction protein ZO-1 (A0A0G2K2P5), BLAST hits converge on tight-junction annotations (GO:0070160), while Foldseek-derived annotations describe junction assembly and barrier regulation. SeqStudio uses the convergent sequence evidence to support the Zonula occludens family and ZO-1 subfamily assignment (self-assessed confidence: 0.98). The structural evidence supplies related process-level context, illustrating that sequence evidence can be primary when it is more specific to the family being assigned.

\textbf{Complementary evidence supports a joint annotation.} For Notum-like protein 1b (Q6DFZ6), Foldseek supplies hydrolase activity evidence and a hydrolase-like structural fold, while InterProScan assigns the NOTUM family. SeqStudio combines the activity and family evidence into a NOTUM/PAE family, carboxylesterase annotation (self-assessed confidence: 0.95). Neither evidence source is designated as uniquely primary in this case.

Across these examples, evidence is weighted according to its \textit{specificity and relevance} to the functional dimension being assigned. Structural or domain evidence supports catalytic identity, sequence consensus supports a specific family assignment, and complementary signals are synthesized. The displayed assessments therefore illustrate context-dependent evidence use, without establishing a fixed tool hierarchy or treating model confidence as independent validation.

\subsection{Scaling annotation to millions of uncharacterized sequences}

\subsubsection{Large-scale functional reconstruction of TrEMBL}
\label{sec:trembl}

\begin{figure}[!htbp]
  \centering
  \includegraphics[width=0.8\linewidth]{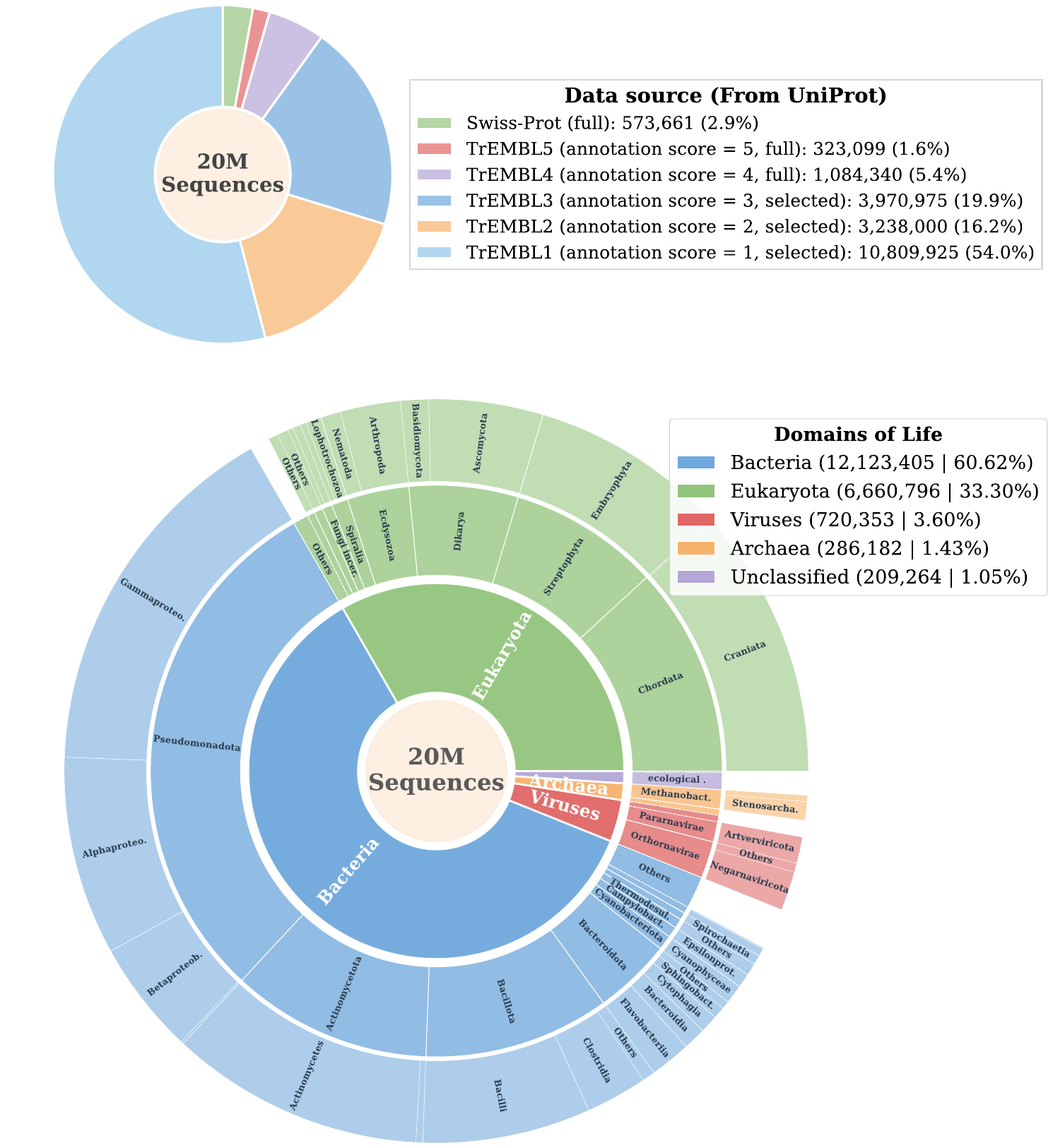}
  \caption{\textbf{Composition and Taxonomic Diversity of the 20M Proteins From UniProt.}
    \textbf{a}, Distribution of data sources. The dataset integrates the entire Swiss-Prot database with 19.42 million TrEMBL sequences, the majority of which possess low annotation scores (1–3).
    \textbf{b}, Taxonomic breadth across the Domains of Life. The 20M sequences encompass Bacteria (60.62\%), Eukaryota (33.30\%), Viruses (3.60\%), and Archaea (1.43\%), representing a comprehensive functional map from diverse evolutionary lineages.
  }
  \label{fig:trembl1}
\end{figure}

To test whether curator-level reasoning scales beyond the gold-standard corpus, we applied SeqStudio to 20 million protein sequences from UniProtKB/TrEMBL. We assessed three aspects of scaling behavior: whether annotation completeness gains observed on Swiss-Prot persist at this scale, whether confidence calibration transfers from the gold-standard benchmark, and whether performance varies across functional dimensions in ways that reveal intrinsic reasoning boundaries. The dataset comprises the entire Swiss-Prot database (575,661 reviewed entries, 2.9\%) and 19.42 million unreviewed TrEMBL sequences (Figure~\ref{fig:trembl1}a). The TrEMBL subset was selected to represent the annotation deficit of current databases: 80.1\% of these sequences carry annotation scores of 1--3, indicating minimal or absent functional characterization. The dataset spans all domains of life, with Bacteria (60.62\%), Eukaryota (33.30\%), Viruses (3.60\%), and Archaea (1.43\%) represented (Figure~\ref{fig:trembl1}b).

\begin{figure}[!htbp]
  \centering
  \includegraphics[width=0.9\linewidth]{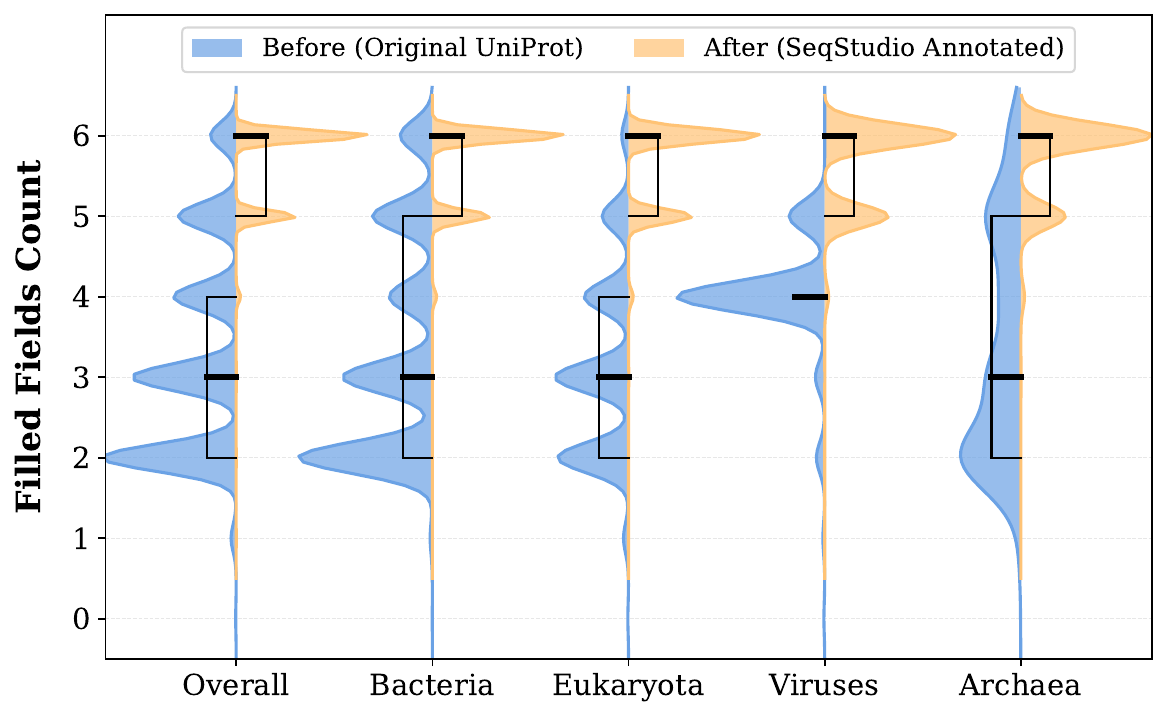}
  \caption{\textbf{Enrichment of functional information density across the 20M dataset.} Comparison of the number of filled functional fields per protein before and after SeqStudio annotation. The analysis covers the entire dataset and is broken down into major taxonomic groups: Bacteria, Eukaryota, Viruses, and Archaea. Across all taxonomic groups, the distribution shifts from 1–3 filled fields per protein to 5–6 fields after annotation.}
  \label{fig:trembl2}
\end{figure}

\begin{figure}[!htbp]
  \centering
  \includegraphics[width=\linewidth]{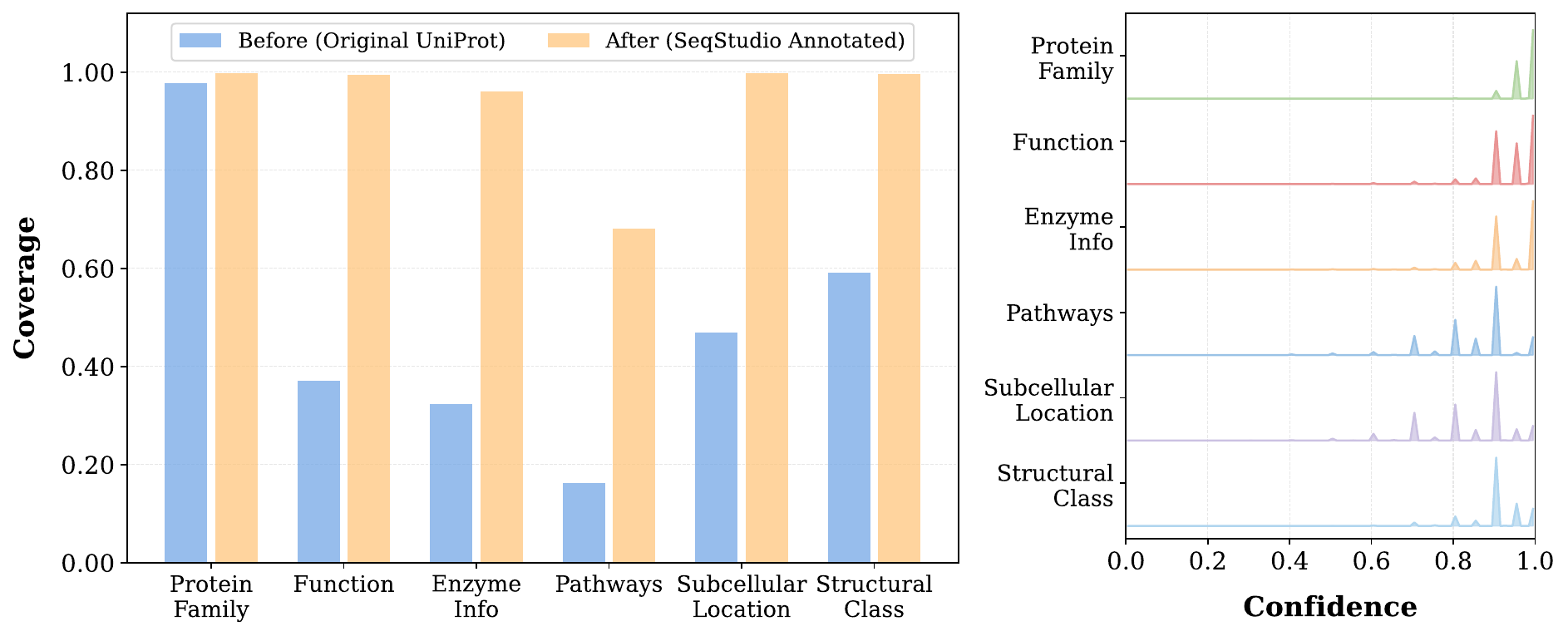}
  \caption{\textbf{Annotation coverage and confidence distribution on the 20M dataset.} \textbf{a}, Comparison of functional dimension coverage before and after SeqStudio annotation. The framework achieves near-saturation coverage across primary fields, specifically elevating sparse categories such as Pathways and Structural Class. \textbf{b}, Confidence score distribution for generated annotations across six functional dimensions. High-density peaks occur in the $0.8$--$1.0$ interval; the relationship between confidence and reference consistency was evaluated separately on Swiss-Prot.
}
  \label{fig:trembl3}
\end{figure}

\textbf{Reasoning-driven annotation uniformly rescues information sparsity across all domains of life.} Across the 20M dataset, the deployment of SeqStudio effectively resolved the severe fragmentation characteristic of unreviewed records. The median number of filled functional fields shifted dramatically from 1--3 per protein in baseline records to 5--6 after annotation (Figure~\ref{fig:trembl2}). This enrichment was consistent across taxonomic divisions: bacterial sequences (\textit{n}~=~12.12M), eukaryotic sequences (\textit{n}~=~6.66M), and the relatively under-represented viral and archaeal proteomes all exhibited comparable gains. The magnitude of this shift was most pronounced for sequences with the lowest baseline scores, demonstrating that the reasoning engine actively compensates for, rather than strictly depends on, pre-existing metadata.

\textbf{Previously sparse functional dimensions reach near-complete coverage.} Baseline TrEMBL coverage for Pathways and Enzyme Info fell below 0.40 (Figure~\ref{fig:trembl3}a). The observation that prediction difficulty varies across functional dimensions underscores the need for quality assurance mechanisms that operate without gold-standard references—precisely the scenario encountered when annotating millions of unreviewed sequences. After annotation, most fields approached saturation ($\approx$1.00), with Pathway coverage increasing to approximately 0.80. Structural Class and Subcellular Location, which already had moderate baseline coverage, reached near-complete annotation.

\textbf{Internal confidence calibration enables autonomous quality auditing at scale.} In the absence of a gold-standard reference for the TrEMBL corpus, ensuring annotation fidelity requires the system to accurately estimate its own uncertainty—especially given the highly divergent nature of unreviewed sequences. Across all six functional dimensions, over 96\% of the generated annotations concentrated in the high-confidence interval (0.8--1.0) (Figure~\ref{fig:trembl3}b). Importantly, the system explicitly flags predictions falling below this threshold ($<$0.8, representing fewer than 4\% of outputs) for downstream human review rather than forcing a definitive label. This discriminative uncertainty estimation acts as an autonomous quality audit: rather than hallucinating when confronted with out-of-distribution sequences lacking sufficient physical or evolutionary signals, the framework accurately recognizes its cognitive boundaries, thereby preventing large-scale database contamination.

\textbf{Annotation gains are not uniform across functional dimensions, revealing intrinsic prediction difficulty.} 
While Structural Class and Subcellular Location achieve near-saturated coverage with high confidence, Pathway annotations show a broader confidence distribution with a heavier tail below 0.8 (Figure~\ref{fig:trembl3}b). This disparity likely reflects the greater inferential distance between sequence-level evidence and pathway-level function: structural class can be directly inferred from domain architecture and topology, whereas pathway membership often requires contextual information like genomic neighborhood that is not available from the sequence alone. This pattern sets an empirical boundary on what single-sequence reasoning can achieve.

\begin{table}[htbp]
\centering
\caption{\textbf{Temporal and Economic Efficiency Analysis.} Comparison of estimated time and cost required to annotate 10,000 protein sequences using traditional manual curation versus the SeqStudio human-in-the-loop framework.}
\label{tab:cost_efficiency}
\setlength{\tabcolsep}{3mm} %
\begin{tabular}{lcc}
\toprule
\textbf{Metric} & \makecell{\textbf{Manual Expert} \\ \textbf{Curation}} & \makecell{\textbf{SeqStudio} \\ \textbf{(Human-in-the-loop)$^\ast$}} \\
\midrule
Turnaround Time (per protein) & $\sim$35 minutes & \textbf{$<$0.3 seconds} \\
Compute Cost (per 10k proteins) & --- & \$8 \\
Expert Review Time (per 10k proteins) & 5,833 hours & \textbf{23.3 hours} \\
Total Estimated Cost (per 10k proteins)$^\dagger$ & \$291,650 & \textbf{\$1,173} \\
\bottomrule
\end{tabular} 
\vspace{1.5mm}
\raggedright
\footnotesize{$^\ast$Based on the high-throughput pipeline configuration. $^\dagger$Assuming a standard bioinformatics consulting rate of \$50/hour. Human review in the SeqStudio workflow is strictly allocated to the $\sim$7\% of predictions falling below the 0.8 confidence threshold.}
\end{table}

\textbf{Confidence calibration enables a highly cost-efficient human-in-the-loop framework.} To quantify the real-world utility of SeqStudio's autonomous quality auditing, we benchmarked its temporal and economic efficiency against traditional manual curation using a standardized sample of 10,000 sequences (Table~\ref{tab:cost_efficiency}). Traditional expert curation—encompassing database searches, literature review, functional reasoning, and standardized data entry—requires approximately 35 minutes per protein. In high-throughput mode, SeqStudio fully automates this workflow in under 0.3 seconds per sequence, yielding a speedup of over 2,300$\times$. Economically, manually annotating 10,000 proteins would require 5,833 expert hours, incurring a prohibitive cost of approximately \$291,650 (assuming a standard bioinformatics consulting rate of \$50 per hour). In contrast, SeqStudio's pure computational cost (LLM tokens and GPU inference) is merely $\sim$\$0.0008 per sequence, totaling \$8 for the cohort. By utilizing the system's confidence scores to triage predictions, human review is strictly limited to the $\sim$7\% of cases falling below the 0.8 confidence threshold. Reviewing these 700 flagged sequences (at an estimated 2 minutes per sequence) requires only 23.3 expert hours, costing \$1,165. The combined SeqStudio and human-in-the-loop approach totals \$1,173—a 99.6\% reduction in overall annotation costs. This efficiency proves that reasoning-driven generative annotation is an economically viable solution for deciphering population-scale genomic dark matter.

\subsubsection{Autonomous quality auditing without gold-standard references}
\label{sec:judge}

The large-scale annotation reported above raises a practical challenge: in the absence of expert-curated references, how can the quality of millions of newly generated annotations be assessed? Confidence scores (Section~\ref{sec:trembl}) provide one proxy, but their reliability ultimately rests on the assumption that the underlying reasoning is grounded in genuine evidence synthesis rather than fluent confabulation. We reasoned that if this assumption holds, the same reasoning process should exhibit a testable property: it should be able to independently judge whether a statement about a protein's function is correct. We constructed a protein-centric question-answering dataset comprising 300 questions derived from high-confidence Swiss-Prot records, evenly distributed across three cognitive dimensions: Core Function Identification (primary biochemical activities), Functional Classification (family and structural assignments), and Functional Specificity (precise pathway and mechanistic details). The corresponding gold-standard answers were synthesized from established molecular metadata such as Gene Ontology terms and InterPro domain signatures, and subjected to strict quality control to ensure multi-database consensus and factual accuracy. Utilizing these meticulously verified answers, human experts conducted an open-book evaluation to establish the ground-truth baseline. This expert-curated benchmark provides a rigorous standard against which SeqStudio's autonomous factual verification can be directly compared. We collected responses from 12 mainstream LLMs spanning a range of model scales and architectures, yielding 3,600 individual protein–answer evaluations. Each answer was scored independently on a ternary scale ($-1$: incorrect, $0$: partially correct, $+1$: correct) by human experts and SeqStudio. Detailed protocols regarding QA generation and quality filtering are provided in Appendix~\ref{sec:app_judge}.

\begin{figure}[htbp]
  \centering
  \includegraphics[width=\linewidth]{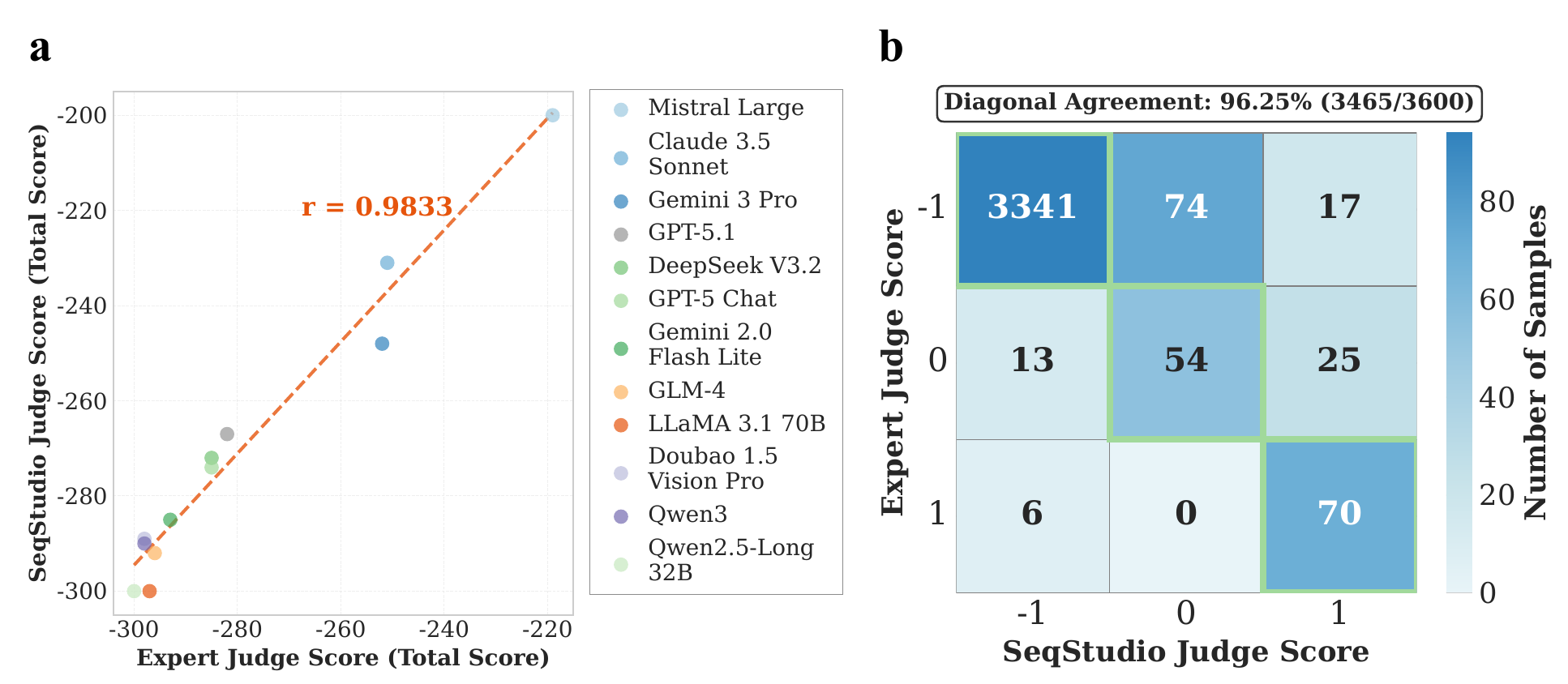}
  \caption{\textbf{Benchmarking SeqStudio as an autonomous judge for annotation.} \textbf{a}, Model-level correlation between SeqStudio and expert scoring. The alignment of average scores across 12 foundation models demonstrates SeqStudio’s capacity to rank model performance, perfectly matching human evaluation trends. \textbf{b}, Confusion matrix of per-sample scoring agreement. High diagonal density confirms consistent error detection and factual verification between SeqStudio’s reasoning-based appraisal and expert evaluation.}
  \label{fig:judge}
\end{figure}

\textbf{SeqStudio demonstrates expert-level consistency in macro-scale model assessment.} As shown in the scatter plot Figure~\ref{fig:judge}a, the average scores assigned to the 12 candidate models by SeqStudio exhibit a remarkably high Pearson correlation ($r = 0.9833$) with the human expert evaluations. This near-perfect alignment suggests that SeqStudio effectively captures the relative strengths and weaknesses of various models, correctly identifying those prone to hallucination versus those providing high-fidelity biological descriptions. Unlike traditional metrics that rely on keyword matching, SeqStudio’s evaluation is rooted in its internal reasoning engine, allowing it to distinguish between nuanced biological truths and plausible-sounding but erroneous claims with expert-like discernment.

\textbf{Micro-scale analysis confirms high granular agreement in error detection.} To verify if SeqStudio accurately identifies specific types of mistakes, we analyzed the per-sample scoring agreement using a confusion matrix heatmap (Figure~\ref{fig:judge}b). Across 3,600 individual evaluations, SeqStudio achieved a 96.25\% diagonal agreement with the expert judgments. Most notably, the high density in the $(-1, -1)$ and $(1, 1)$ quadrants proves that SeqStudio is exceptionally sensitive to both factual hallucinations and precise valid descriptions. Even without an answer key, SeqStudio successfully identifies critical errors such as mismatched protein families, incorrect functional assignments, and biological process hallucinations. This capability establishes SeqStudio as a scalable, autonomous auditor for the protein universe, providing a reliable mechanism for blind quality control in genomic discovery pipelines where gold-standard annotations are inherently unavailable.

\begin{figure}[htbp]
  \centering
  \includegraphics[width=\linewidth]{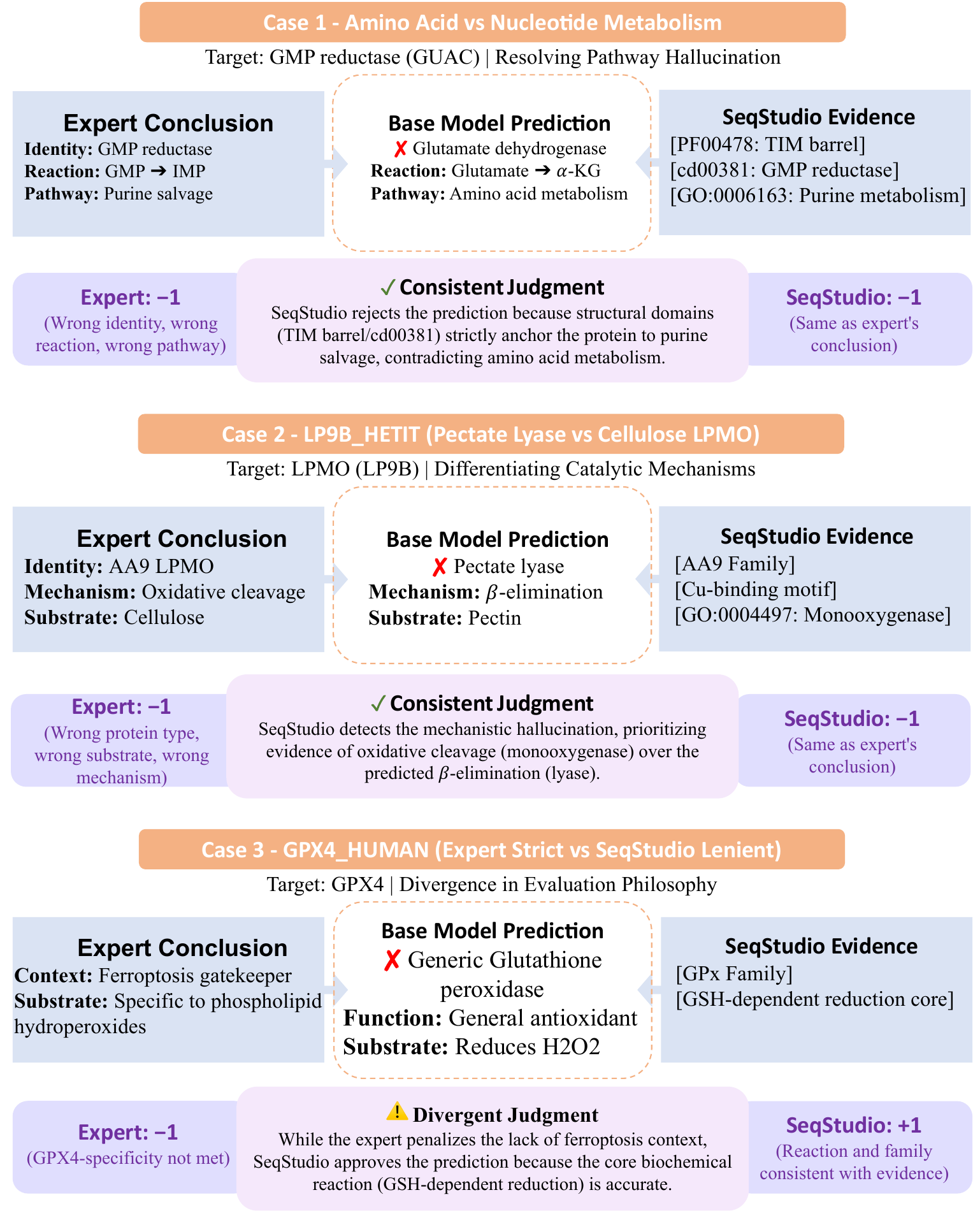}
  \caption{\textbf{Case studies of SeqStudio's reasoning-based autonomous auditing.} 
    \textbf{Case 1}, Correction of a biological pathway hallucination. SeqStudio uses structural domain evidence to reject the base model's misclassification of a purine salvage enzyme as an amino acid metabolic enzyme. 
    \textbf{Case 2}, Correction of a catalytic mechanism hallucination. SeqStudio distinguishes between plant cell wall degradation mechanisms, correctly prioritizing oxidative cleavage (monooxygenase) over the hallucinated $\beta$-elimination (lyase). 
    \textbf{Case 3}, Analysis of a divergent judgment. While experts penalize a generic functional description for missing specific biological context (ferroptosis gatekeeper), SeqStudio validates the answer based on the accuracy of the core biochemical reaction (GSH-dependent reduction), reflecting differences in evaluation criteria rather than reasoning errors.
  }
  \label{fig:judge_cases}
\end{figure}

\textbf{Qualitative analysis reveals biochemically grounded reasoning mechanisms.} To elucidate how SeqStudio achieves this alignment and navigates edge cases, we analyzed representative examples of its reasoning process (Figure~\ref{fig:judge_cases}). In cases of severe hallucination, SeqStudio successfully leverages domain constraints to reject plausible-sounding but fundamentally flawed narratives. For instance, when evaluating the GMP reductase GUAC, the base model incorrectly predicted a glutamate dehydrogenase, effectively shifting the protein to amino acid metabolism. SeqStudio correctly anchored its judgment on the TIM barrel and GMP reductase domains to strictly assign the protein to purine salvage, contradicting the amino acid hypothesis and yielding a consistent $-1$ score alongside the experts. Even more challenging are mechanistic hallucinations within similar biological contexts, such as the cell-wall targeting LP9B. Here, the base model predicted a pectate lyase operating via $\beta$-elimination. SeqStudio distinguished this from the true oxidative cleavage mechanism, prioritizing evidence of monooxygenase activity to detect the mechanistic hallucination. 

Crucially, analyzing the minor fraction of divergent scores highlights a fundamental difference in evaluation philosophy rather than a failure in reasoning. For GPX4—a protein heavily involved as a ferroptosis gatekeeper—the candidate model provided a generic description of a glutathione peroxidase. Human experts penalized this response ($-1$) for failing to capture the unique ferroptosis context. In contrast, SeqStudio awarded a positive score ($+1$), reasoning that the core enzymatic reaction (GSH-dependent reduction) was fundamentally accurate based on the GPx family structural evidence. This divergence suggests that expert rubrics emphasize broader biological context, whereas SeqStudio focuses on core catalytic and biochemical fidelity constrained by computational evidence.

These results support reasoning-derived confidence as a scalable proxy for annotation quality. The concordance between human experts and SeqStudio indicates that the reasoning process exhibits internal consistency, enabling quality judgments based on independent evidence evaluation rather than answer comparison. This provides a principled basis for using SeqStudio self-assigned confidence scores to assess the millions of TrEMBL annotations lacking gold standards. At the same time, the concentration of expert scores at -1 underscores the difficulty of the underlying task and highlights the importance of tool-grounded evidence in bridging the gap between LLM capabilities and the precision required for protein functional annotation.

\subsection{Biological insights at the boundaries of reasoning}

\subsubsection{Integrative reasoning overcomes the homology trap in functional discovery}
\label{sec:rhodopsin}

Discovering specific functional proteins, such as rhodopsins, from uncharacterized sequence pools is a fundamental challenge in genome mining. Traditional sequence-based search tools tend to emphasize local similarity or single high-scoring hits. This makes them vulnerable to retrieving candidates that contain short, high-identity fragments but lack the global structural prerequisites for functionality—a phenomenon we term the homology trap. This leads to high false-positive rates and increases the downstream burden of wet-lab characterization. To demonstrate SeqStudio's capacity to overcome this bottleneck, we evaluated its performance against baseline tools on a highly deceptive, adversarial screening task.

\textbf{Expert-curated and validated positives.} The positive set comprises 1,209 high-fidelity microbial rhodopsin sequences. This ground-truth cohort was established through a combination of experimental validation and expert-guided curation. Specifically, the sequences were systematically compiled from public databases and literature, anchored by experimentally verified targets, and rigorously filtered to ensure the presence of full functional architecture, including confirmed proton-pumping signatures, seven predicted transmembrane helices, and conserved functional motifs.

\textbf{Adversarial decoy sequences.} To construct a maximally deceptive negative set, we engineered 1,600 synthetic ``decoy'' sequences (pseudo-rhodopsins) at two similarity tiers: 800 high-identity fakes and 800 low-identity fakes. These decoys were deliberately generated by concatenating randomly extracted k-mers (90-aa for high-identity, 50-aa for low-identity) from true rhodopsins. This chimeric design simulates realistic fragmentary assemblies: the decoys preserve local sequence similarity and trigger partial domain hits (e.g., the PF01036 Bacteriorhodopsin-like domain) but disrupt the global structural integrity required for true biological function.

\begin{figure}[!htbp]
  \centering
  \vspace{-4mm}
  \includegraphics[width=\linewidth]{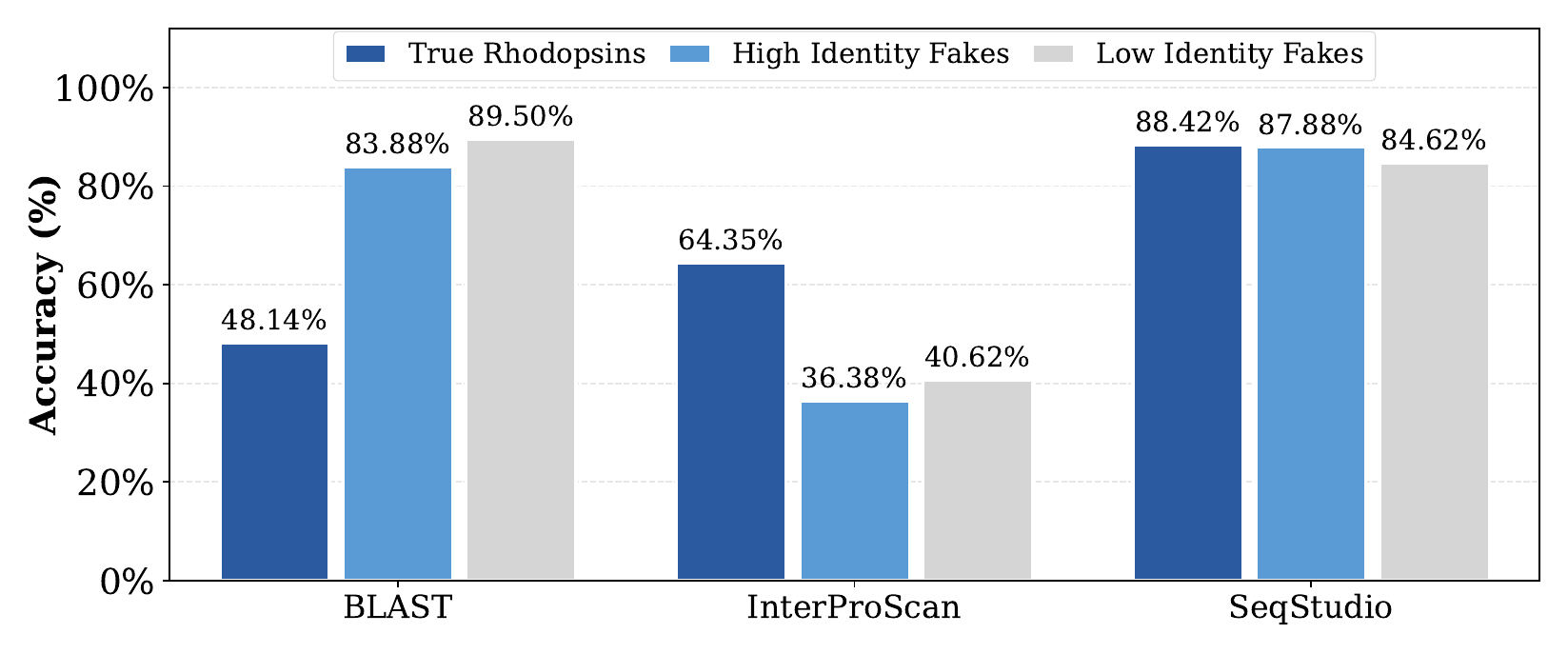}
  \vspace{-4mm}
  \caption{Visualization of prediction accuracy across true rhodopsins and decoy sequence sub-datasets. SeqStudio robustly distinguishes true functional proteins from deceptive chimeric decoys through multi-modal reasoning.}
  \label{fig:rhodopsin_accuracy}
\end{figure}

This design creates an adversarial setting that isolates the specific capability under investigation: distinguishing genuine functional architecture from deceptive sequence similarity. The screening results are summarized in Table~\ref{tab:discovery_method_comparison} and Figure~\ref{fig:rhodopsin_accuracy}.

\begin{table}[!htbp]
\centering
\caption{Performance comparison for rhodopsin discovery on a mixed sequence pool.}
\label{tab:discovery_method_comparison}
\setlength{\tabcolsep}{5mm}
\begin{tabular}{lcccccc}
\toprule
\textbf{Method} & \textbf{Precision} & \textbf{Recall} & \textbf{Accuracy} & \textbf{F1} & \textbf{FPR$^\dagger$} \\
\midrule
SeqStudio      & \textbf{0.829} & \textbf{0.884} & \textbf{0.872} & \textbf{0.856} & 0.138 \\
InterProScan   & 0.442 & 0.644 & 0.496 & 0.524 & 0.615 \\
BLAST          & 0.732 & 0.481 & 0.701 & 0.581 & \textbf{0.133} \\
\bottomrule
\end{tabular} 
\raggedright
\footnotesize $^\dagger$FPR: fraction of negatives predicted as positive.
\end{table}

The evaluation reveals the fundamental limitations of traditional pattern-matching paradigms. BLAST exhibited a notably low recall (0.481). This underperformance highlights the inherent dilemma of alignment-centric searches: achieving an acceptable false-positive rate requires strict empirical thresholds (in our baseline, sequence identity $\ge$ 40\% and coverage $\ge$ 50\%). However, such rigid cutoffs inevitably reject genuine but divergent novel sequences (yielding low recall). Conversely, relaxing these thresholds would cause the system to fixate on local fragmentary matches, resulting in a surge of false positives. Similarly, domain-level annotation tools like InterProScan struggled to capture the full structural context, exhibiting suboptimal precision (0.442) and a high FPR (0.615). Partial or low-confidence structural domain hits are simply insufficient to distinguish fully functional rhodopsins from chimeric decoys.

In contrast, SeqStudio achieved the optimal discovery balance with an F1 score of 0.856 and a precision of 0.829. Rather than relying solely on local alignment scores or single domain matches, SeqStudio's reasoning engine dynamically integrated sequence homology with membrane topology predictions (via TMHMM) and stringent structural rules, such as evaluating the presence of seven transmembrane helices and conserved residues in TM3/TM7. 

To illustrate this mechanistic advantage, we highlight two representative cases demonstrating SeqStudio's multi-modal reasoning capabilities:
\begin{enumerate}
    \item \textbf{Rescuing true positives from ambiguous domain signals (ID: true\_4):} In the case of a true rhodopsin, InterProScan falsely predicted a negative outcome due to the inherent ambiguity of isolated domain annotations (e.g., the PF01036 domain match lacked optimal confidence). SeqStudio successfully rescued this prediction by synthesizing the 7-transmembrane (7TM) topology from TMHMM, strong BLAST functional annotations, and relaxed motif constraints, concluding that the comprehensive evidence outweighed a single motif's weakness.
    \item \textbf{Rejecting high-identity decoys via structural constraints (ID: fake\_90\_673-5488-638):} When presented with a highly deceptive chimeric decoy sharing substantial sequence identity, traditional tools were easily misled by the overwhelming local homology and superficial domain matches. SeqStudio correctly rejected this decoy: by mapping conserved sequence motifs onto the predicted topology, it recognized the critical absence of the essential retinal-binding motif (DxxxK) on TM7, effectively overriding the deceptive sequence similarity.
\end{enumerate}

By computationally resolving these evidential conflicts and filtering out structurally unviable chimeric candidates before physical testing, SeqStudio ensures a high-purity candidate pool. This multi-evidence reasoning approach significantly reduces the downstream burden of wet-lab characterization, accelerating the experimental discovery of novel functional proteins.

\subsubsection{Annotating orphan proteins beyond the sequence barrier}
\label{sec:remote_homologs}

The most stringent test of an evidence integration framework occurs when the dominant source of evidence, sequence homology, is entirely absent. Across UniProtKB/TrEMBL, a substantial fraction of proteins yield no significant BLAST hits against curated reference databases, a challenge that intensifies as sequence identity enters the twilight zone~\cite{rost1999twilight}. We refer to them as orphan proteins, for which no alignable homolog can be detected. Instead, functional hypotheses must be inferred from the remaining sources of evidence, most notably structural fold similarity. We therefore examined whether, under this minimal evidential setting, structural similarity alone can ground the reasoning framework sufficiently to produce coherent functional annotations.

To ensure that the analysis focuses on genuinely intractable cases, we assembled a benchmark of 99 sequence orphans from two previously established and independent datasets originally designed to benchmark single-sequence protein structure prediction. The first dataset, Orphan25~\cite{wang2022single}, contains 25 protein sequences with an effective homolog count of fewer than 10, indicating an almost complete absence of co-evolutionary signal in sequence databases. The second dataset, OrphanRGN2~\cite{chowdhury2022single}, comprises 74 proteins and adopts a stricter definition of orphanhood, requiring a multiple sequence alignment depth of exactly one across UniRef30, PDB70, and MGnify, meaning that no alignable homolog was detected in any of these three major repositories. To confirm their orphan status, we performed BLAST searches against Swiss-Prot and found no significant hits ($E$-value $< 10^{-3}$) for any of the 99 proteins. These sequences therefore represent the extreme of sequence isolation, in which functional inference must rely entirely on evidence sources other than sequence homology. While these targets lack sequence homologs, SeqStudio's structural pipeline queried their 3D topologies against structural databases to retrieve structural neighbors. We then examined whether generative reasoning grounded in these structural anchors could converge on consistent biological interpretations.

\begin{figure}[htbp]
\centering
\includegraphics[width=\textwidth]{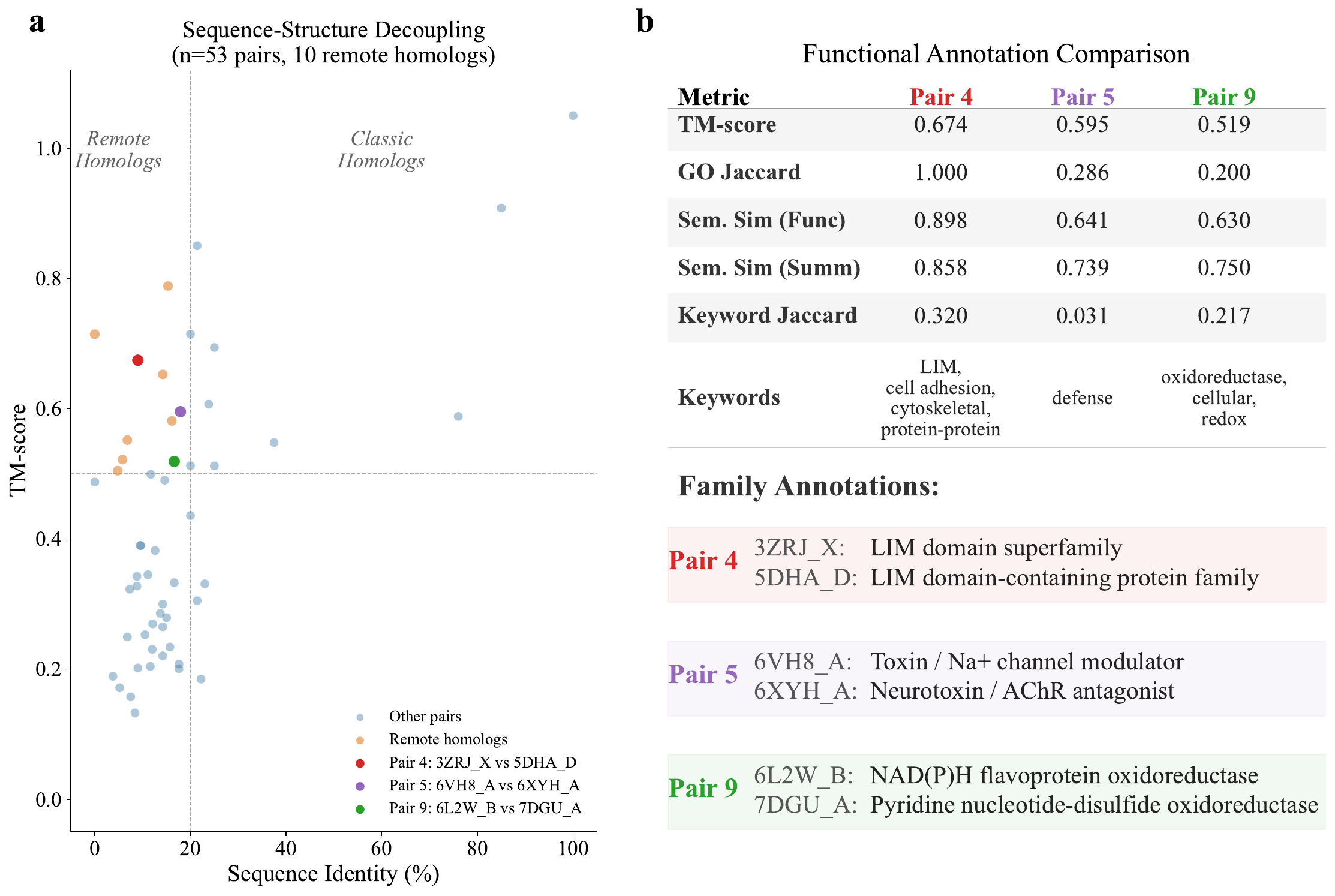}
\caption{\textbf{Structure-grounded annotation of sequence-orphan proteins.}
(A) Scatter plot of sequence identity versus TM-score for 53 protein pairs identified by Foldseek.
Points in the upper-left quadrant (orange) represent remote homologs with low sequence identity ($<$20\%) but high structural similarity (TM-score $>$0.5). (B) Functional annotation comparison for highlighted pairs.
Metrics include TM-score, GO term Jaccard, semantic similarity of functional descriptions (Sem.\ Sim).}
\label{fig:seqid_tmscore}
\end{figure}

To assess whether structural fold conservation drives annotation consistency in the absence of sequence signal, we adopted a pairwise comparison design. We conducted an all-against-all structural comparison of the 99 orphan proteins using Foldseek, which identified 53 pairs with significant structural similarity. Among these, 10 pairs fall within the twilight zone, defined by sequence identity below 20\% and TM-score above 0.5 (Figure~\ref{fig:seqid_tmscore}a). We compared SeqStudio's annotations within each pair using two metrics: GO term Jaccard similarity for discrete ontology overlap, and semantic similarity of the generated functional descriptions and summaries. Across the 10 twilight-zone pairs (Figure~\ref{fig:seqid_tmscore}b), GO Jaccard similarity varied widely, ranging from 0.200 to 1.000,  whereas semantic similarity of the functional descriptions remained consistently high (0.630 to 0.898 for function; 0.739 to 0.858 for summaries). These results indicate that the reasoning framework generates functionally convergent annotations for structurally related orphan proteins whose relationship is undetectable through sequence-based search. The consistently high semantic similarity, despite variable GO term overlap, further suggests that free-text functional descriptions capture graded mechanistic relatedness that discrete ontology terms do not fully resolve.

In Pair 4 (3ZRJ\_X vs 5DHA\_D; sequence identity < 20\%), Foldseek established a structural anchor with a TM-score of 0.674. Building on this structural signal, SeqStudio correctly converged on the LIM domain superfamily for both targets, achieving perfect metric alignment (GO Jaccard 1.000; Semantic Function 0.898) and producing identical mechanistic keywords, “cell adhesion” and “cytoskeletal.” More importantly, Pairs 5 and 9 demonstrate the system’s ability to infer abstract functional relationships even when direct ontology overlap is limited. In Pair 5 (6VH8\_A vs 6XYH\_A; TM-score 0.595; GO Jaccard 0.286), SeqStudio moved beyond mismatched ontology labels and synthesized a coherent mechanistic interpretation. It correctly identified both proteins as neurotoxins and ion channel modulators involved in defense mechanisms, resulting in a semantic similarity score of 0.641. A similar pattern appears in Pair 9 (6L2W\_B vs 7DGU\_A; TM-score 0.519). Despite a GO Jaccard of only 0.200, the reasoning module captured the shared structural basis of NAD(P)H-dependent catalysis and independently annotated both sequences with the mechanistic keywords “oxidoreductase” and “redox,” yielding a semantic summary score of 0.750. 

These results establish that the system does not merely propagate patterns. When the sequence barrier prohibits classical homology transfer, SeqStudio dynamically shifts its inferential weight to structural morphology, using generative reasoning to bridge the gap between divergent sequences and convergent biological mechanisms.

\subsubsection{Reconstructing multi-domain functional logic}
\label{sec:molecular_logic}

Many proteins are composed of multiple functional domains whose coordinated activities depend on specific inter-domain relationships that determine the overall molecular mechanism. A critical question is whether a reasoning framework can reconstruct this inter-domain functional logic from structured tool outputs (domain lists, homolog annotations, structural matches), and to what extent this reconstruction relies on the LLM's parametric knowledge versus being genuinely evidence-driven. To assess this, we designed a controlled experiment using seven multi-domain proteins selected for their well-characterized mechanisms and diverse functional architectures (Table~\ref{tab:molecular_logic_proteins}). We compared the performance of a Naive LLM (relying solely on parametric memory), an InterPro-only baseline, and the full SeqStudio framework (Table~\ref{tab:multidomain_performance}).

\begin{table}[!htbp]
\centering
\caption{Proteins selected for molecular logic analysis.}
\label{tab:molecular_logic_proteins}
\setlength{\tabcolsep}{1.3mm}
\begin{tabular}{llcll}
\toprule
\textbf{Protein} & \textbf{UniProt ID} & \textbf{Length} & \textbf{Class} & \textbf{Key Domains} \\
\midrule
Cas9 & Q99ZW2 & 1,368 aa & CRISPR endonuclease & REC, HNH, RuvC, PI \\
Cas12a & A0Q7Q2 & 1,220 aa & CRISPR endonuclease & REC1/2, RuvC, Nuc \\
Cas13a & P0DOC6 & 1,389 aa & CRISPR ribonuclease & HEPN$\times$2 \\
Dicer & Q9UPY3 & 1,922 aa & RNase III enzyme & Helicase, PAZ, RNaseIII \\
Diphtheria Toxin & P00588 & 567 aa & Bacterial toxin & C, T, R domains \\
CAD & P27708 & 2,225 aa & Multifunctional enzyme & CPSase, ATCase, DHOase \\
Anthrax LF & P15917 & 809 aa & Metalloprotease & PA-binding, Catalytic \\
\bottomrule
\end{tabular}
\end{table}

\begin{table}[!htbp]
\centering
\caption{\textbf{Performance Comparison on Multi-Domain Proteins.} Evaluation of functional logic reconstruction across different annotation methods. SeqStudio successfully identifies all functional mechanisms, whereas the Naive LLM consistently misclassifies them.}
\label{tab:multidomain_performance}
\setlength{\tabcolsep}{0.4mm}
\begin{tabular}{lccccccc}
\toprule
\textbf{Method / Feature} & \makecell{\textbf{Anthrax} \\ \textbf{LF}} & \textbf{CAD} & \textbf{Cas12a} & \textbf{Cas13a} & \textbf{Cas9} & \textbf{Dicer} & \makecell{\textbf{Diphtheria} \\ \textbf{Toxin}} \\
\midrule
\textbf{SeqStudio} & \checkmark & \checkmark & \checkmark & \checkmark & \checkmark & \checkmark & \checkmark \\
\textbf{InterPro Only}    & \checkmark & \checkmark & \checkmark & $\times$   & \checkmark & $\times$   & \checkmark \\
\textbf{Naive LLM}        & $\times$   & $\times$   & $\times$   & $\times$   & $\times$   & $\times$   & $\times$   \\
\midrule
\makecell[l]{\textbf{Naive LLM} \\ \textbf{Misclassification}} & 
\makecell[c]{Scaffold /\\ coiled-coil} & 
\makecell[c]{Receptor\\ tyrosine\\ kinase} & 
\makecell[c]{Ser/Thr\\ kinase} & 
\makecell[c]{Mechanism\\ missed$^\dagger$} & 
\makecell[c]{Ser/Thr\\ kinase} & 
\makecell[c]{AGC Ser/Thr\\ kinase} & 
\makecell[c]{Zinc metallo-\\ protease} \\
\bottomrule
\end{tabular} 
\raggedright
\footnotesize{$^\dagger$For Cas13a, both the Naive LLM and InterPro failed to correctly identify the protein's identity and mechanism. Although they retrieved generic "RNA"-related keywords, they lacked the capacity to reconstruct the specific HEPN domain architecture and its RNA-targeting activity.}
\vspace{-4mm}
\end{table}

The controlled experiment reveals a fundamental limitation of parametric memory. The Naive LLM failed to accurately identify the domains or reconstruct the mechanisms for any of the seven proteins, even though several, especially Cas9, are among the most extensively documented proteins in the scientific literature. This outcome is informative rather than surprising. It indicates that LLM parametric knowledge of protein mechanisms is indexed by conceptual entities such as protein names or family descriptions rather than by raw amino acid sequences. When a sequence is presented without identifying context, the relevant stored knowledge is not activated. The specific error patterns are also non-random: the model consistently defaulted to kinase-family classifications for long sequences, including Cas9, Cas12a, and Dicer, likely reflecting the overrepresentation of eukaryotic kinases in training corpora for proteins exceeding 1,000 residues. These results show that multi-domain functional reconstruction cannot be achieved through sequence-to-text generation alone.

\begin{figure}[htbp]
\centering
\includegraphics[width=\textwidth]{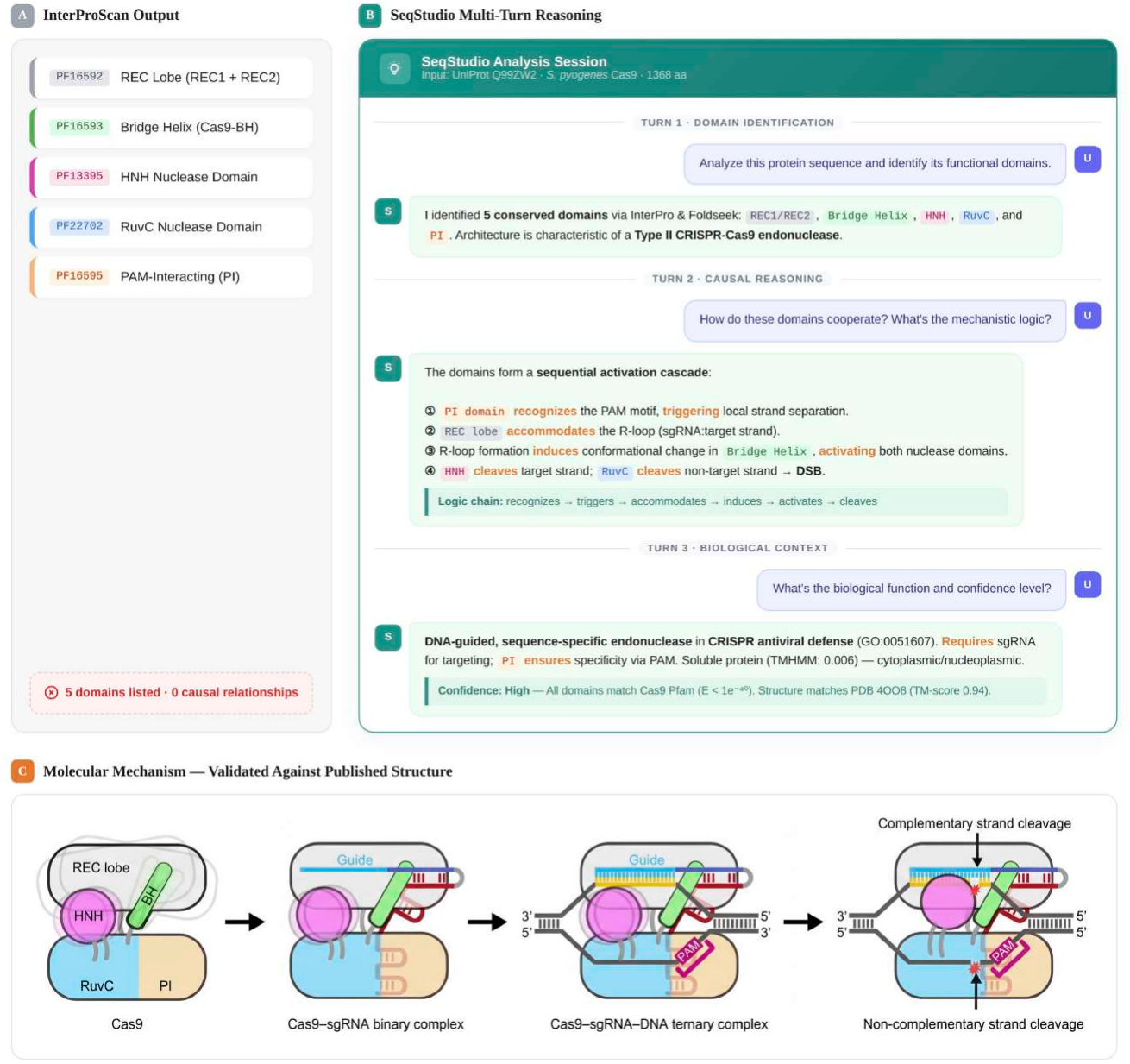}
\vspace{-4mm}
\caption{Comparison of InterProScan and SeqStudio outputs for \textit{S.\ pyogenes} Cas9.
(\textbf{A})~InterProScan output: five Pfam domains listed without functional relationships.
(\textbf{B})~SeqStudio output: multi-turn reasoning reconstructs the causal activation cascade, specifying HNH/RuvC substrate assignment and PAM recognition as a prerequisite for cleavage. Key relationship terms (\textit{recognizes}, \textit{triggers}, \textit{accommodates}, \textit{induces}, \textit{activates}, \textit{cleaves}) are highlighted in orange.
(\textbf{C})~Established Cas9 cleavage mechanism adapted from Nishimasu et al.~\cite{nishimasu2014crystal}, validating the inter-domain logic inferred by SeqStudio. Domain colors are matched across all panels.}
\vspace{-4mm}
\label{fig:molecular_logic}
\end{figure}

Domain annotation tools such as InterProScan successfully identified conserved domains in five of the seven proteins, with domain counts ranging from 2 to 9. However, the outputs consisted solely of isolated domain identifiers without elucidating their functional relationships. For Cas9, InterProScan listed five Pfam domains (REC lobe, Bridge Helix, HNH nuclease, RuvC nuclease, PAM-interacting) but did not specify which domain cleaves which DNA strand, nor the sequential requirements for target recognition (Figure~\ref{fig:molecular_logic}A). Furthermore, InterProScan failed to detect any structural domains for Cas13a and Dicer, representing a critical gap in signature-based matching.

In contrast, SeqStudio successfully generated comprehensive mechanistic descriptions for all seven proteins. By integrating InterProScan's domain lists with its reasoning engine, SeqStudio reconstructed the causal logic between the isolated parts. For Cas9, SeqStudio specified the division of labor between nuclease domains (HNH for the complementary strand, RuvC for the non-complementary strand) and, crucially, inferred the sequential activation cascade: identifying PAM recognition by the PI domain as the triggering event that induces Bridge Helix conformational changes, ultimately activating the nuclease domains (Figure~\ref{fig:molecular_logic}B). These complex relationships, absent from the raw InterProScan output, are highly consistent with the established Cas9 mechanism~\cite{jinek2012programmable,nishimasu2014crystal} (Figure~\ref{fig:molecular_logic}C). 

Crucially, when primary domain annotations were absent, as in the InterPro-only baseline for Cas13a and Dicer, SeqStudio compensated by synthesizing supplementary evidence from sequence homology and structural comparisons. For Cas13a, it successfully leveraged alternative tool outputs to deduce the HEPN domain architecture and its RNA-targeting activity, consistent with structural studies~\cite{liu2017molecular}. These results provide strong evidence that SeqStudio’s functional logic reconstruction is not merely a stylistic reformulation of InterProScan domain lists. When the domain inventory is incomplete, the system integrates multiple sources of evidence to recover missing information, effectively performing multi-source triangulation similar to the strategy employed by expert curators working with incomplete data.

\subsection{Interactive annotation via the SeqStudio web server}

\begin{figure}[htbp]
\centering
\vspace{-5mm}
\includegraphics[width=\textwidth]{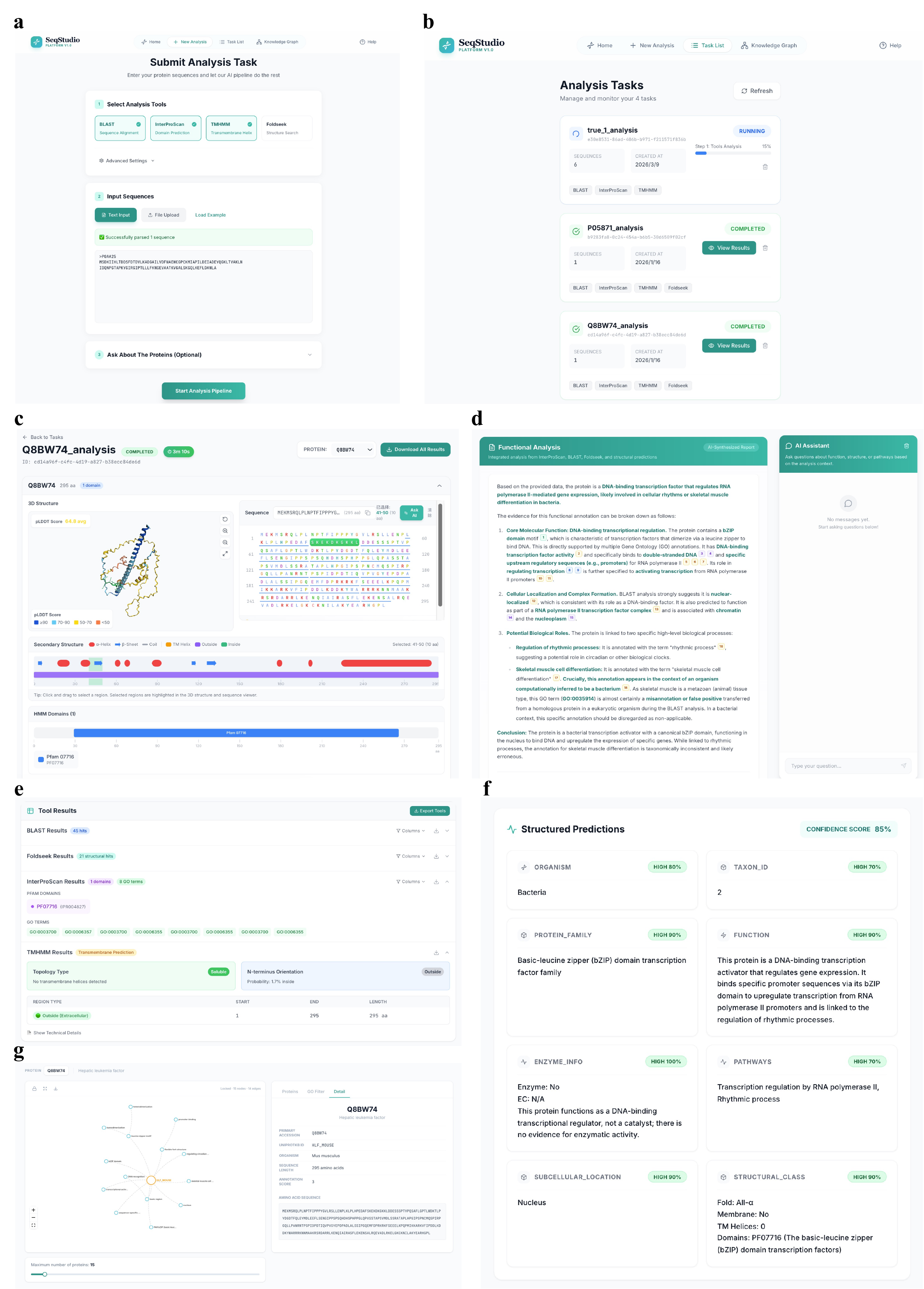}
\caption{\textbf{The SeqStudio web server. }
  \textbf{a-b}, Task submission interface with customizable tool selection and the asynchronous task management dashboard.
  \textbf{c}, Interactive 3D structure and 1D sequence viewer.
  \textbf{d-f}, Comprehensive results panels displaying the AI-synthesized functional report with a context-aware chat assistant, raw tool outputs, and structured predictions.
  \textbf{g}, Knowledge graph visualizing key biological entities and causal relationships extracted from the functional summary.
}
\label{fig:web1}
\end{figure}

To democratize our automated reasoning framework into an accessible tool for the broader biological community, we developed the SeqStudio web server, a highly interactive platform that facilitates real-time protein annotation (Figure \ref{fig:web1}). The platform's interface is designed to seamlessly integrate tool execution, multi-modal visualization, and dynamic LLM reasoning into a unified human-in-the-loop workflow.

The analytical workflow begins with a versatile submission interface (Figure \ref{fig:web1}\textbf{a}), where researchers can submit multi-sequence FASTA files or upload custom PDB coordinates to enable precise structure-based searches. This functionality is particularly important for the characterization of sequence orphans. Users can dynamically configure the analysis pipeline by selecting specific modules, including BLAST, InterProScan, TMHMM, and Foldseek, according to their research objectives. Because comprehensive computational annotations can be time-intensive, tasks are executed asynchronously. Task progress can be monitored through a dedicated management dashboard (Figure \ref{fig:web1}\textbf{b}), which provides real-time status updates and allows access to previously completed analyses.

Upon accessing a completed task, the platform presents a highly interpretable suite of results panels. To facilitate an intuitive understanding of protein architecture, the platform features a synchronized 3D structure and 1D sequence viewer (Figure \ref{fig:web1}\textbf{c}). This interactive module allows users to visually map predicted Pfam domains and secondary structures directly onto the folded conformation, with color-coded pLDDT confidence scores indicating structural reliability. 

The core of the analysis is presented in the AI-synthesized functional report (Figure \ref{fig:web1}\textbf{d}). To ensure strict epistemic traceability, the generated text is annotated with inline citation tags that link directly to specific computational evidence. Adjacent to this report is a context-aware AI chat assistant, empowering users to ask dynamic follow-up questions. The LLM responds by reasoning over the protein's curated evidence context, effectively mitigating hallucination risks.

To guarantee complete data transparency, the Tool Results panel (Figure \ref{fig:web1}\textbf{e}) allows users to inspect the granular, raw outputs from all executed bioinformatics tools, encompassing detailed BLAST/Foldseek hit lists, InterProScan GO terms, and TMHMM topological predictions. These multi-modal clues are ultimately distilled into the Structured Predictions module (Figure \ref{fig:web1}\textbf{f}), which standardizes the inferred functional dimensions (e.g., Protein Family, Subcellular Location, Pathways) alongside discrete confidence scores.

Finally, to assist users in grasping complex biological mechanisms, SeqStudio features an automated Knowledge Graph module (Figure \ref{fig:web1}\textbf{g}). Driven by a predefined domain ontology, the LLM parses the unstructured textual summary to extract logic-dense entities and causal relationships. It dynamically renders an interactive network, mapping the target protein to its multi-dimensional biological context (e.g., interacting metabolites, localized compartments, and regulatory pathways). All synthesized reports, visualizations, and raw data are fully downloadable to support downstream bioinformatic integration.

\clearpage
\section{Discussion}
\label{sec:discussion}

\textbf{This study investigated whether evidence-grounded generative reasoning can approximate expert curation of protein function}. The results indicate that it can within defined limits. SeqStudio achieved 91\% consistency with expert-curated Swiss-Prot records across six functional dimensions and maintained stable performance on Swiss-Prot entries first released in 2025, while prior exposure to the underlying sequences or related annotations cannot be excluded. The framework also demonstrated practical utility through experimental validation of rhodopsin predictions and large-scale annotation of more than 20 million UniProtKB entries. Separately, the protein-centric QA benchmark evaluated SeqStudio as an autonomous judge against expert assessments; it did not independently validate the full 20-million-protein annotation corpus. Importantly, our economic efficiency analysis reveals that SeqStudio directly overcomes the traditional scalability barrier of expert curation, delivering a 2,300-fold temporal acceleration and slashing total annotation costs by 99.6\%. Ablation analysis showed that this performance depends on the structured evidence object rather than the language model alone. When frontier models were given sequences without tool-derived evidence, annotation quality declined markedly.

This finding has implications for the broader protein annotation landscape. The gap between Swiss-Prot, with about 570{,}000 reviewed entries, and TrEMBL, which contains roughly 250 million unreviewed sequences, is often attributed to limited curation capacity. However, the underlying analytical tools such as BLAST, InterProScan, and Foldseek already operate at database scale. The principal bottleneck lies in evidence integration: \textbf{evaluating heterogeneous signals and synthesizing them into coherent functional interpretations}. Our results suggest that this step can be approximated by conditioning a language model on structured and semantically enriched evidence. In this framework, generative reasoning functions as a complementary layer that integrates fragmented signals rather than replacing existing pattern-matching tools. The increase in annotation completeness from one to three populated fields under rule-based pipelines to five or six fields under SeqStudio indicates that the primary benefit arises from integrative synthesis.

Several limitations should be noted. The remaining 9\% inconsistency is concentrated in enzyme and pathway predictions, where correct annotation often depends on metabolic context or experimental evidence that is not included in the current pipeline. Genomic context, co-expression data, and protein interaction networks are not yet incorporated in the current pipeline. In addition, the confidence scores produced by the model are self-assessed rather than externally calibrated. The case studies on remote homologs and multi-domain proteins are illustrative but limited in scale.

The modular architecture of SeqStudio allows straightforward extension. Additional evidence sources such as genomic context, interaction networks, and expression profiles could be incorporated as new views without modifying the reasoning engine. The confidence estimates generated by the system could also support active learning strategies that direct limited curation resources toward uncertain predictions. More broadly, the evidence-grounded framework presented here, in which a model integrates externally validated signals rather than serving as an independent knowledge source, may generalize to other scientific interpretation tasks. 
These results also speak to the design principles of tool-augmented AI systems that are now being explored across the sciences~\cite{boiko2023autonomous,m2024augmenting, huang2025biomni,gao2025democratizing}. SeqStudio employs a deliberately minimal architecture: a fixed tool repertoire, a two-round reasoning protocol, and no autonomous planning or self-correction loops. This constrained design achieves over 91\% consistency with expert curation on Swiss-Prot and has been applied to more than 20 million UniProtKB proteins. The consistency result and evidence-ablation findings suggest that, for structured scientific interpretation tasks, the primary performance determinant is \textit{evidence quality and integration fidelity} rather than architectural sophistication. Whether richer agentic capabilities---such as dynamic tool selection, iterative self-verification, or multi-agent debate---would further improve annotation quality is an open empirical question; the evaluation framework and temporal benchmarks established here are designed to facilitate such comparisons. 
\textbf{The central challenge in functional annotation is not computational scale but the ability to approximate expert reasoning over diverse evidence within feasible economic constraints.} Our results suggest that such approximation is now achievable, dramatically lowering the temporal and financial barriers to high-fidelity annotation, and offering a path toward narrowing the gap between the volume of sequenced proteins and our capacity to interpret them.

\section{Methods}

\subsection{Preliminaries}
\subsubsection{Protein functional annotation as a multi-dimensional characterization task}

Protein functional annotation assigns biological meaning to a protein sequence by specifying \textit{what the protein does}, \textit{how it performs this function}, and \textit{where it operates within the cell}. In computational settings, this problem is commonly formulated as a classification task. Given a protein sequence $x = (a_1, a_2, \dots, a_n)$ over the amino acid alphabet $\mathcal{A}$, the objective is to predict a set of discrete labels $L \subseteq \mathcal{L}$ drawn from a controlled vocabulary $\mathcal{L}$ such as the Gene Ontology:
\begin{equation}
f \colon x \longrightarrow L \subseteq \mathcal{L}.
\label{eq:classification}
\end{equation}
This formulation underlies most automated annotation tools and community benchmarks, including the Critical Assessment of Functional Annotation (CAFA)~\cite{zhou2019cafa}. However, expert-curated entries in databases such as UniProtKB/Swiss-Prot~\cite{uniprot2025uniprot} contain substantially richer information than a set of labels. Each record provides a structured description of a protein that includes family membership, molecular function, enzymatic activity, pathway involvement, subcellular localization, and structural architecture, together with mechanistic explanations and explicit evidence.

To better reflect the structure of expert curation, we adopt a more comprehensive formulation. Let $Y = (y_1, y_2, \dots, y_K)$ denote a structured annotation consisting of $K$ functional dimensions. Each $y_k$ represents a composite entry that includes a functional description, a confidence assessment, and supporting evidence. The annotation task can therefore be defined as:
\begin{equation}
g \colon x \longrightarrow Y = (y_1, y_2, \dots, y_K),
\label{eq:structured}
\end{equation}
where each $y_k = (\text{description}_k, \text{confidence}_k, \text{evidence}_k)$.
In this study, we consider $K = 6$ dimensions: Protein Family, Function, Enzyme Information, Pathways, Subcellular Location, and Structural Class. This formulation more closely reflects the structure of expert-curated records and serves as the evaluation target throughout this work.

\subsubsection{Existing paradigms: pattern matching and its variants}

Current automated annotation methods can be grouped into three major paradigms. Each paradigm detects a particular type of molecular signal by matching a query sequence against curated reference resources. Their outputs collectively form the evidence base used by most automated annotation pipelines, including the system developed in this work.

\paragraph{Sequence similarity search (global homology)}

Given a query sequence $x$ and a reference database $\mathcal{D}$ of annotated proteins, a similarity search operator $\mathcal{S}$ retrieves a ranked list of homologous sequences:
\begin{equation}
\mathcal{E}_{\mathrm{sim}}
= \{(h_i, \alpha_i, L_i)
\mid h_i = \mathcal{S}(x, \mathcal{D}), \;
\alpha_i \geq \tau \},
\label{eq:similarity}
\end{equation}
where $h_i$ is a retrieved homolog, $\alpha_i$ is a similarity score such as sequence identity or $E$-value, $L_i$ denotes the annotation associated with $h_i$, and $\tau$ is a significance threshold. Tools such as BLAST~\cite{blast} implement this paradigm at the sequence level, while structural search tools such as Foldseek~\cite{van2024fast} extend it to three-dimensional structures using similarity metrics such as TM-score. The underlying assumption is that evolutionary or structural similarity implies functional similarity. This relationship is reliable at high similarity but weakens substantially in the sequence identity \textit{twilight zone} below roughly 25–30\%~\cite{rost1999twilight}. It may also fail in cases of convergent evolution where proteins share structural folds but perform different functions~\cite{gherardini2007convergent}.

\paragraph{Domain and motif signature matching (local features)}

A second class of methods detects conserved domains or motifs rather than comparing full-length sequences. Let $\Phi = \{\phi_1, \phi_2, \dots\}$ denote a library of signatures represented by hidden Markov models or regular expression patterns. A signature matching operator $\mathcal{M}$ identifies local matches within the query sequence:
\begin{equation}
\mathcal{E}_{\mathrm{motif}}
= \{(\phi_j, r_j, L_j)
\mid \mathcal{M}(x, \Phi) \rightarrow \text{match}_j \},
\label{eq:motif}
\end{equation}
where $\phi_j$ denotes the matched signature, $r_j$ specifies the matched region in $x$, and $L_j$ is the functional annotation associated with that signature. InterProScan~\cite{jones2014interproscan} is the primary meta-tool in this paradigm, integrating multiple signature databases including Pfam~\cite{paysan2025pfam}, CDD~\cite{wang2023conserved}, and PANTHER~\cite{thomas2022panther}. Domain signatures provide modular and position-specific functional information that complements global homology searches. A protein may have low overall similarity to known sequences yet still contain a well-characterized catalytic domain. However, domain matches alone rarely determine how multiple domains cooperate within a protein or resolve cases where the same domain participates in different functional contexts.

\paragraph{Rule-based annotation transfer (expert-encoded logic)}

Rule-based systems encode expert knowledge as conditional rules that combine evidence from similarity searches and domain detection. Systems such as UniRule~\cite{unirule} and ARBA~\cite{saidi2017rule} apply deterministic rules of the form
\begin{equation}
\mathcal{R}\bigl(\mathcal{E}_{\mathrm{sim}},
\mathcal{E}_{\mathrm{motif}}\bigr) \longrightarrow L_{\mathrm{rule}},
\label{eq:rules}
\end{equation}
where $\mathcal{R}$ represents a predefined rule set and $L_{\mathrm{rule}}$ is the resulting annotation. Such rules encode high-confidence biological conditions. For example, a rule may assign an enzyme classification when a sequence contains a specific Pfam domain with strong statistical support and belongs to a particular taxonomic group. Rule-based systems achieve high precision within the conditions they encode, but they are inherently limited. They cannot easily accommodate new domain combinations, integrate graded evidence, or reconcile conflicting signals.

\paragraph{Shared characteristics and limitations}

Despite their methodological differences, all three paradigms follow the same operational logic: detect a molecular pattern and transfer the associated annotation. Each method captures only one aspect of the protein’s functional signal. When used independently, they provide partial views of functional identity. When combined, their outputs are typically merged as a simple union of predicted labels:
\begin{equation}
L_{\mathrm{combined}}
= L_{\mathrm{sim}} \cup L_{\mathrm{motif}} \cup L_{\mathrm{rule}} .
\label{eq:union}
\end{equation}
This strategy increases coverage but does not resolve conflicts among evidence sources, explain how different signals relate mechanistically, or evaluate the overall coherence of the annotation. In this study, we refer to this operational logic collectively as \textbf{pattern matching}: the detection and transfer of individual functional signals without an integrative synthesis of evidence.

\subsubsection{From pattern matching to evidence-grounded generative reasoning}

The gap identified above, between detecting individual signals and integrating them into a coherent functional characterization, is precisely the step performed by expert curators during manual annotation. When annotating a protein, a Swiss-Prot curator does not simply combine BLAST hits, InterProScan domains, and UniRule assignments. Instead, the curator evaluates the reliability of each signal, reconciles contradictions when different analyses suggest different interpretations, and synthesizes a unified mechanistic explanation that accounts for all available evidence. The final result is not a simple set of labels $L_{\mathrm{combined}}$, but a structured and internally consistent annotation $Y$ with graded confidence and explicit evidence.

To investigate whether this integrative process can be approximated computationally, we developed an experimental system called SeqStudio. Instead of aggregating outputs through set union as in Eq.~\ref{eq:union}, SeqStudio performs conditional generation using a large language model. The system uses the same evidence sources described in the previous section but organizes them into four complementary views that represent different levels of molecular information. For each query sequence $x$, we construct a multi-view evidence object $\mathcal{O}_x$:
\begin{equation}
\mathcal{O}_x
= \langle
v_{\mathrm{seq}},
v_{\mathrm{dom}},
v_{\mathrm{struc}},
v_{\mathrm{topo}}
\rangle,
\label{eq:evidence_object}
\end{equation}
where each view corresponds to a distinct source of evidence:
\begin{itemize}
\item $v_{\mathrm{seq}}$ derives from sequence similarity search (Eq.~\ref{eq:similarity}), implemented using BLAST~\cite{blast} against the Swiss-Prot database. It reflects evolutionary relationships and recent functional conservation.
\item $v_{\mathrm{dom}}$ aggregates domain and motif signatures detected by InterProScan~\cite{jones2014interproscan} together with rule-based assignments from UniRule~\cite{unirule}. This view represents modular functional units and expert-curated biological rules.
\item $v_{\mathrm{struc}}$ is obtained from Foldseek~\cite{van2024fast}. Structural comparisons reveal fold-level relationships that may not be detectable at the sequence level.
\item $v_{\mathrm{topo}}$ describes predicted membrane topology generated by TMHMM~\cite{hallgren2022deeptmhmm}, including the number and orientation of transmembrane helices. This information provides physical constraints that help determine localization and functional class.
\end{itemize}
Each view is enriched with semantic information from curated databases. Rather than providing the language model with identifiers or statistical scores alone, every evidence hit is accompanied by its associated biological description, such as Gene Ontology term definitions~\cite{gene2023gene} or Pfam domain descriptions~\cite{paysan2025pfam}. This enrichment allows the generative process to operate on biological meaning rather than opaque identifiers.

The final annotation is generated as:
\begin{equation}
\hat{Y} = \underset{Y}{\arg\max}\; P_{\mathrm{LLM}}\!\bigl(Y \;\big|\; \mathrm{Prompt}(\mathcal{O}_x)\bigr),
\label{eq:generation}
\end{equation}
where $P_{\mathrm{LLM}}$ denotes the conditional distribution of a LLM and $\mathrm{Prompt}(\mathcal{O}_x)$ is a structured prompt containing the evidence object and instructions to produce a multi-dimensional annotation. The predicted annotation $\hat{Y} = (\hat{y}_1, \dots, \hat{y}_K)$ follows the structure defined in Eq.~\ref{eq:structured}, where each component $\hat{y}_k$ includes a functional description, a confidence estimate, and supporting evidence.

\paragraph{Conceptual distinction from pattern matching.}

The distinction in Figure~\ref{fig:teaser} concerns the integration step. Sequence retrieval, signature matching, and rule-based assignment provide signals that can subsequently be interpreted by expert curators or annotation pipelines. Equation~\ref{eq:union} represents a simple combination of these outputs, without modeling their relationships; it is an operational comparator rather than a description of every conventional annotation workflow.

In contrast, the generative formulation in Eq.~\ref{eq:generation} performs evidence-level integration. The language model receives all four evidence views simultaneously and produces a single structured annotation $\hat{Y}$. The generated output must reconcile overlapping signals, redundant information, and potential contradictions across the evidence sources. This difference can be summarized as
\begin{equation}
  \underbrace{
    L_{\mathrm{sim}} \cup L_{\mathrm{motif}} \cup L_{\mathrm{rule}}
  }_{\text{Pattern matching: union of fragments}}
  \quad\longrightarrow\quad
  \underbrace{
    P_{\mathrm{LLM}}\!\bigl(Y \mid
      v_{\mathrm{seq}},\, v_{\mathrm{dom}},\,
      v_{\mathrm{struc}},\, v_{\mathrm{topo}}\bigr)
  }_{\text{Generative reasoning: conditional generation}}.
  \label{eq:contrast}
\end{equation}
Whether this form of conditional generation can approximate expert curation in terms of consistency, generalization, and scalability is the central empirical question of this study. SeqStudio provides the experimental platform used to evaluate this hypothesis, and its implementation is described in the following sections.

\subsection{SeqStudio Implementation}

\begin{figure}[!htbp]
    \centering
    \includegraphics[width=\textwidth]{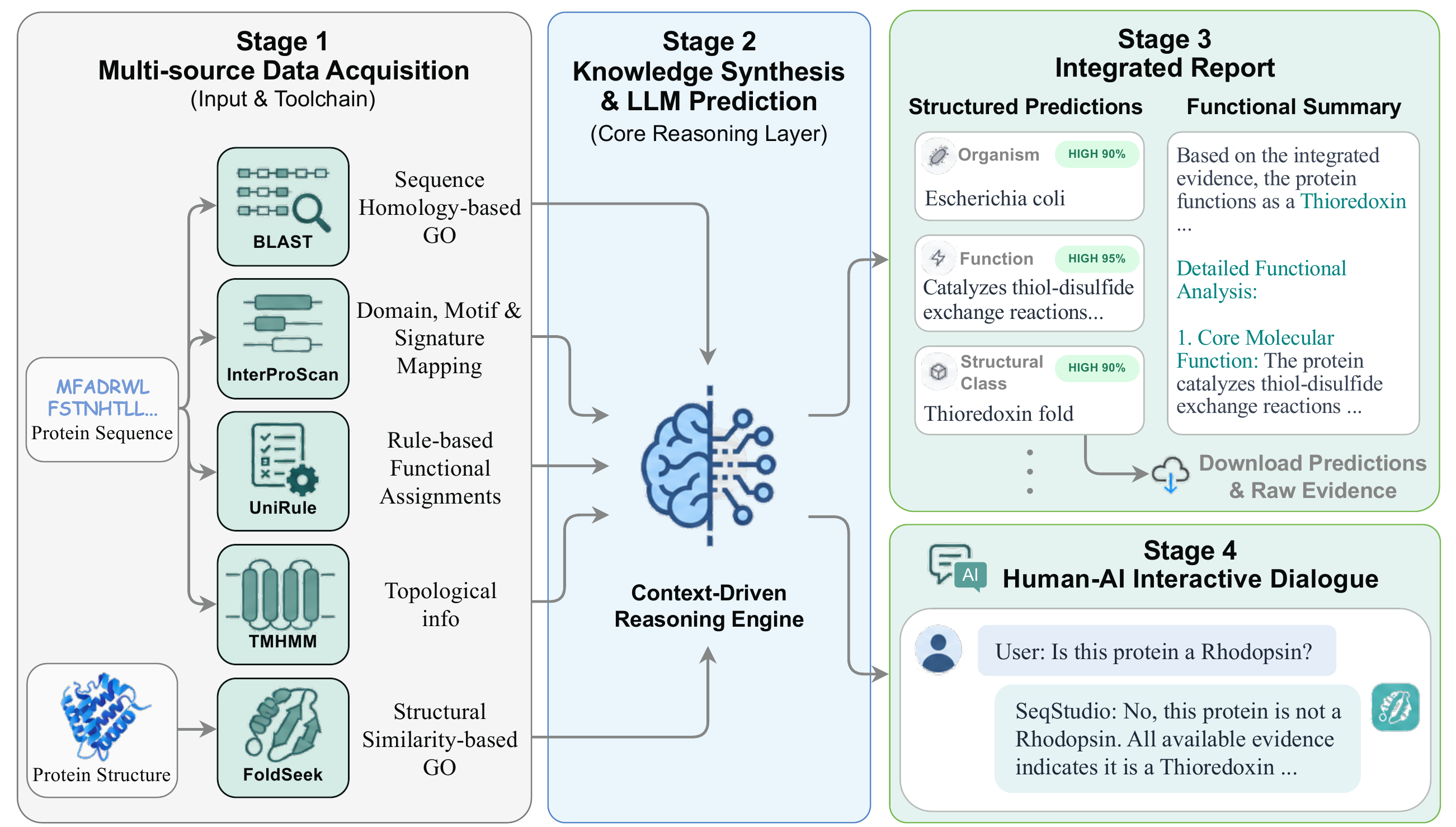}
    \caption{\textbf{The four-stage SeqStudio workflow.} Stage 1 acquires sequence-derived evidence with BLAST, InterProScan, UniRule, and TMHMM, and structure-derived evidence with Foldseek when a structure is available. Stage 2 integrates these sources through a context-driven LLM reasoning engine. Stage 3 presents structured predictions and a functional summary, with predictions and raw evidence available for download. Stage 4 supports follow-up dialogue grounded in the evidence and report. The thioredoxin annotation, organism field, confidence badges, and rhodopsin question are illustrative interface examples. Confidence badges display model self-assessments, not experimentally measured certainty.}
    \label{fig:frame}
\end{figure}

To evaluate whether evidence-grounded generative reasoning can approximate expert curation, we implemented SeqStudio as an experimental platform that operationalizes the formulation in Eq.~\ref{eq:generation}. As revealed in Figure~\ref{fig:frame}, SeqStudio operates through a four-stage modular workflow designed to transform heterogeneous protein features into a synthesized biological narrative~\cite{huo2024multi,zhou2025decoding}. This integrative approach allows for the systematic adjudication of complex, multi-scale evidence with expert-level logic, yielding functional interpretations that go beyond simple label assignment~\cite{jha2025gaia}.

\subsubsection{Stage 1: Multi-source evidence acquisition}

SeqStudio collects molecular evidence from four complementary levels of biological organization, chosen to approximate the diverse sources of biological information consulted by Swiss-Prot curators when performing expert protein annotation~\cite{uniprot2025uniprot}.

\textit{Evolutionary lineage} is obtained through BLAST~\cite{blast} searches against the reviewed Swiss-Prot database. This step retrieves homologous proteins whose annotations can be transferred with high reliability at strong sequence identity.

\textit{Modular functional units} are identified using InterProScan~\cite{jones2014interproscan}, which detects conserved domains and motifs. These results are supplemented with rule-based assignments from UniRule~\cite{unirule}. Together, they reveal local functional elements that may not be captured by full-length sequence comparison.

\textit{Three-dimensional fold relationships} are detected using Foldseek~\cite{van2024fast}, which performs structural similarity searches. Structural comparison enables the identification of remote homologs whose sequence similarity is too low for reliable detection.

\textit{Biophysical constraints} are inferred by TMHMM~\cite{hallgren2022deeptmhmm}, which predicts membrane topology, including the number and orientation of transmembrane helices, providing information that help determine subcellular localization and functional class.

These five tools provide four evidence views: sequence homology, domains and rules, structural similarity, and membrane topology. Sequence-based analyses use the amino acid sequence, while Foldseek is run when a protein structure is available. Tool versions, target databases, and core execution parameters are summarized in Supplementary Table~\ref{tab:seqstudio-tools}. Each evidence hit is further enriched with biological descriptions obtained from the corresponding source databases. Examples include Gene Ontology term definitions~\cite{gene2023gene} and Pfam domain functional summaries~\cite{paysan2025pfam}. This enrichment ensures that the reasoning stage operates on interpretable biological concepts rather than identifiers or statistical scores alone.

\subsubsection{Stage 2: Knowledge synthesis and LLM prediction}

The core of SeqStudio is a generative reasoning engine that acts as a semantic integrator for multi-modal molecular data~\cite{huo2024multi}. The enriched evidence object $\mathcal{O}_x$ (Eq.~\ref{eq:evidence_object}) is processed by DeepSeek-V3.2 using a two-round dialogue strategy. In the first round, the model integrates all evidence into a coherent natural-language summary describing the protein’s molecular mechanism, biological role, and major structural characteristics. This stage encourages comprehensive synthesis of the available information without structural constraints. In the second round, the generated summary is provided again together with the original evidence. The model is then instructed to produce structured predictions across six functional dimensions using a predefined output schema that requires confidence scores and explicit evidence provenance. 

This two-round design separates reasoning from structured extraction. The first round supports integrative interpretation, while the second converts the reasoning into a consistent, machine-readable annotation. Both stages operate under a shared system instruction that defines an evidence reliability hierarchy~\cite{farquhar2024detecting}. Experimentally validated annotations and high-identity homologs are prioritized over lower-confidence computational predictions, and the model must explicitly justify its interpretation when evidence sources conflict. Prompt templates are in Supplementary Note~\ref{app:prompt_engineering}.

\subsubsection{Stage 3: Integrated Report Generation}

The reasoning engine produces a structured annotation
$\hat{Y} = (\hat{y}_1, \dots, \hat{y}_6)$ following the formulation in Eq.~\ref{eq:generation}. Each prediction includes a self-assessed confidence score between 0.0 and 1.0, intended to summarize the model's assessment of the strength and consistency of supporting evidence. This score is distinct from the empirical consistency measured against reference annotations. Every claim is also accompanied by explicit provenance tags that link the prediction to specific evidence sources such as Pfam domains, Gene Ontology terms, BLAST homologs, or TMHMM topology predictions. This traceability provides a transparent audit trail that enables independent verification of the generated annotations.

The integrated report presents the natural-language functional summary alongside structured predictions. Predictions and raw tool outputs can be downloaded for inspection, as shown in Figure~\ref{fig:frame}. The complete report is serialized into a standardized JSON schema that preserves the final predictions and intermediate tool outputs for traceability. The schema and an example output are provided in Supplementary Note~\ref{app:seqstudio_output}. SeqStudio also supports task-specific adaptation: by modifying only the instructions while retaining the identical evidence pipeline, the reasoning objective can be redirected to specialized tasks such as the rhodopsin screening described in Section~\ref{sec:rhodopsin}.

\subsubsection{Stage 4: Human-AI Interactive Dialogue}

Following report generation, researchers can engage in free-form dialogue with the reasoning engine, which retains access to both the complete evidence set and the structured annotation produced in the previous stages~\cite{zhou2025decoding}. This allows targeted follow-up queries such as requesting mechanistic explanations for a specific domain arrangement, evaluating alternative functional hypotheses, or examining the basis of a low-confidence prediction. This capability supports hypothesis testing and targeted inquiry but is not systematically evaluated in the present study.

\section*{Data Availability}
All SeqStudio-generated annotations are publicly available under the MIT license on Hugging Face at \href{https://huggingface.co/datasets/OpenRaiser/SeqStudio}{https://huggingface.co/datasets/OpenRaiser/SeqStudio}. The repository (approximately 74.8\,GB total) contains the complete 20-million-protein annotation corpus (\texttt{seqstudio\_uniprot\_20m}) in Apache Parquet format, organized into six files stratified by curation status and annotation score: \texttt{swiss\_57w.parquet} (575,661 reviewed Swiss-Prot entries; 3\,GB), \texttt{trembl1\_1081w.parquet} (TrEMBL annotation score\,=\,1; 32.9\,GB), \texttt{trembl2\_324w.parquet} (score\,=\,2; 12.5\,GB), \texttt{trembl3\_397w.parquet} (score\,=\,3; 17.6\,GB), \texttt{trembl4\_108w.parquet} (score\,=\,4; 6.74\,GB), and \texttt{trembl5\_32w.parquet} (score\,=\,5; 2.02\,GB). Each record preserves the original UniProt metadata, raw bioinformatics tool execution results, and the complete SeqStudio-generated annotations comprising a natural-language functional summary and six-field structured predictions with confidence scores and evidence provenance. The datasets are compatible with standard data science libraries including Hugging Face Datasets, Dask, and Polars. Auxiliary reference files (Pfam descriptions, Gene Ontology definitions, and InterPro metadata) used for semantic enrichment during annotation are distributed within the code repository.

\section*{Code Availability}

The source code for the SeqStudio annotation framework is available at \href{https://github.com/OpenRaiser/SeqStudio}{https://github.com/OpenRaiser/SeqStudio} under the MIT license. The repository provides a reproducible annotation stack comprising two command-line pipelines: \texttt{seq\_annotation} for annotating protein sequences supplied as FASTA files with optional PDB structures, and \texttt{seqstudio\_pipeline} for processing UniProt-style JSON.gz archives at database scale. The codebase includes all prompt templates, tool orchestration scripts, GO/Pfam semantic enrichment modules, structured output schemas, and evaluation protocols described in this work.

An interactive web server for real-time protein functional annotation is publicly accessible at \href{https://seqstudio.opendatalab.org.cn/home}{https://seqstudio.opendatalab.org.cn/home}, enabling researchers to submit individual or batch protein sequences (in FASTA format) or custom PDB structures and obtain evidence-grounded functional reports, structured predictions with confidence scores, interactive 3D structure visualization, context-aware dialogue for follow-up queries, and automated knowledge graphs---all without local installation or computational infrastructure (see Section~2.5 for a detailed description).

\section*{Acknowledgments}
\label{sec:ack}
This work was supported by 2025 Local Supporting Project (PJ-PRJ24DATA001) by Shanghai Artificial Intelligence Laboratory, National Natural Science Foundation of China Project (No. 624B2115).

\section*{Author Contributions}

Y.L. designed and implemented the core SeqStudio computational framework, executed the large-scale UniProtKB annotations, conducted the majority of the performance evaluations, and wrote the original draft of the manuscript. T.X. contributed to the autonomous LLM-as-a-Judge evaluation pipeline and performed specific experimental case studies. Y.W. developed and deployed the SeqStudio interactive web server. J.X. designed and conducted the sequence-orphan and remote homology detection analysis. C.P. carried out the downstream foundation model instruction-tuning experiments. J.P. developed the automated knowledge graph visualization module for the web platform. L.S. and Q.L. designed and executed the adversarial rhodopsin screening experiments, including the construction of pseudo-rhodopsin decoy datasets and the blinded evaluation protocol. Y.Z., J.W., S.L., and J.C. provided critical revisions to the manuscript. X.Z., C.H., and J.N. supervised the research and revised the manuscript. C.T. conceived and supervised the research, provided strategic guidance and revised the manuscript. All authors read and approved the final manuscript.

\bibliography{ref}

\newpage

\begin{appendices}
\section{Implementation Details of SeqStudio}
\subsection{Data Input and Preprocessing}
The standard SeqStudio pipeline utilizes the \textbf{FASTA} format as the universal entry point. The string preceding the first space in the header line (starting with \texttt{>}) is parsed as the unique protein identifier, driving the execution of BLAST, InterProScan, and TMHMM. 

For structural comparisons, \textbf{PDB} files (e.g., AlphaFold predicted structures) serve as the input for Foldseek. The structural search is conditionally executed only if a PDB file is provided. If only a PDB file is provided without a corresponding FASTA sequence, the pipeline automatically extracts the amino acid sequence from the PDB coordinates to generate a temporary FASTA file, ensuring that sequence-based tools can still run seamlessly. In large-scale annotation scenarios, sequences and metadata can also be extracted directly from UniProtKB JSON entries to generate the necessary FASTA inputs.

\subsection{Configuration of Bioinformatics Tools}
To construct the multi-scale semantic context, we define specific roles and hyperparameters for each bioinformatics tool to ensure computational efficiency and high-fidelity evidence retrieval. A comprehensive overview of these tool versions, target databases, and core execution parameters is summarized in Table~\ref{tab:seqstudio-tools}.

\begin{table*}[htbp]
  \centering
  \caption{\textbf{Configuration and scope of core bioinformatics tools integrated within the SeqStudio pipeline.} Software versions, target databases, and primary execution parameters utilized for multi-modal evidence retrieval.}
  \label{tab:seqstudio-tools}
  \small
  \setlength{\tabcolsep}{1mm} %
  
  \begin{tabular}{@{}ll>{\raggedright\arraybackslash}p{4.5cm}>{\raggedright\arraybackslash}p{5cm}@{}}
    \toprule
    \textbf{Tool} & \textbf{Version} & \textbf{Target Database / Scope} & \textbf{Execution Configuration} \\
    \midrule
    BLAST+\\(\texttt{blastp}) & 2.16.0 & 
    \textbf{UniProtKB/Swiss-Prot}: Protein sequence homology search. & 
    E-value threshold $\leq 10^{-2}$; XML output format (\texttt{-outfmt 5}) for downstream metric parsing. \\
    \addlinespace
    InterProScan & 5.75-106.0 & 
    \textbf{Comprehensive Signature Libraries}: Domain architecture and motif identification (e.g., Pfam). & 
    JSON output mode (\texttt{-f JSON}); GO terms (\texttt{-goterms}) and Pathway (\texttt{-pa}) annotations enabled. \\
    \addlinespace
    Foldseek &8.2& 
    \textbf{AlphaFold Swiss-Prot Database}: Three-dimensional structural similarity and fold comparison. & 
    \texttt{easy-search} mode; E-value threshold $\leq 10^{-2}$; normalized tabular output. \\
    \addlinespace
    TMHMM & 2.0c & 
    \textbf{Transmembrane Topology}: Prediction of membrane-helix layout and physical localization constraints. & 
    Textual prediction mode (\texttt{-noplot}) to extract helical coordinates and topology status. \\
    \bottomrule
  \end{tabular}
  \raggedright
\end{table*}

\subsubsection{Sequence Homology (BLAST)}
\begin{itemize}
    \item \textbf{Database}: The manually reviewed Swiss-Prot database (\texttt{uniprot\_swissprot}).
    \item \textbf{Hyperparameters}: E-value threshold set to $0.01$. The output format is set to XML (\texttt{outfmt=5}) to accurately parse the ID, Identity\%, Coverage\%, E-value, and Bit Score of each hit.
    \item \textbf{Integration}: The top-$k$ hits are utilized for Gene Ontology (GO) propagation. GO terms associated with these homologous proteins are aggregated into the evidence pool, explicitly tagged with "BLAST" as the source and their corresponding E-values.
\end{itemize}

\subsubsection{Domain and Motif Analysis (InterProScan)}
\begin{itemize}
    \item \textbf{Hyperparameters}: Executed with \texttt{-f JSON} to extract specific IDs (e.g., Pfam, InterPro). We enable GO and Pathway annotations (\texttt{-goterms}, \texttt{-pa}). To accelerate large-scale processing, online pre-calculated matches (\texttt{-dp}) and residue-level annotations (\texttt{-dra}) are disabled. The sequence type is explicitly defined as protein (\texttt{-t p}).
    \item \textbf{Databases}: The standard pipeline activates comprehensive signature libraries including Pfam, PANTHER, CDD, SUPERFAMILY, PRINTS, SMART, etc. For lightweight discovery tasks, the pipeline can be restricted to Pfam only (\texttt{-appl Pfam}).
\end{itemize}

\subsubsection{Membrane Topology (TMHMM)}
\begin{itemize}
    \item \textbf{Hyperparameters}: Executed with the \texttt{-noplot} flag to generate text-only predictions. The textual output detailing the number of predicted transmembrane helices (TMHs) and topology coordinates is directly parsed into the LLM context to constrain spatial localization predictions.
\end{itemize}

\subsubsection{Structural Homology (Foldseek)}
\begin{itemize}
    \item \textbf{Hyperparameters}: Executed in the \texttt{easy-search} mode against the Swiss-Prot structure database (\texttt{foldseek\_db/sp}). The E-value threshold is set to $0.01$.
    \item \textbf{Integration}: Tabular outputs are normalized into metrics parallel to BLAST (Identity\%, Coverage\%, E-value) and mapped to protein IDs. These hits contribute to structural GO propagation.
\end{itemize}

\subsubsection{Gene Ontology (GO) Integration}
The GO integration module acts as the bridge between raw tool outputs and the LLM reasoning engine. It aggregates GO terms from BLAST, InterProScan, and Foldseek into a unified evidence list for each protein. To ensure semantic grounding, each GO term is appended with its official ontological definition, evidence type, reliability score, and source origin.

\subsection{Hierarchical Prompt Engineering Templates}
\label{app:prompt_engineering}

SeqStudio employs a two-round dialogue strategy. The first round synthesizes a natural language functional summary, while the second round forces the model to generate structured predictions grounded in the provided evidence.

\subsubsection{Round 1: Functional Summary Generation}
The model is provided with the aggregated evidence pool (grouped by tool origin) and prompted with the following instructions:

\begin{center}
\begin{promptbox}[The Prompt for Functional Summary Generation]
\small\ttfamily
You are a senior systems biologist specializing in protein function analysis. Your task is to analyze the provided information and write a concise functional summary for this protein.

\textbf{Analysis Guidelines}\\
1. \textbf{Review all evidence}: Domain annotations, GO terms, homology results, and membrane predictions. \\
2. \textbf{Synthesize information}: Identify consistent functional signals, resolve conflicts using majority evidence. \\
3. \textbf{Write a concise summary} covering:
\begin{itemize}
\small
    \item \textbf{Function}: What does this protein do? (molecular activity and biological role)
    \item \textbf{Enzyme activity}: If an enzyme, what reaction does it catalyze?
    \item \textbf{Localization}: Where in the cell does it function?
    \item \textbf{Key structural features}: Important domains, membrane association, etc.
\end{itemize}

\textbf{Output Format}\\
Write a well-structured paragraph that starts with the protein's identity/family and function(s), describes the molecular mechanism, mentions subcellular localization, notes key structural features, and indicates confidence level based on evidence consistency. Do NOT output step-by-step analysis. Write a direct, professional summary suitable for a protein database annotation.

\end{promptbox}
\label{prompt1_seqstudio}
\end{center}

\subsubsection{Round 2: Structured Prediction and Evidence Grounding}

Building upon the functional summary generated in Round 1, the second round enforces a strict schema for downstream parsing and auditability:

\begin{center}
\begin{promptbox}[The Prompt for Structured Prediction]

\small
You are a senior systems biologist. 
Based on the input information from various tools and the previous LLM analysis, extract structured predictions for the protein. You need to provide predictions for the following fields, each with a confidence score (0.0 to 1.0) and supporting evidence:

1. \textbf{PROTEIN\_FAMILY}: The protein family or superfamily name.\\
2. \textbf{FUNCTION}: A concise description of the protein's biological functions.\\
3. \textbf{ENZYME\_INFO}: Provide a JSON object with: "isEnzyme" (Yes/No/Unknown), "ecNumber" (using the latest post-2018 EC classification), and "description".\\
4. \textbf{PATHWAYS}: A list of biological pathways. Use [] if not determinable.\\
5. \textbf{SUBCELLULAR\_LOCATION}: The subcellular location(s).\\
6. \textbf{STRUCTURAL\_CLASS}: JSON with: "domains", "foldType" (All-$\alpha$, All-$\beta$, $\alpha$/$\beta$, $\alpha$+$\beta$, Small, Disordered, Unknown), "isMembrane" (Yes/No/Unknown), "tmHelices" (number from TMHMM).

\textbf{Output Format:} \\
You must output your predictions in the following exact format at the end of your response:\\
--- \\
\texttt{\#\#\# [STRUCTURED\_PREDICTION\_WITH\_EVIDENCE]} \\
\texttt{PROTEIN\_FAMILY: <value> | CONFIDENCE: <0.0-1.0> | SUPPORT: <evidence\_json>} \\
\texttt{FUNCTION: <value> | CONFIDENCE: <0.0-1.0> | SUPPORT: <evidence\_json>} \\
\texttt{ENZYME\_INFO: <json\_object> | CONFIDENCE: <0.0-1.0> | SUPPORT: <evidence\_json>} \\
\texttt{PATHWAYS: <JSON array> | CONFIDENCE: <0.0-1.0> | SUPPORT: <evidence\_json>} \\
\texttt{SUBCELLULAR\_LOCATION: <value> | CONFIDENCE: <0.0-1.0> | SUPPORT: <evidence\_json>} \\
\texttt{STRUCTURAL\_CLASS: <json\_object> | CONFIDENCE: <0.0-1.0> | SUPPORT: <evidence\_json>} \\
---

\textbf{Evidence Format:}\\
SUPPORT is a JSON object with: "motifs" (Pfam/InterPro IDs), "go\_terms" (GO IDs), "sources" (e.g., "InterProScan", "BLAST", "TMHMM").

\end{promptbox}
\end{center}

\subsection{Output Formatting}
\label{app:seqstudio_output}
To ensure that the generated annotations are both machine-readable and fully auditable, SeqStudio encapsulates all outputs, intermediate results, and metadata into a comprehensive, standardized JSON format. 

\textbf{Core Annotations:} The ultimate deliverables of the SeqStudio pipeline consist of two main components:
\begin{enumerate}
    \item \textbf{Functional Summary (\texttt{seqStudioSummary})}: A cohesive, expert-level natural language paragraph that synthesizes the protein's molecular mechanism, biological role, localization, and key structural features.
    \item \textbf{Structured Predictions (\texttt{seqStudioComments})}: A detailed, six-field prediction object encompassing \textit{Protein Family, Function, Enzyme Info, Pathways, Subcellular Location}, and \textit{Protein Structure}. Each field is rigorously quantified with a \texttt{confidence} score (ranging from 0.0 to 1.0) and backed by a \texttt{support} object that traces the specific evidence back to its source (e.g., specific motifs, GO terms, and tool sources like InterProScan or TMHMM).
\end{enumerate}

\textbf{Data Retention for Reproducibility:} Beyond the final annotations, SeqStudio emphasizes complete transparency. The final JSON object intentionally preserves the raw execution results from all bioinformatics tools (BLAST, InterProScan, TMHMM, Foldseek) alongside the complete LLM invocation history.

\textbf{Task-Specific Output Customization:} While the aforementioned functional summary and six-field structured predictions serve as the standard output, the SeqStudio framework is intrinsically adaptable to specific analytical objectives. By simply modifying the prompt instructions while utilizing the exact same underlying tool execution results, the LLM's reasoning objective can be redirected. For targeted biological discovery tasks—such as the rhodopsin screening application described in the main text—the prompt is dynamically tailored. In such scenarios, instead of generating a full functional profile, the LLM is instructed to return a definitive Boolean classification (e.g., \texttt{Answer: Yes/No}), an overall confidence score, and a concise mechanistic rationale citing specific structural or motif evidence. This zero-shot adaptability allows SeqStudio to seamlessly pivot from generalized annotation to high-throughput, specific protein mining without altering the core evidence-gathering pipeline.

\textbf{Large-Scale Database Annotation Schema:} For large-scale deployments (such as the 20M UniProtKB/TrEMBL dataset processing), the output is serialized into a highly structured JSON schema optimized for database integration. At the top level, the schema records metadata such as the pipeline \texttt{version} and the \texttt{generatedAt} timestamp. The \texttt{predictions} block houses the six structured fields, where each field inherently contains a \texttt{value} (which can be a string, list, or nested JSON object like \texttt{enzymeInfo}), its corresponding \texttt{confidence} metric, and the \texttt{support} evidence trail. This schema seamlessly bridges the gap between readable biological narratives and strict, queryable relational database structures.

\begin{center}
\begin{jsonbox}[Example of Large-Scale Database Annotation]
{
  "seqStudioComments": {
    "version": "1.0",
    "generatedAt": "2025-12-15T07:24:02.293117Z",
    "predictions": {
      "proteinFamily": {
        "value": "L-type lectin receptor kinase (LecRK) family",
        "confidence": 0.95,
        "support": {
          "motifs": [
            "IPR050528",
            "PF00139",
            "PF00069"
          ],
          "sources": [
            "InterProScan",
            "Pfam",
            "FunFam"
          ],
          "goTerms": [
            "GO:0016301",
            "GO:0030246"
          ]
        }
      },
      "function": {
        "value": "Functions as a transmembrane receptor kinase involved in carbohydrate-mediated signaling; biological roles include pollen development, pollen aperture formation, and defense responses against pathogens such as oomycetes.",
        "confidence": 0.85,
        "support": {
          "motifs": [
            "PF00139",
            "PF00069"
          ],
          "sources": [
            "InterProScan",
            "BLAST",
            "UniProt"
          ],
          "goTerms": [
            "GO:0009555",
            "GO:0062075",
            "GO:0002229",
            "GO:0046777"
          ]
        }
      },
      "enzymeInfo": {
        "value": {
          "isEnzyme": "Yes",
          "ecNumber": "2.7.11.1",
          "description": "Transferase (Protein-serine/threonine kinase) that catalyzes the phosphorylation of serine or threonine residues on target proteins using ATP."
        },
        "confidence": 0.95,
        "support": {
          "motifs": [
            "PF00069",
            "IPR000719",
            "IPR008271"
          ],
          "sources": [
            "InterProScan",
            "UniProt",
            "BLAST"
          ],
          "goTerms": [
            "GO:0004674",
            "GO:0016301",
            "GO:0106310",
            "GO:0005524"
          ]
        }
      },
      "pathways": {
        "value": [
          "Pollen development",
          "Plant defense signaling"
        ],
        "confidence": 0.75,
        "support": {
          "motifs": [],
          "sources": [
            "UniProt",
            "BLAST"
          ],
          "goTerms": [
            "GO:0009555",
            "GO:0062075",
            "GO:0002229"
          ]
        }
      },
      "subcellularLocation": {
        "value": "Plasma membrane",
        "confidence": 0.9,
        "support": {
          "motifs": [],
          "sources": [
            "UniProt",
            "TMHMM"
          ],
          "goTerms": [
            "GO:0005886"
          ]
        }
      },
      "proteinStructure": {
        "value": {
          "domains": [
            "PF00139 (Legume lectin domain)",
            "PF00069 (Protein kinase domain)"
          ],
          "foldType": "$\alpha$/$\beta$",
          "isMembrane": "Yes",
          "tmHelices": 2
        },
        "confidence": 0.95,
        "support": {
          "motifs": [
            "PF00139",
            "PF00069"
          ],
          "sources": [
            "InterProScan",
            "TMHMM",
            "SUPERFAMILY"
          ],
          "goTerms": [
            "GO:0016020"
          ]
        }
      }
    }
  },
  "seqStudioSummary": "This protein is an L-type lectin receptor kinase (LecRK), a transmembrane signaling protein that integrates carbohydrate recognition with intracellular kinase activity. It functions as a serine/threonine protein kinase, catalyzing the transfer of a phosphate group from ATP to serine or threonine residues on target proteins, including itself (autophosphorylation). Its extracellular legume lectin domain mediates carbohydrate binding, likely involved in sensing pathogen-associated molecular patterns or cell wall components, linking to biological roles in pollen development, aperture formation, and defense responses against oomycetes. The protein localizes to the plasma membrane, supported by a transmembrane topology prediction showing two transmembrane helices flanking an intracellular region containing the kinase domain and an extracellular region containing the lectin domain. Key structural features include an N-terminal legume lectin domain (ConA-like) for carbohydrate recognition and a C-terminal serine/threonine protein kinase domain for signal transduction. The evidence is highly consistent, with strong domain support and congruent functional annotations, indicating high confidence in this functional summary."
}
\end{jsonbox}
\end{center}

\section{Dataset Details}
\subsection{Large-Scale UniProt Annotation Dataset}
\label{app:uniprot_datasets}

To demonstrate the scalability, robustness, and discovery potential of our framework, we applied a specialized, high-throughput SeqStudio pipeline to annotate sequences from the UniProt knowledgebase at scale (the engineering and deployment details of this process are described in Appendix \ref{Uniprot pipeline}). All resulting datasets are formatted as enriched JSON records. These records preserve the original UniProt metadata (e.g., accession, features, existing GO terms) while appending SeqStudio's generative outputs—specifically the natural language \texttt{seqStudioSummary} and the evidence-grounded \texttt{seqStudioComments}. Furthermore, the complete raw execution results from all integrated bioinformatics tools (\texttt{toolResult}) are explicitly retained to ensure full reproducibility and auditability.

\begin{table}[h]
\centering
\caption{Summary of SeqStudio-annotated datasets.}
\label{tab:seqstudio_datasets_alt}
\setlength{\tabcolsep}{3mm}
\begin{tabular}{llr>{\raggedright\arraybackslash}p{5.8cm}@{}}
\toprule
\textbf{Dataset} & \textbf{Subsets} & \textbf{Count} & \textbf{Source / Description} \\
\midrule
10k & Swiss-Prot (part 0) & 10,000 & First partition of Swiss-Prot \cite{boutet2007uniprotkb}. \\
\midrule
0.57M & Swiss-Prot (full) & 575,661 & Full Swiss-Prot database.\\
\midrule
\multirow{12}{*}[2.5\baselineskip]{20M} & Swiss-Prot & 575,661 & Full Swiss-Prot (same as 0.57M). \\
 & TrEMBL5 & 323,099 & TrEMBL \cite{uniprot2025uniprot}, annotation\_score=5, full subset. \\
 & TrEMBL4 & 1,084,340 & TrEMBL, annotation\_score=4, full subset. \\
 & TrEMBL3 & 3,970,975 & TrEMBL, annotation\_score=3, length 50--400, non-fragment. \\
 & TrEMBL2 & 3,238,000 & TrEMBL,   annotation\_score=2, random from $\sim$12M filtered. \\
 & TrEMBL1 & 10,807,925 & TrEMBL, annotation\_score=1, random from $\sim$12M filtered. \\
\cmidrule(lr){2-3}
\multicolumn{2}{r}{Total 20M} & 20,000,000 & --- \\
\bottomrule
\end{tabular}
\end{table}

Based on the evaluation requirements and deployment stages, the annotation results are categorized into three progressive scales:

\textbf{1. The 10k Subset (Rapid Evaluation Set):} 
This subset comprises approximately 10,000 sequences sampled from the first data partition of the Swiss-Prot database. It was annotated using the high-quality, two-round dialogue configuration powered by the DeepSeek-V3.2 model. Due to its manageable size, this subset was primarily utilized for rapid framework evaluation, prompt tuning, and the ablation studies presented in the main text.

\textbf{2. The 570k Dataset (Complete Swiss-Prot Benchmark):} 
Encompassing the entirety of the manually reviewed UniProtKB/Swiss-Prot database ($\sim$575,000 entries), this dataset represents the gold-standard benchmark. It was processed utilizing the same high-quality, two-round reasoning pipeline (DeepSeek-V3.2) as the 10k subset. This comprehensive dataset serves as the foundation for the rigorous consistency evaluation and the macro-scale biological benchmarking discussed in the main text.

\textbf{3. The 20M Dataset (Large-Scale "Dark Matter" Corpus):} 
Representing a phased milestone in our ongoing effort to illuminate the unreviewed protein universe, this massive dataset merges the 570k Swiss-Prot benchmark with over 19.42 million unreviewed sequences from UniProtKB/TrEMBL. To optimize computational efficiency and minimize latency at this unprecedented scale, the TrEMBL subsets were processed using a streamlined, single-turn dialogue configuration powered by Gemini 2.5 Flash Lite. Importantly, this single-turn approach merely compresses the conversational flow; the final generated outputs adhere strictly to the identical comprehensive schema (functional summary and six-field structured predictions) as the two-turn method.

Crucially, the UniProtKB/TrEMBL database contains an overwhelming volume of sequences with extremely low annotation scores (scores of 1, 2, or 3). Due to this massive scale, we prioritized sequences by applying strict rule-based filtering prior to generative annotation to focus on high-value functional "dark matter." Specifically, we applied the following criteria to select uncharacterized but biologically plausible targets: \texttt{(reviewed:false) AND (length:[50 TO 400]) AND (annotation\_score:3) AND (fragment:false) AND (existence:1 OR 2 OR 3)}. As SeqStudio continues to annotate new TrEMBL sequences, this 20M dataset serves as a foundational, continuously expanding resource for large-scale protein functional discovery.

\subsection{Orphan Protein Dataset for Remote Homology Detection}
\label{sec:appendix_orphan}

The orphan protein dataset used in this study comprises 99 proteins compiled from two published benchmarks for single-sequence protein structure prediction (Table~\ref{tab:orphan_dataset}).

\begin{table}[h]
\centering
\caption{Summary of orphan protein datasets}
\label{tab:orphan_dataset}
\setlength{\tabcolsep}{1.8mm}
\begin{tabular}{lcccc}
\toprule
\textbf{Dataset} & \textbf{Proteins} & \textbf{Length (aa)} & \textbf{Selection criterion} & \textbf{Source} \\
\midrule
Orphan25 & 25 & 22--239 (median 42) & $N_{\mathrm{eff}} < 10$ & Wang et al.~\cite{wang2022single} \\
OrphanRGN2 & 74 & 6--97 (median 20) & MSA depth $= 1$ & Chowdhury et al.~\cite{chowdhury2022single} \\
\midrule
Combined & 99 & 6--239 (median 22) & --- & --- \\
\bottomrule
\end{tabular}
\end{table}

The Orphan25 dataset was curated by Wang et al. from PDB structures released after May 2020, selecting chains with effective sequence number $N_{\mathrm{eff}} < 10$, indicating few or no detectable homologs in sequence databases~\cite{wang2022single}. The OrphanRGN2 dataset was compiled by Chowdhury et al., consisting of 77 proteins with MSA depth of 1 (i.e., only the query sequence itself) when searched against UniRef30, PDB70, and MGnify databases~\cite{chowdhury2022single}. After removing three chains overlapping with Orphan25, 74 proteins from OrphanRGN2 were retained, yielding a combined set of 99 proteins.

Each protein has a corresponding experimentally determined structure deposited in the PDB. Of the 99 proteins, 56 (56.6\%) have UniProt accession mappings; the remaining 43 lack UniProt entries, primarily comprising viral, phage, or synthetic proteins.

To characterize the annotation status of these orphan proteins, we submitted all 99 sequences to SeqStudio, which integrates BLAST, Foldseek, and InterProScan as upstream search tools. Based on tool outputs, proteins were classified into two groups (Table~\ref{tab:orphan_annotation}):

\begin{table}[h]
\centering
\caption{Annotation coverage of orphan proteins by tool}
\label{tab:orphan_annotation}
\setlength{\tabcolsep}{10.5mm}
\begin{tabular}{lrr}
\toprule
\textbf{Category} & \textbf{Count} & \textbf{Percentage} \\
\midrule
Tool hits (total) & 32 & 32.3\% \\
\quad BLAST & 29 & 29.3\% \\
\quad Foldseek & 2 & 2.0\% \\
\quad InterProScan & 1 & 1.0\% \\
\midrule
No tool hits (Genuine Orphans) & 67 & 67.7\% \\
\midrule
Total & 99 & 100\% \\
\bottomrule
\end{tabular}
\end{table}

Proteins with at least one tool returning a hit (32/99, 32.3\%) were classified as ``Resolved Orphans''; the remaining 67 proteins (67.7\%) received no hits from any conventional tool and were classified as ``Genuine Orphans.'' For all 99 proteins, SeqStudio generated functional annotations by integrating tool outputs with LLM-based reasoning.

AlphaFold3 structure predictions were obtained for 98 of 99 proteins. Among these, 93 (95\%) achieved mean pLDDT scores $\geq$70, indicating high-confidence predicted structures. Structural validation against experimental PDB structures using US-align yielded TM-scores $>$0.5 for 41 proteins (41.8\%).

\subsection{Target and Decoy Sequence Pools for Rhodopsin Screening}
This section details the dataset construction, adversarial decoy generation, and the blinded execution protocol for the rhodopsin discovery experiment discussed in Section~\ref{sec:rhodopsin}.

\textbf{Positive dataset construction.} To construct a high-fidelity positive dataset of microbial rhodopsins, we initially compiled 183 experimentally validated type-I microbial rhodopsin sequences from UniProt and published literature. These seed sequences were utilized as anchors to expand the dataset via BLASTp searches (E-value cutoff $1 \times 10^{-5}$) against comprehensive databases including UniProt, Pfam, NCBI, KEGG, eggNOG, and MicRhoDE. From the pooled candidates, we retained only sequences functionally annotated as proton-pumping rhodopsins. To ensure strict structural integrity, the pooled candidates underwent rigorous expert-guided curation and validation to retain only those possessing the full functional architecture. The final curated positive set comprised 1,209 high-fidelity sequences. Domain annotation confirmed that all retained sequences contained the Bacteriorhodopsin-like domain (PF01036).

\textbf{Adversarial decoy sequence construction.} To evaluate the ability of SeqStudio to discriminate true functional rhodopsins from chimeric sequences containing rhodopsin-like fragments, we constructed two synthetic pseudo-rhodopsin datasets: High-identity fakes ($n=800$) and Low-identity fakes ($n=800$). For the high-identity fakes, the procedure for building each pseudo-sequence was as follows: (1) randomly sample three non-redundant sequences from the curated positive database; (2) extract a single random k-mer of length 90 amino acids from each sampled sequence; (3) concatenate the three 90-aa k-mers to produce one 270-aa pseudo-rhodopsin sequence. This process was repeated independently 800 times. For the low-identity fakes, we generated 800 sequences using a more fragmented approach: each pseudo-sequence was formed by concatenating five independently sampled 50-aa k-mers (extracted from five randomly selected positive sequences), yielding sequences of 250 amino acids in length. These pseudo-datasets simulate realistic chimeric cases encountered in genomic assemblies and short reads, where sequence segments share domain architecture with \textit{bona fide} rhodopsins, but the sequence as a whole lacks physiological functionality. Tool-based annotation confirmed that these decoys could weakly or partially match the PF01036 domain, making them highly deceptive negative controls for evaluating alignment-based and machine-learning classifiers.

\section{Large-scale Deployment and Inference Pipeline}
\label{Uniprot pipeline}
Processing millions of protein sequences requires transitioning from a standard, single-input integrated pipeline to a robust, high-throughput distributed architecture. This section details the engineering optimizations, parallelization strategies, and data formatting modifications implemented to scale SeqStudio for the 20M UniProtKB dataset, followed by a quantitative analysis of its temporal and economic efficiency.

\subsection{Architecture Transition and Deployment Modes}
The foundational SeqStudio pipeline was designed for single-file FASTA inputs, loading all sequences into memory and executing tools sequentially—a paradigm unsuitable for database-scale operations. To accommodate the UniProtKB corpus, we engineered a scalable \texttt{uniprot\_pipeline} characterized by batched processing and streaming execution. Depending on the downstream task, the pipeline dynamically switches between two deployment modes:
\begin{itemize}
    \item \textbf{High-Precision Mode (Swiss-Prot benchmark):} Utilizes DeepSeek-V3.2 with a two-turn reasoning flow. It operates with a batch size of 500 and an LLM concurrency of 50, achieving an average processing time of $<0.9$ seconds per sequence.
    \item \textbf{High-Throughput Mode (TrEMBL completion):} Employs Gemini 2.5 Flash Lite with a compressed single-turn reasoning flow. Operating with a larger batch size of 2000 and an LLM concurrency of 100, it reduces the processing time to $<0.3$ seconds per sequence, optimizing latency for massive-scale datasets.
\end{itemize}

\subsection{I/O Formatting and Fault-Tolerant Checkpointing}
To manage massive datasets without exceeding memory limits, the large-scale pipeline fundamentally restructures data input and output (I/O) handling. 
Instead of loading entire FASTA files, the pipeline utilizes iterative streaming (via \texttt{ijson}) to read compressed UniProt \texttt{JSON.gz} dumps. Temporary FASTA chunks are dynamically generated on-the-fly for bioinformatics tool execution. 

For outputs, rather than scattering results across multiple tool-specific directories, the pipeline aggregates the raw UniProt metadata, intermediate tool results, and the final LLM-generated \texttt{seqStudioComments} into a unified, self-contained \texttt{JSONL} (JSON Lines) format. To ensure fault tolerance during weeks-long processing runs, outputs are sharded (e.g., 2,000 records per file). A rigorous checkpointing mechanism records the completed batch indices and processed sequence counts. In the event of an API failure or node crash, the pipeline seamlessly resumes from the exact boundary of the last completed shard, skipping previously processed records and eliminating redundant computational overhead.

\subsection{Parallelization and Compute Optimization}
To maximize hardware utilization and eliminate redundant computations, the pipeline implements a multi-tiered optimization strategy:
\begin{enumerate}
    \item \textbf{Tool-Chain Multiprocessing:} Within each batch, CPU-intensive bioinformatics tools are heavily parallelized. BLAST searches are distributed across multiple processes (e.g., 4 processes $\times$ 256 threads), Foldseek conducts concurrent PDB downloads and structural alignments, and TMHMM topology predictions are executed across 32 parallel workers. When processing standard raw FASTA inputs via the integrated pipeline, InterProScan is also fully executed utilizing a built-in sequence chunking and multi-processing wrapper.
    \item \textbf{Pre-computed Data Extraction (InterProScan Bypass):} As a massive computational optimization specifically engineered for the large-scale UniProt pipeline, the system deliberately bypasses the execution of InterProScan. Because the downloaded UniProt \texttt{JSON.gz} dumps natively encompass comprehensively pre-computed InterPro domain signatures and GO annotations, the pipeline directly extracts and harmonizes these fields. This zero-compute retrieval saves thousands of CPU-hours per batch without compromising the depth of structural and functional evidence.
    \item \textbf{Adaptive LLM Concurrency:} Network-bound LLM invocations are managed via asynchronous concurrent requests. The pipeline features an adaptive concurrency controller that dynamically adjusts the number of simultaneous API requests (ranging from 50 to 100) based on real-time API rate limits and latency feedback. Failed requests are caught by an exponential backoff retry mechanism, significantly reducing the impact of transient network or server errors on large batches.
\end{enumerate}

\subsection{Temporal and Economic Efficiency Analysis}
To quantify the real-world utility of SeqStudio, we benchmarked its efficiency against traditional manual expert curation using a standardized sample of 10,000 sequences. 

\textbf{Temporal Acceleration:} Traditional manual curation—encompassing database searches, literature review, functional reasoning, and standardized data entry—requires approximately 35 minutes per protein. SeqStudio fully automates this workflow, reducing the turnaround time to $<0.9$ seconds (in High-Precision mode) or $<0.3$ seconds (in High-Throughput mode). This represents an order-of-magnitude acceleration, yielding a speedup of over 2,300$\times$.

\textbf{Cost-Benefit Analysis:} Assuming a standard bioinformatics consulting rate of \$50 per hour, manually annotating 10,000 proteins (5,833 hours) incurs a prohibitive cost of approximately \$291,650. In contrast, the SeqStudio hybrid model drastically reduces this financial barrier. The pure computational cost (LLM tokens and GPU inference) is approximately \$0.0008 per sequence, totaling \$8 for the cohort. By utilizing SeqStudio's confidence calibration to triage predictions, human review is only required for the $\sim$7\% of cases with confidence scores below 0.8. Reviewing these 700 flagged sequences (at 2 minutes per sequence) requires 23.3 expert hours, costing \$1,165. The combined SeqStudio and human-in-the-loop approach totals \$1,173—a 99.6\% reduction in overall annotation costs.

\section{Experimental and Evaluation Protocols}
\label{sec:evaluation_protocols}

\subsection{Swiss-Prot Consistency Evaluation and t-SNE Visualization Setup}
\label{app:swissprot_setup}
This section details the experimental design, ground-truth (GT) mapping, evaluation criteria, and dimensionality reduction hyperparameters used for the systematic evaluation of the Swiss-Prot database ($\sim$570,000 entries), corresponding to Section \ref{sec:swissprot} and Figure \ref{fig:swissprot_bench} in the main text.

\textbf{Annotation Pipeline and Self-Exclusion Protocol.} The full-database annotation was executed using the DeepSeek-V3.2 model with a two-turn reasoning flow and a batch size of 500. Crucially, to rigorously evaluate the system's reasoning capacity and prevent self-referential bias, the query sequence was strictly excluded from its own search results during all evidence-gathering steps (e.g., BLAST and Foldseek). 

\textbf{Ground-Truth Extraction and Field Mapping.} To establish a rigorous baseline for evaluation, the six structured prediction fields were dynamically mapped to specific metadata extracted from the raw UniProt JSON records. As detailed in Table \ref{tab:gt_mapping}, each evaluation field relies on designated extraction sources. The extraction process evaluates the presence of these annotations to assign a GT status: \texttt{AVAILABLE} (comprehensive primary annotation exists), \texttt{PARTIAL} (only secondary or partial sources exist, e.g., only Pfam cross-references), or \texttt{ABSENT} (no corresponding annotation). Only fields with an \texttt{AVAILABLE} or \texttt{PARTIAL} status were included in the denominator for consistency calculations, ensuring the system was not penalized for lacking GT data.

\begin{table}[!htbp]
\centering
\caption{Mapping and extraction rules for Swiss-Prot Ground Truth (GT).}
\label{tab:gt_mapping}
\setlength{\tabcolsep}{8mm}
\begin{tabular}{l>{\raggedright\arraybackslash}m{7cm}}
\toprule
\textbf{Evaluation Field} & \textbf{UniProt JSON Extraction Source} \\
\midrule
Protein Family & \texttt{comments[type='SIMILARITY']} \newline \texttt{keywords[category='Domain']} \\
\midrule
Function & \texttt{comments[type='FUNCTION']} \\
\midrule
Enzyme Info & \texttt{ecNumbers} \newline \texttt{comments[type='CATALYTIC ACTIVITY']} \\
\midrule
Pathways & \texttt{comments[type='PATHWAY']} \\
\midrule
Subcellular Location & \texttt{comments[type='SUBCELLULAR LOCATION']} \\
\midrule
Structural Class & \texttt{features[type='Domain']} \newline \texttt{uniProtKBCrossReferences[database='Pfam']} \\
\bottomrule
\end{tabular}
\end{table}

\textbf{Rating Criteria for Consistency.} Building upon this extraction framework, the consistency evaluation was performed exclusively on the fields possessing a valid GT (\texttt{AVAILABLE} or \texttt{PARTIAL}). For each qualifying field, an LLM-based evaluator compared the SeqStudio prediction against the extracted UniProt reference text and classified the output into one of five categories:
\begin{itemize}
    \item \textbf{MATCH:} Highly consistent with the reference; core information is identical.
    \item \textbf{CORRECT:} Fundamentally accurate, though minor details or phrasing may differ.
    \item \textbf{INCOMPLETE:} Factually correct but omits important information present in the GT.
    \item \textbf{VAGUE:} Overly broad or generic; lacks sufficient actionable information.
    \item \textbf{ERROR:} Contains factual contradictions or clear hallucinations.
\end{itemize}
A prediction was deemed ``Correct'' if classified as MATCH, CORRECT, or INCOMPLETE. Predictions classified as VAGUE or ERROR were deemed ``Incorrect''. If the system returned an ``Unknown'' prediction (or confidence = 0) despite a GT existing, it was classified as an omission (\texttt{UNKNOWN\_PRED}) and penalized as an error. Overall \textit{Consistency} was mathematically defined as the ratio of Correct predictions to the total number of fields possessing a GT. Cases where the system made a confident prediction but no GT existed were classified as \texttt{EXTENSIONs}; these were excluded from consistency calculations but tracked to quantify novel functional discoveries.

\textbf{Completeness and Hard Cases Definition.} Annotation completeness (Figure \ref{fig:swissprot_bench}a) was quantified by comparing the number of filled fields before and after annotation. The baseline (Before) was defined as the number of available GT fields (out of 6) in the original Swiss-Prot record. The annotated density (After) included all valid SeqStudio predictions (confidence $>0$, excluding ``Unknown''), encompassing both matching GTs and newly generated \texttt{EXTENSIONs}. To stress-test the system, ``Hard Cases'' were strictly defined as proteins possessing $\le 3$ filled GT fields in their original Swiss-Prot records, forcing the model to rely predominantly on computed computational evidence rather than pre-existing metadata.

\textbf{Semantic Embedding and t-SNE Hyperparameters.} For the semantic space visualization (Figure \ref{fig:swissprot_bench}g), we extracted and concatenated the text from the six structured fields for both the GT records and the SeqStudio predictions (excluding the free-text summary to ensure a strict comparison of structured knowledge). Both sets of texts were embedded using the \texttt{Qwen3-Embedding-8B} model. The resulting high-dimensional embeddings were standardized (zero mean, unit variance via \texttt{StandardScaler}) and projected into a 2D space using t-Distributed Stochastic Neighbor Embedding (t-SNE). To guarantee exact reproducibility, the t-SNE algorithm was initialized with PCA and executed with the following fixed hyperparameters: \texttt{n\_components=2}, \texttt{perplexity=40}, \texttt{learning\_rate='auto'}, \texttt{early\_exaggeration=20.0}, \texttt{max\_iter=3000}, \texttt{method='barnes\_hut'}, and \texttt{random\_state=42}.

\subsection{Baselines and Setup for the Evidence Ablation Study}
\label{app:ablation_setup}

To quantitatively dissect the marginal contribution of individual multi-modal evidence sources, we conducted an extensive evidence ablation study on a 10,000-sequence Swiss-Prot subset (the 10k dataset). This section details the experimental configurations, fallback strategies, and the comprehensive numerical results corresponding to the radar chart (Figure \ref{fig:ablation1}) in the main text.

\textbf{Experimental Design and Fallback Strategy.} The ablation framework isolates the impact of specific bioinformatics tools (BLAST, UniRule, InterProScan, Foldseek, and TMHMM) by systematically masking their outputs in the LLM prompt. To prevent confounding variables, all ablation groups utilized the same underlying model (DeepSeek-V3.2), the identical two-turn reasoning prompt templates, and the exact same set of ground-truth (GT) sequences as the main evaluation. 

During the prompt construction phase for each ablation group, a strict masking filter was applied. If a sequence yielded no outputs from the permitted tools for a specific group, the system employed a fallback strategy: it provided the LLM solely with the raw amino acid sequence and basic derived physicochemical features (sequence length, molecular weight, and amino acid composition). Consequently, the \textit{LLM Only} baseline relies exclusively on this fallback information without any external database retrieval, establishing the lower bound of performance derived purely from the foundation model's internal parametric memory.

\textbf{Ablation Results and Analysis.} The evaluation criteria for accuracy remain strictly identical to the consistency metrics defined in Appendix \ref{app:swissprot_setup} (i.e., the percentage of MATCH, CORRECT, or INCOMPLETE predictions among fields possessing a valid GT). The comprehensive numerical results across all six functional dimensions are presented in Table \ref{tab:ablation-accuracy}.

\begin{table}[!htbp]
\centering
\caption{\textbf{Evidence ablation study: accuracy (\%) across annotation fields.} The table details the step-wise performance gains achieved by incrementally providing the reasoning engine with multi-source molecular evidence.}
\label{tab:ablation-accuracy}
\setlength{\tabcolsep}{1.7mm}
\begin{tabular}{@{} lcccccc @{}}
\toprule
\textbf{Configuration} & 
\makecell{\textbf{Protein} \\ \textbf{Family}} & 
\textbf{Function} & 
\makecell{\textbf{Enzyme} \\ \textbf{Info}} & 
\textbf{Pathways} & 
\makecell{\textbf{Subcellular} \\ \textbf{Location}} & 
\makecell{\textbf{Structural} \\ \textbf{Class}} \\
\midrule
LLM Only & 8.68 & 12.76 & 4.54 & 0.82 & 61.45 & 34.21 \\ 
\addlinespace %
LLM + BLAST & 30.99 & 54.54 & 21.16 & 25.51 & 90.03 & 37.67 \\ 
\addlinespace
LLM + InterPro & 92.98 & 90.61 & 73.64 & 77.35 & 84.60 & 99.87 \\ 
\addlinespace
LLM + UniRule & 38.08 & 57.16 & 50.77 & 54.09 & 57.87 & 31.49 \\ 
\addlinespace
\makecell[l]{LLM + BLAST + \\ UniRule} & 57.30 & 81.68 & 63.19 & 77.60 & 90.81 & 51.43 \\ 
\addlinespace
\makecell[l]{LLM + BLAST + \\ UniRule + Foldseek} & 62.39 & 83.96 & 65.36 & 79.39 & 92.30 & 60.24 \\ 
\addlinespace
\makecell[l]{LLM + BLAST + \\ UniRule + InterPro} & 93.37 & 92.78 & 77.47 & 86.51 & 94.03 & 99.70 \\ 
\addlinespace
\makecell[l]{LLM + BLAST + \\ UniRule + InterPro \\ + Foldseek} & \textbf{93.36} & 92.91 & 77.87 & 86.94 & 93.94 & \textbf{99.87} \\ 
\midrule
\textbf{SeqStudio (Full)} & 93.23 & \textbf{94.81} & \textbf{80.27} & \textbf{89.41} & \textbf{96.94} & 99.70 \\
\bottomrule
\end{tabular}
\end{table}

The numerical data clearly illustrates the synergistic necessity of multi-modal evidence. The \textit{LLM Only} baseline exhibits near-random performance in mechanistic fields such as Pathways (0.82\%) and Enzyme Info (4.54\%). While single-tool augmentations like \textit{LLM + InterPro} or \textit{LLM + BLAST} provide substantial improvements in specific domains, they fail to achieve uniform high fidelity. The full \textit{SeqStudio} configuration—which uniquely incorporates TMHMM for membrane topology and orchestrates all available evidence streams—achieves the global maximum across nearly all categories, culminating in a 96.94\% accuracy in Subcellular Location and 89.41\% in Pathways.

\subsection{Temporal Evaluation on Newly Released Swiss-Prot Entries}
\label{app:temporal_generalization}

A critical challenge in evaluating Large Language Models (LLMs) on established biological databases is the risk of data leakage—high performance might stem from the model memorizing training data rather than performing genuine mechanistic reasoning. To assess performance on newly released Swiss-Prot entries and compare tool-augmented prediction with sequence-only baselines, we designed a temporal evaluation, corresponding to Section \ref{sec:ablation2} and Figure \ref{fig:ablation2} in the main text.

\textbf{Dataset Curation (The 2025 Swiss-Prot Benchmark).} We constructed a temporal benchmark by querying the UniProtKB/Swiss-Prot database exclusively for newly reviewed entries created in the year 2025. The exact query utilized was: \texttt{reviewed:true AND date\_created:[2025-01-01 TO 2025-12-31]}. This filtering yielded 1,057 protein sequences associated with Swiss-Prot entries first released in 2025. The selection criterion concerns entry creation dates, not verified absence from model training corpora. The same sequences or related annotations may have been available from other sources before these entries were created; this evaluation therefore does not establish freedom from training-data exposure.

\textbf{Experimental Setup and Strict Tool-Deprivation.} To quantify the reliance on external tool evidence versus internal parametric memory, we evaluated two distinct experimental paradigms on the exact same 1,057 sequences:
\begin{enumerate}
    \item \textbf{SeqStudio (Ours):} The complete integrated pipeline, executing the full suite of bioinformatics tools (BLAST, InterProScan, Foldseek, TMHMM) and performing the standard two-turn reasoning flow utilizing DeepSeek-V3.2.
    \item \textbf{Naive LLM Baselines:} A strict tool-deprived setting where three state-of-the-art frontier models (GPT-5, Gemini-3-Pro-Preview, and DeepSeek-V3.2) were provided \textit{only} with the raw amino acid sequence. All external bioinformatics retrievals were disabled. 
\end{enumerate}

\textbf{Prompt Design for Naive LLMs.} To ensure that the baseline models were fully guided to leverage their internal biochemistry knowledge without relying on external tool inputs, we designed a specific zero-shot reasoning prompt. The core instructions provided to the Naive LLMs were structured as follows:

\begin{center}
\begin{promptbox}[The Prompt for Naive LLMs]
\textbf{System Role:} You are a senior systems biologist specializing in protein function analysis. Your task is to analyze the provided protein sequence and write a \textbf{concise functional summary} for this protein.
\textbf{Analysis Guidelines:}
\begin{enumerate}
    \item \textbf{Analyze the sequence directly:} Examine the amino acid sequence patterns, composition, and potential motifs.
    \item \textbf{Infer from sequence properties:} Use your knowledge of protein biochemistry to identify potential functional domains based on sequence patterns, amino acid composition hints (e.g., hydrophobic regions, charged clusters), sequence length and complexity implications, and potential post-translational modification sites.
    \item \textbf{Apply biological knowledge:} Leverage your understanding of protein structure-function relationships to predict molecular activities based on sequence characteristics, potential enzyme activities from catalytic motifs, subcellular localization signals, and structural features and domain organization.
    \item \textbf{Write a concise summary covering:} \textbf{Function} (molecular activity and biological role), \textbf{Enzyme activity} (if it appears to be an enzyme, what type of reaction might it catalyze), \textbf{Localization} (membrane, cytoplasm, nucleus, etc.), and \textbf{Key structural features} (inferred domains or structural elements).
\end{enumerate}
\end{promptbox}
\end{center}

\textbf{Evaluation Metrics.} To guarantee a fair and standardized comparison, the outputs generated by these naive baselines were formatted into the identical JSON schema used by SeqStudio. The predicted annotations from all four configurations (SeqStudio and the three Naive LLMs) were then evaluated against the official 2025 Swiss-Prot records using the exact same automated consistency rating criteria detailed in Appendix \ref{app:swissprot_setup}.

\subsection{Protein-Centric QA Dataset and Autonomous Judge Evaluation}
\label{sec:app_judge}

This section details the construction of the protein-centric question-answering (QA) dataset and the evaluation protocols used to benchmark SeqStudio's capacity as an autonomous quality auditor (as discussed in Section~\ref{sec:judge}). To prevent circularity during evaluation, all sequences selected for this benchmark were strictly excluded from SeqStudio's training or prompt-tuning data, and any SeqStudio-generated fields (e.g., \texttt{seqStudioComments}, \texttt{seqStudioSummary}) were excluded from the ground-truth construction process.

\textbf{Dataset construction and cognitive dimensions.} The QA dataset was derived from a curated subset of high-confidence UniProtKB/Swiss-Prot entries (manually reviewed, annotation score of 3 or higher). We generated a total of 300 open-ended questions, balanced evenly across three distinct cognitive dimensions (100 questions each) to test different facets of functional understanding:
\begin{enumerate}
    \item \textbf{Core Function Identification:} Evaluates the understanding of primary biochemical activities (e.g., ``What is the primary function of this protein?''). Ground truth was synthesized from \texttt{proteinDescription} (recommended name, EC numbers), \texttt{comments} (FUNCTION), and InterProScan Gene Ontology molecular function terms (\texttt{GO:F}).
    \item \textbf{Functional Classification:} Evaluates accurate assignment to protein families and structural classes (e.g., ``Which protein family or superfamily does this protein belong to?''). Ground truth relied on PANTHER, NCBIfam, CATH/Gene3D, and SUPERFAMILY/SCOP classifications.
    \item \textbf{Functional Specificity:} Evaluates precise knowledge of biological pathways and mechanistic details (e.g., ``Which specific pathway or biological process is this protein involved in?''). Ground truth was extracted from \texttt{GO:P} terms, Pfam domain descriptions, and \texttt{comments} (PATHWAY, CATALYTIC ACTIVITY).
\end{enumerate}
Each prompt was formatted identically, beginning with the raw amino acid sequence: \textit{``Given the protein sequence \{sequence\}, [Question]''}.

\textbf{Gold-standard synthesis and quality control.} To establish a rigorous baseline, gold-standard answers for each of the 300 questions were manually synthesized exclusively from established molecular metadata (e.g., Swiss-Prot comments, GO terms, InterProScan signatures). These reference answers were subjected to strict quality control to ensure multi-database consensus, hierarchical completeness, and factual accuracy. Crucially, these gold-standard answers were used \textit{only} to establish the human expert baseline; they were completely hidden from SeqStudio during its autonomous auditing phase.

\textbf{Evaluation protocol and scoring scale.} To generate a diverse pool of responses, the 300 questions were submitted to 12 mainstream foundation models (spanning various scales and architectures), using only the raw sequence as input with no additional system prompts. This yielded a total of 3,600 unique (question, model answer) pairs. Each of the 3,600 answers was scored independently by two distinct judges using a strict ternary scale:
\begin{itemize}
    \item \textbf{+1 (Correct/Valid):} Factually accurate, complete, and precise identification of the protein's name, catalytic activity, substrates, and pathway.
    \item \textbf{0 (Partially Correct):} No clear factual errors, but missing key details (e.g., identifying the broad enzyme class but missing substrate specificity or localization).
    \item \textbf{-1 (Incorrect/Hallucination):} Contains factual errors, severe confabulations, or misidentifications (e.g., wrong protein family, invented domains, or incorrect metabolic pathways).
\end{itemize}

\textbf{Expert (human) evaluation arm.} Human experts (assisted by a strict, fixed rubric) conducted an ``open-book'' evaluation, scoring the foundation models' answers against the synthesized gold-standard reference answers. This yielded the \texttt{score\_gold} baseline. 

\textbf{SeqStudio autonomous evaluation arm.} SeqStudio evaluated the same 3,600 answers in a ``blind'' manner relative to the gold standard. Instead of comparing the model's answer to an answer key, SeqStudio utilized its internal reasoning engine, grounding its judgment purely in the raw evidence retrieved from its integrated tools (BLAST, InterProScan, TMHMM, Foldseek). This yielded the \texttt{score\_seqstudio} evaluations. The resulting scores were aggregated to compute both the macro-level Pearson correlation across the 12 models and the micro-level confusion matrix for per-sample agreement, as reported in the main text.

\subsection{Rhodopsin Discovery Evaluation}

\textbf{Blinded execution and evaluation protocol.} To ensure a rigorous and unbiased evaluation, the screening task was executed using a strict blinding protocol. First, the 1,209 true rhodopsins and 1,600 decoys were merged into a single FASTA file. During this process, their original identifiers were completely masked and replaced with anonymous headers (e.g., \texttt{Protein\_1}, \texttt{Protein\_2}). The mapping between the anonymous ID and the ground-truth label (\texttt{true}, \texttt{fake\_90}, \texttt{fake\_50}) was stored in an encrypted manifest file that was strictly inaccessible to the reasoning model. Next, for each anonymized sequence, the standard bioinformatics suite (BLAST, InterProScan, TMHMM) was executed. The outputs—including top homologous hits, Gene Ontology terms, Pfam domain signatures, and predicted transmembrane coordinates—were aggregated into a standardized evidence block. The language model was then prompted with the anonymized sequence and the aggregated evidence block. The system prompt explicitly instructed the model to perform a Boolean classification task, outputting a definitive \texttt{Answer: Yes/No} regarding whether the sequence is a true functional rhodopsin, accompanied by a \texttt{Confidence} score (0.0--1.0) and a concise mechanistic \texttt{Reason}. The framework evaluated the presence of the 7TM topology and specifically checked for conserved motifs (e.g., the DxxxK motif on TM7) using the TMHMM coordinates. Following the completion of the generative predictions, the outputs were unblinded against the manifest file. Performance metrics (Precision, Recall, Accuracy, F1 Score, and False Positive Rate) were calculated based on the ground-truth labels. The Predicted Positive Rate (PPR) was also tracked across the three sub-groups to ensure the model successfully accepted the true positives while rejecting both tiers of adversarial fakes.

\subsection{Evaluation of Remote Homology and Multi-Domain Logic}
\label{sec:appendix_remote_homology_logic}

This section describes the experimental procedures and evaluation metrics used in Sections~\ref{sec:remote_homologs} and~\ref{sec:molecular_logic}.

\subsubsection{Remote Homology Detection}

\textbf{Dataset and structure prediction.}
The evaluation used 99 orphan proteins compiled from two published benchmarks (Appendix~\ref{sec:appendix_orphan}). AlphaFold3 structure predictions were obtained for 98 of the 99 proteins. Predicted structures with mean pLDDT scores $\geq$70 (93/98, 95\%) were retained for downstream structural comparison.

\textbf{All-versus-all structural comparison.}
Pairwise structural similarity among the orphan proteins was assessed using Foldseek. All predicted structures were compared in an all-versus-all manner using the following command:

\begin{verbatim}
foldseek easy-search structures/ structures/ result.m8 tmp \
    --format-output "query,target,fident,alntmscore"
\end{verbatim}

\noindent The output fields include query identifier, target identifier, fractional sequence identity (\texttt{fident}), and alignment TM-score (\texttt{alntmscore}). Self-hits were excluded, and redundant pairs (A--B and B--A) were merged by retaining only one entry per unordered pair. This procedure yielded 53 unique protein pairs with detectable structural similarity.

\textbf{Remote homolog identification.}
Protein pairs were classified based on two thresholds: sequence identity $<$20\% and TM-score $>$0.5. Pairs satisfying both criteria were designated as remote homologs, reflecting high structural similarity despite low sequence conservation. A TM-score of 0.5 is a widely used threshold indicating that two structures share the same overall fold. A sequence identity of 20\% corresponds to the ``twilight zone'' below which sequence-based methods have limited sensitivity.

\textbf{Functional annotation comparison metrics.}
To quantify the consistency of SeqStudio annotations between structurally similar protein pairs, the following metrics were computed:

\begin{itemize}
    \item \textbf{GO term Jaccard similarity.} For each protein pair, the Jaccard index was calculated as $J_{\mathrm{GO}} = |G_A \cap G_B| / |G_A \cup G_B|$, where $G_A$ and $G_B$ denote the sets of Gene Ontology terms assigned by SeqStudio to proteins A and B, respectively.

    \item \textbf{Semantic similarity.} Functional descriptions and summary texts generated by SeqStudio were encoded using the \texttt{all-MiniLM-L6-v2} sentence transformer model. Pairwise cosine similarity was computed between the embedding vectors of corresponding text fields. Two variants were reported: Sem.\ Sim (Func) for functional descriptions and Sem.\ Sim (Summ) for summary annotations.

    \item \textbf{Keyword Jaccard similarity.} Keywords were extracted from SeqStudio functional annotations for each protein. The Jaccard index was calculated as $J_{\mathrm{KW}} = |K_A \cap K_B| / |K_A \cup K_B|$, where $K_A$ and $K_B$ denote the keyword sets for proteins A and B.
\end{itemize}

\subsubsection{Molecular Logic Analysis}

\textbf{Protein selection.}
Seven multi-domain proteins were selected for molecular logic evaluation (Table~\ref{tab:molecular_logic_proteins}). Selection criteria included: (1) presence of multiple functionally distinct domains, (2) availability of experimentally validated mechanistic models in the literature, and (3) representation of diverse functional classes (CRISPR nucleases, bacterial toxins, metabolic enzymes). Protein lengths ranged from 567 to 2,225 amino acids.

\textbf{Experimental setup and baselines.}
To determine the extent to which functional logic reconstruction relies on parametric memory versus evidence-driven reasoning, we established a controlled experiment with three configurations:
\begin{itemize}
    \item \textit{Naive LLM baseline:} The backbone large language model was provided with only the raw amino acid sequence and asked to predict the protein's identity and mechanism. This tests the limits of the model's internal parametric knowledge without external tool grounding.
    \item \textit{InterPro-only baseline:} Each protein sequence was submitted to InterProScan (v5.x). InterProScan integrates multiple signature databases (Pfam, SMART, CDD, etc.) to identify conserved domains. The output consists of domain identifiers and positional coordinates, which were evaluated for their ability to elucidate functional relationships.
    \item \textit{SeqStudio (Full):} The complete framework integrates upstream tool outputs (BLAST, Foldseek, TMHMM, InterProScan) with LLM-based reasoning. This tests the capacity of the system to synthesize multi-modal evidence into coherent inter-domain mechanistic narratives.
\end{itemize}

\textbf{Evaluation criteria.}
The outputs from all three configurations were evaluated for the presence of accurate mechanistic relationship terms. Four categories of logic keywords were defined: (1) \textit{causal} (triggers, enables, activates, inhibits, mediates); (2) \textit{sequential} (first, then, subsequently, following); (3) \textit{mechanistic} (via, through, by, whereby); and (4) \textit{conditional} (upon, when, requires, dependent on). A prediction was marked as successful (indicated by a checkmark) only if it correctly identified the protein's primary function and explicitly described the accurate inter-domain logic. The accuracy of these mechanistic descriptions was rigorously verified by manual comparison with published structural and biochemical studies for each protein.

\subsection{Utility of SeqStudio Annotations for Downstream Model Fine-Tuning}
\label{app:finetuning}

To demonstrate the utility and value of the large-scale annotations generated by SeqStudio, we conducted an instruction-tuning experiment to evaluate whether this data can enhance the domain-specific reasoning capabilities of foundation models. 

\textbf{Instruction-tuning dataset construction.} We transformed the high-confidence annotations from the 570k Swiss-Prot dataset into a structured Question-Answering (QA) format. For each protein, we generated up to six questions corresponding to the core functional dimensions (Protein Family, Function, Enzyme Info, Pathways, Subcellular Location, and Structural Class). To ensure strictly high-quality training signals, we applied rigorous filtering: any predictions with a confidence score below 0.8, empty values, or ``Unknown'' labels were discarded. Crucially, to prevent data leakage during evaluation, any sequences belonging to the 2025 Swiss-Prot release were explicitly excluded from the training corpus based on their unique accessions. 

\textbf{Experimental setup and evaluation.} We utilized the Qwen3-32B foundation model for this fine-tuning experiment. The temporally held-out 2025 Swiss-Prot subset (1,057 proteins) served as the strict zero-shot test set, yielding 5,005 valid QA pairs after filtering. We evaluated both the Base Qwen3-32B model and the Fine-tuned Qwen3-32B model in a tool-grounded context setting (i.e., providing the raw amino acid sequence alongside the upstream tool execution results as context). The outputs were autonomously evaluated using GPT-4o as a judge, applying the identical strict ternary rubric ($-1$: Error/Hallucination, $0$: Incomplete/Vague, $+1$: Valid/Accurate) defined in Section~\ref{sec:app_judge}.

\begin{table}[!htbp]
\centering
\caption{\textbf{Impact of SeqStudio-derived Fine-Tuning on Model Performance.} Evaluation on the temporally held-out 2025 Swiss-Prot QA dataset ($n=5,005$). Fine-tuning the Qwen3-32B model on high-confidence SeqStudio annotations significantly improves its capacity to generate accurate, valid biological descriptions when provided with tool-grounded context.}
\label{tab:finetune_results}
\setlength{\tabcolsep}{5mm}
\begin{tabular}{lcccc}
\toprule
\textbf{Model Configuration}  & \textbf{Valid} & \textbf{Incomplete} & \textbf{Error} & \textbf{Avg Score} \\
\midrule
Base Qwen3-32B       &  3,005 & 1,799 & \textbf{201} & 0.562 \\
Fine-tuned Qwen3-32B  & \textbf{3,635} & \textbf{1,132} & 238 & \textbf{0.679} \\
\bottomrule
\end{tabular}
\end{table}

\textbf{Results.} As summarized in Table~\ref{tab:finetune_results}, training on SeqStudio's annotations yielded substantial improvements. The fine-tuned model generated significantly more fully valid and accurate responses (3,635 vs. 3,005) while markedly reducing the number of vague or incomplete answers (1,132 vs. 1,799). Consequently, the overall average score increased from 0.562 to 0.679. These results empirically validate that SeqStudio serves not only as an annotation pipeline but also as a high-fidelity data engine. Its evidence-grounded outputs can be directly utilized as high-quality instruction-tuning corpora to align foundation models with expert-level biological reasoning.

\end{appendices}

\end{document}